\documentclass[
	aps, prd, reprint,
	10pt, notitlepage, a4paper,
        floats, floatfix,
	amsmath, amssymb, amsfonts, eqsecnum,
	superscriptaddress,
	showpacs, showkeys,
	nofootinbib,
 	longbibliography,
]{revtex4-2}

\usepackage{wrapfig}
\usepackage{graphicx} 
\usepackage[usenames,dvipsnames]{xcolor}
\usepackage{xspace} 
\usepackage{bm} 
\usepackage{amsmath,amsfonts,amssymb,amsthm,mathtools}
\usepackage[utf8]{inputenc} 

\xdefinecolor{mylinkcolor}{rgb}{0,0,0.5}
\usepackage[
	bookmarksnumbered, bookmarksopen, bookmarksopenlevel=2,
	breaklinks=true,
	colorlinks=true, filecolor=mylinkcolor, citecolor=mylinkcolor,
	linkcolor=mylinkcolor, urlcolor=mylinkcolor, menucolor=mylinkcolor,
]{hyperref}

\usepackage{verbatim}
\usepackage{mathrsfs}
\usepackage{color}
\usepackage{enumitem}
\usepackage[dvipsnames]{xcolor}
\usepackage{comment}
\usepackage{tikz}
\usetikzlibrary{decorations.pathmorphing}
\usetikzlibrary{arrows}
\newcommand{\midarrow}{\tikz \draw[-triangle 90] (0,0) -- +(.1,0);}
\tikzset{snake it/.style={decorate, decoration=snake}}

\usepackage[normalem]{ulem}

\newcommand{\nnm}{\nonumber}
\newcommand{\doe}{\partial}
\newcommand{\be}{\begin{equation}}
\newcommand{\ee}{\end{equation}}
\newcommand{\bse}{\begin{subequations}}
\newcommand{\ese}{\end{subequations}}

\newcommand{\mr}{\mathrm}

\newcommand{\mc}{\mathcal}
\newcommand{\py}{\phantom{yo}}
\newcommand{\bs}{\boldsymbol}

\newcommand{\bpm}{\begin{pmatrix}}
\newcommand{\epm}{\end{pmatrix}}

\newcommand{\AEI}{\affiliation{Max Planck Institute for Gravitational Physics (Albert Einstein Institute), Am M\"uhlenberg 1, Potsdam 14476, Germany}}
\newcommand{\Maryland}{\affiliation{Department of Physics, University of Maryland, College Park, MD 20742, USA}}

\newcommand{\comm}[1]{\textcolor{red}{[#1]}}
\renewcommand{\comm}[1]{}

\begin{document}

\title{Conservative and radiative dynamics in classical relativistic scattering \\ and bound systems}

\author{M. V. S. Saketh}
\email{msaketh@aei.mpg.de}\Maryland\AEI
\author{Justin Vines}\email{justin.vines@aei.mpg.de}\AEI
\author{Jan Steinhoff}\email{jan.steinhoff@aei.mpg.de}\AEI
\author{Alessandra Buonanno}\email{alessandra.buonanno@aei.mpg.de}\AEI\Maryland

\date{\today}

\begin{abstract}
As recent work continues to demonstrate, the study of relativistic  scattering  processes leads to valuable insights and computational tools applicable to the relativistic bound-orbit two-body problem.  This is particularly relevant in the post-Minkowskian approach to the gravitational two-body problem, where the field has only recently reached a full description of certain physical observables for scattering orbits, including radiative effects, at the third post-Minkowskian (3PM) order.  As an historically instructive simpler example, we consider here the analogous problem in electromagnetism in flat spacetime.  We compute for the first time the changes in linear momentum of each particle and the total radiated linear momentum, in the relativistic classical scattering of two point-charges, at sixth order in the charges (analogous to 3PM order in gravity).  We accomplish this here via direct iteration of the classical equations of motion, while making comparisons where possible to results from quantum scattering amplitudes, with the aim of contributing to the elucidation of conceptual issues and scalability on both sides.  We also discuss further extensions to radiative quantities of recently established relations which analytically continue certain observables from the scattering regime to the regime of bound orbits, applicable for both the electromagnetic and gravitational cases.
\end{abstract}

 
\maketitle

\section{Introduction}

The dawn of gravitational-wave astronomy \cite{Abbott:2016blz,TheLIGOScientific:2016pea,TheLIGOScientific:2017qsa,LIGOScientific:2018mvr,Abbott:2020niy} and the promise of more sensitive future detectors \cite{Punturo:2010zz, Audley:2017drz, Reitze:2019iox} have renewed interest in varied approaches to solving the two-body problem in general relativity (GR). In particular, much recent work has focused on importing advanced tools from quantum field theory to treat classical scattering of massive bodies in the post-Minkowskian (PM) regime, with large impact parameters, but unconstrained speeds.  Knowledge gained in this regime may also be used to develop a better understanding of inspiraling bound systems, with the aim of constructing more precise waveform models for detection and analysis of gravitational waves from compact binaries.

Alongside gravitational scattering, there is significant interest in (and overlap with) analogous but simpler problems in Yang-Mills theories, including the Abelian case, electromagnetism, in flat spacetime.  In spite of the lesser nonlinearity, making calculations more easily tractable, scattering problems in gauge theories still share many of the same technical difficulties encountered in gravity, and thus can serve as instructive toy models.
A further reason for this interest stems from the study of double-copy relations between gauge and gravity theories \cite{Kawai:1985xq,Bern:2008qj,Bern:2010ue,Bern:2019prr} and their uses in accomplishing gravity calculations with simpler gauge-theory building blocks.  

A significant milestone in the gravitational case has been the recent completion of the calculation of the relativistic impulse (net change in momentum) of each body, in the scattering of two spinless massive bodies, at the third post-Minkowskian (3PM) order.  Following important works such as Refs.~\cite{Damour:2017zjx,Bjerrum-Bohr:2018xdl,Cheung:2018wkq}, which established connections between scattering amplitudes and classical dynamics at the 2PM level, the study of the 3PM level began in Refs.~\cite{Bern:2019nnu,Bern:2019crd}, which determined the conservative sector of the 3PM dynamics by matching to amplitudes computed via modern on-shell techniques, such as generalized unitarity~\cite{Bern:1994zx,Bern:1994cg,Britto:2004nc} and the double copy~\cite{Bern:2008qj,Bern:2010ue,Bern:2019prr}, employing the effective-field-theory matching procedure set up in Ref.~\cite{Cheung:2018wkq}.  These results have been confirmed using various complementary methods \cite{Cheung:2020gyp,Kalin:2020fhe,DiVecchia:2021bdo,Bjerrum-Bohr:2021din}, including calculations based on classical worldlines instead of quantum fields \cite{Kalin:2020fhe}.  The resultant 3PM conservative contribution to the scattering angle function presented a puzzle \cite{Damour:2019lcq} in that it did not have a well-behaved high-energy limit, and did not smoothly connect to previous results for scattering of massless particles first derived in Ref.~\cite{Amati:1990xe} and more recently confirmed in Refs.~\cite{DiVecchia:2019kta,Bern:2020gjj,DiVecchia:2020ymx}.  
 
The resolution of this tension was first suggested in Ref.~\cite{DiVecchia:2020ymx}, which demonstrated, in the case of $\mc N=8$ supergravity at two-loop order, the importance of including radiative effects in order to obtain a well-behaved ultrarelativistic limit.  Subsequently, in the case of GR, Ref.~\cite{Damour:2020tta} showed that a smooth high-energy limit (and a match to the massless results from Ref.~\cite{Amati:1990xe} in that limit) is restored by including the radiative contribution to the scattering angle in addition to the conservative one, with the former being determined (via a relation derived in \cite{Bini:2012ji}) by the total radiated angular momentum at 2PM order.  As we will detail below, the complete (conservative plus radiative) 3PM impulses are determined by the complete 3PM scattering angle along with the total linear (energy-)momentum radiated away in gravitational waves, first appearing at 3PM order.  This last missing piece, the radiated momentum, was calculated in Ref.~\cite{Herrmann:2021lqe}, by employing the formalism of Ref.~\cite{Kosower:2018adc} (KMOC) for computing classical observables from on-shell amplitudes.  Finally, Ref.~\cite{Herrmann:2021tct} also used the KMOC formalism to directly compute the impulses through 3PM order, thereby confirming and combining all the results mentioned above.  In further confirmation, the radiative contributions to the 3PM impulse have been reproduced by a classical variation of constants method in Ref.~\cite{Bini:2021gat}, and the full 3PM scattering angle has been reproduced by a complementary quantum double-copy method in Ref.~\cite{Brandhuber:2021eyq}.  Meanwhile, the analysis of the 4PM level has been initiated in Refs.~\cite{Bern:2021dqo,Dlapa:2021npj}.

Another strategy to solve the scattering problem in a way that includes both conservative and radiative dynamics is direct iteration of the classical equations of motion, in the situation where one lets two particles (representing compact bodies) perform a fly-by with a large impact parameter.  One then constructs the particles' worldlines as expansions in the coupling strength, in the weak field (large impact parameter, small deflection) regime, with the zeroth-order worldlines given by uniform (straight-line) motion. The field equations and the equations of motion are then solved iteratively, informing each other to the required order. Any radiative/dissipative effects are included at each order by employing retarded boundary conditions (for fields) and using a regularization technique to evaluate the effect of each particle's field on itself.
Historically, this route was followed in Ref. \cite{Westpfahl:1985tsl} where the impulse was calculated for both electromagnetism (EM) and GR up to $2^{\mr{nd}}$ order. 

Here we push this method to the $3^{\mr{rd}}$ order in the EM case.  In addition to including radiative effects, another advantage of the classical method is that it is relatively simple to automate the iteration to go to higher orders with the only potential challenge being computation of one-dimensional integrals. This may be useful for efficiently computing scattering observables, which can then be used for subsequently understanding bound orbits.  It would also be highly interesting to connect this efficient classical iteration to perturbative scattering-amplitude calculations, for instance by expressing amplitudes using the so-called worldline quantum field theory~\cite{Mogull:2020sak,Jakobsen:2021smu}.  Furthermore, 
the classical method is also perhaps the simplest to justify, being most closely related to the actual situation of classical scattering of compact objects, and it gives detailed information about the time-dependent worldlines at each order as another benefit.

Having in hand results for the scattering problem, we investigate the possibility of mapping observables from unbound to bound orbits via analytic continuation, following Refs.~\cite{Kalin:2019rwq,Kalin:2019inp}, where the basis of the mapping procedure was explained and used to relate the scattering angle for unbound orbits to the periastron advance angle for bound orbits.  An analogous map between energy losses was presented in Ref.~\cite{Bini:2020hmy}, and was subsequently verified from the results of Ref.~\cite{Herrmann:2021lqe} for the radiated linear momentum in the GR case. Here we provide a general relation between observables (satisfying certain criteria) for bound and unbound orbits following the method given in Ref.~\cite{Kalin:2019rwq}, and we explicitly relate angular momentum losses between bound and unbound orbits.
We verify the map between energy losses using the expression for leading order radiated momentum in EM derived earlier, and we use the map between angular momentum losses to uncover an error in the expression for angular momentum losses for 1 post-Newtonian (PN) unbound orbits in GR given in Ref. \cite{Junker:1992kle}. We also comment on the scope of using these relations in computing resummed expressions for observables in the bound case.

Section~\ref{sec:exec} summarizes our results for EM scattering while comparing them to analogous results for the GR case obtained elsewhere, and 
Sec.~\ref{sec:cont} summarizes our investigations of unbound-to-bound continuation.

\subsection{Anatomy of relativistic scattering to $3^{\mr{rd}}$ order}\label{sec:exec}
\begin{figure}[h]
\includegraphics[width=\columnwidth]{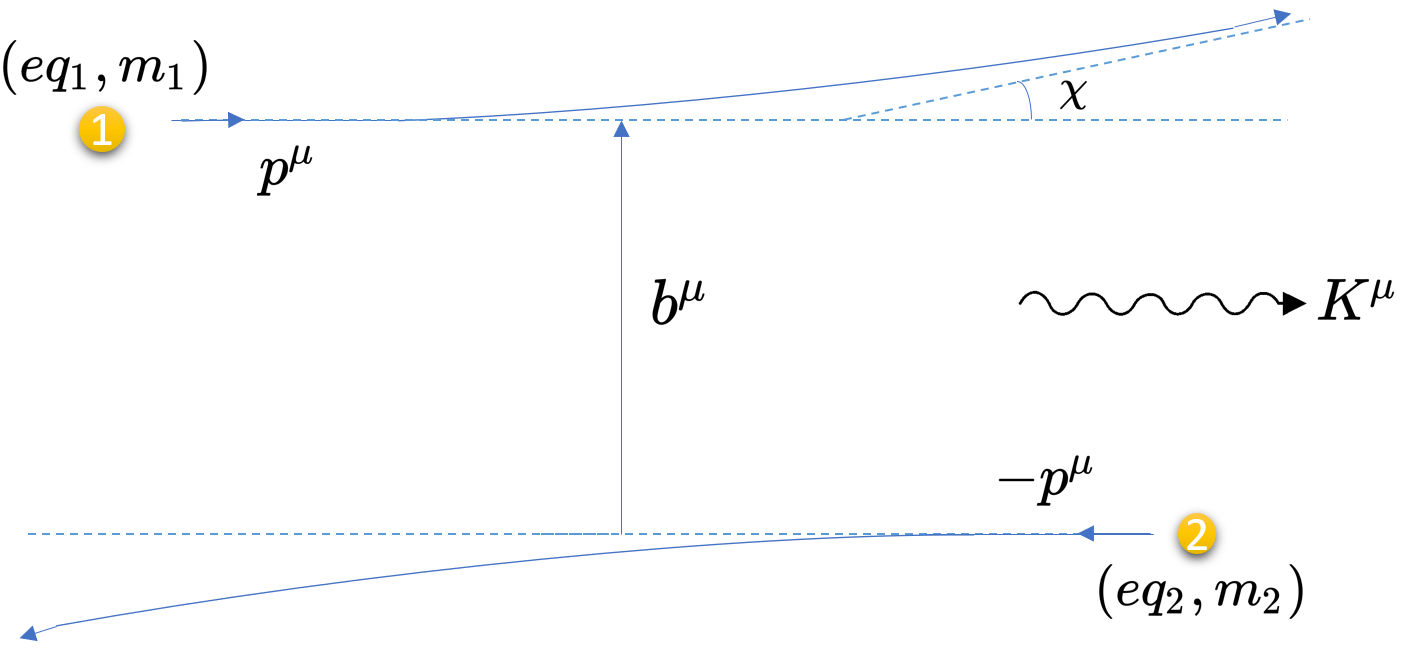}
\caption{ Classical scattering of two charged particles in the COM frame. We denote with $eq_1$ and $e q_2$ the electric charges, where we introduce the factor $e$ to keep track of the order of calculation since we will be solving the scattering problem in a weak-field/coupling expansion. We denote with $m_1$, $m_2$ the masses, $p^{\mu}$ is the initial spatial momentum in the COM frame, and $\sqrt{-b^2}=|b|$ is the impact parameter. Finally, $\chi$ is the scattering angle and $K^{\mu}$ is the radiated momentum.}
\centering
\end{figure}

For both the EM and GR cases, the net results of a two-body scattering encounter can be expressed as functions of the asymptotic incoming state at past infinity, where interactions are perturbatively negligible, with the two bodies moving uniformly on straight-line trajectories in (asymptotically) Minkowski spacetime.  We identify the initial state with the zeroth-order state in our perturbative expansion.  It is specified by the initial momenta $p_1^\mu=m_1 u_1^\mu$ and $p_2^\mu=m_2u_2^\mu$, where $m_a$ are the rest masses and $u_a^\mu$ are the 4-velocities with $u_a^2=1$, and by the initial impact parameter vector $b^\mu$ (pointing 2$\to$1) orthogonally separating the two initial uniform-motion worldlines.  The initial relative velocity $v$, the corresponding Lorentz factor $\gamma$, and the initial total energy $E$ in the center-of-momentum (COM)\ frame (frame of reference in which the spatial components of total momentum is 0) are defined by
\begin{alignat}{3}
\begin{aligned}
\gamma&=\frac{1}{\sqrt{1-v^2}}=\frac{p_1\cdot p_2}{m_1m_2},
\\
E^2 &=(p_1+p_2)^2=m_1^2+m_2^2+2m_1m_2 \gamma,
\end{aligned}
\end{alignat}
where here and in the rest of the paper, we use $c=1$ and employ dot products and squares as in flat spacetime, and the $(+,-,-,-)$ metric signature.  The COM\ frame has velocity $(p_1+p_2)^\mu/E$.  The ``relative momentum'' $p^\mu$, giving (minus) the spatial momentum of body 1 (body 2) in the COM frame, and its magnitude $|p|$ are given by
\begin{alignat}{3}
\begin{aligned}
p^\mu&=\frac{m_1m_2}{E^2}[(m_2+m_1\gamma)u_1^\mu-(m_1+m_2\gamma)u_2^\mu], 
\\ 
|p| &=\frac{m_1 m_2 \gamma v}{E}.
\end{aligned}
\end{alignat}
The magnitude of the initial COM-frame total angular momentum is $J=|p||b|$.

The perturbations of the worldlines due to EM or GR interactions can be computed by iteratively solving the relevant field equations and equations of motion, working perturbatively in Newton's constant $G$ in the GR case, or in $e^2$ in the EM case, where $e$ is an order-counting parameter such that the two charges are $eq_1$ and $eq_2$.  We work in Gaussian units in the EM case.  While we use $G$ and $e^2$ as formal expansion parameters, the true dimensionless quantities, which we assume to be small, are essentially the leading terms in the scattering angles in Eqs.~(\ref{chi12}) and (\ref{chi12GR}) below; our fundamental assumption is that the scattering particles' trajectories are small perturbations of straight-line inertial motion.
All quantities of interest (impulse, radiated energy and angular momentum) can be obtained from the perturbed worldlines computed to the relevant order.

The impulses up to $2^{\mr{nd}}$ order for both the EM and GR cases were first computed by Westpfahl in Ref.~\cite{Westpfahl:1985tsl}. Through this order, no net 4-momentum is radiated away, so the impulses on the two bodies are equal and opposite, $\Delta p_2^\mu=-\Delta p_1^\mu$.  They are given in terms of the COM-frame scattering angle $\chi$ by
\begin{alignat}{3}
\label{cons2nd}
\Delta p_{1}^\mu&=|p|\sin \chi\frac{b^\mu}{|b|}+(\cos\chi-1)p^\mu + {\cal O}(e^6,G^3),
\end{alignat}
in each case, where the second-order scattering angles are
\begin{alignat}{3}\label{chi12}
 \chi_{\mr{EM}}&=\frac{ 2 e^2q_1 q_2 E   }{ m_1 m_2 \gamma  v^2 |b|} -  \frac{\pi e^4 q_1^2 q_2^2  M E }{2 m_1^2 m_2^2 \gamma^2  v^2 |b|^2}+\mc {\cal O}(e^6), 
\\\label{chi12GR}
\chi_\mr{GR} &=-\frac{2G E (2 \gamma^2 -1 ) }{\gamma^2 v^2 |b|}- \frac{3\pi G^2 M E(5 \gamma^2 -1)}{4 \gamma^2 v^2 |b|^2}+\mc {\cal O}(G^3). 
\end{alignat}
Here,  $M=m_1+m_2$ is the sum of rest masses. 

While there is no radiation of linear momentum up to the $2^{\mr{nd}}$ order, the dynamics is not purely conservative at these orders, as there is a loss of angular momentum which is encoded in the second-order trajectories.  We compute this directly from the trajectories for the EM case in Sec.~\ref{angmomsec}, finding the radiated angular momentum (minus the change in the particles' orbital angular momentum) in the COM frame to be
\begin{alignat}{3}
\label{JradEM}
J_{\mr{rad,EM}} &= 2\frac{e^4 q_1^2 q_2^2  }{   E |b| }I_\mr{EM}(v) + {\cal O}(e^6),
\\
I_\mr{EM}(v)&=  -\frac{2}{3}\gamma\left (\frac{q_1/m_1}{q_2/m_2}+ \frac{q_2/m_2}{q_1/m_1}\right )  +\frac{2}{ v^2} - \frac{2\,\mr{arctanh}\,v}{\gamma^2 v^3}.
\end{alignat}
The analogous result for GR,
\begin{alignat}{3}
\label{JradGR}
& J_{\mr{rad,GR}} = 2\frac{G^2 m_1^2 m_2^2 (2 \gamma^2 -1)}{E |b|}I_\mr{GR}(v)  + {\cal O}(G^3),\\&
I_\mr{GR}(v)=-\frac{16}{3} + \frac{2}{v^2}+\frac{2(3 v^2 -1)}{v^3}\mr{arctanh}\,v.
\end{alignat}
was derived in Ref.~\cite{Damour:2020tta} by directly computing the total angular momentum radiated away in the gravitational field.  Note that both results feature a factor of the `rapidity,' $\mr{arctanh}\,v=\mr{arccosh}\,\gamma=2\,\mr{arcsinh}\,\sqrt{(\gamma-1)/{2}}$.  Also note that $J_{\mr{rad,GR}}$ is always positive, so that the orbital angular momentum always decreases in magnitude, while $J_{\mr{rad,EM}}$ is positive for opposite-sign charges (attraction) but can be negative for same-sign charges (repulsion), so that the orbital angular momentum can increase in magnitude in the latter case.

At the $3^{\mr{rd}}$ order, in both the EM and GR cases, linear momentum is radiated away as well.  The impulse on body 1 (with results for body 2 obtained by exchanging identities, $1\leftrightarrow2$) can be written in both cases as follows,
\begin{alignat}{3}
\Delta p_1^\mu& =\Delta p_{1,\mr{cons}}^\mu+\Delta p_{1,\mr{rad}}^\mu, \\
 \label{cons}
\Delta p_{1,\mr{cons}}^\mu&=|p|\sin \chi_\mr{cons}\frac{b^\mu}{|b|}+(\cos\chi_\mr{cons}-1)p^\mu, \\
\label{dissp}
\Delta p_{1,\text{rad}}^{\mu} &= \frac{K\cdot u_2}{(\gamma v)^2}(u_2^{\mu} - \gamma u_{1}^{\mu} ) + |p| \chi_{\text{rad}} \frac{b^{\mu}}{|b|},
\end{alignat}
where $\chi_{\mr{cons}}$ is the conservative part of the scattering angle, $\chi_{\mr{rad}}$ is the radiative part of the scattering angle, and $K^{\mu}$ is the radiated momentum.  In Sec.~\ref{sec:kchi}, we argue for this structure based on general grounds, and also confirm it explicitly by directly computing the full impulse from the $3^{\mr{rd}}$ order force in the EM case.  

For the conservative scattering angles, we have $\chi_\mr{cons}=\chi^{(1)}+\chi^{(2)}+\chi_\mr{cons}^{(3)}+\mc {\cal O}(G^4,e^8)$, where the first- and second-order parts are given in Eq.~(\ref{chi12}). At the $3^{\mr{rd}}$ order in the EM case, as we compute below,
we have
\begin{alignat}{3}\label{chi3consEM}
&\chi^{(3)}_{\text{cons,EM}} = \nonumber \\
 &\quad \frac{e^6 q_1^3 q_2^3 E  [(m_1^2+m_2^2)(4 \gamma^2-6)- 4 m_1 m_2 \gamma (  \gamma^2-3 v^2 )]}{3 m_1^3 m_2^3  \gamma^5 v^6 |b|^3}.
\end{alignat}
This matches the result derived from potential-region integration of the two-loop amplitude in Ref.~\cite{BernPrivate}.  The analogous result for GR,
\begin{alignat}{3}
&\chi^{(3)}_{\text{cons,GR}} = -\frac{G^3E^3}{|b|^3}\bigg[2\frac{64\gamma^6-120\gamma^4+60\gamma^2-5}{3(\gamma v)^6}
\nonumber \\
&\quad-\frac{8m_1m_2}{E^2}\bigg(\frac{14\gamma^2+25}{3\gamma v^2}+\frac{4\gamma^4-12\gamma^2-3}{(\gamma v)^3}\mr{arctanh}\,v\bigg)\bigg],
\end{alignat}
was first derived in Ref.~\cite{Bern:2019nnu}, also from potential-region integration of the two-loop amplitude.The GR result has a logarithmic divergence in the high-energy limit  due to the $\mr{arctanh}\,v$ term, a feature which is not present in the EM result (\ref{chi3consEM}).  This divergence is removed by adding the radiative contribution to the scattering angle $\chi_\mr{rad}$, which was first computed in Ref.~\cite{Damour:2020tta} by using the relation derived in Ref.~\cite{Bini:2012ji},
\be\label{chiradfromJ}
\chi_\mr{rad}=-\frac{1}{2}\frac{\doe\chi_\mr{cons}}{\doe J}J_\mr{rad}+\mc {\cal O}(G^4,e^8),
\ee
where $J_\mr{rad}$ is the radiated angular momentum (and with $\doe/\doe J=(1/|p|)(\doe/\doe|b|)$).  This yields
\begin{alignat}{3}
\chi_\mr{rad,EM}&=\frac{ 2 e^6 q_1^3 q_2^3 E  }{ m_1^2 m_2^2 \gamma^2  v^3 |b|^3}I_\mr{EM}(v)+\mc {\cal O}(e^8), 
\\ \label{chiradEM}
\chi_\mr{rad,GR}&=-\frac{2 G^3 m_1 m_2 E }{|b|^3}\frac{(2\gamma^2-1)^2}{\gamma^3 v^3}I_\mr{GR}(v)+\mc {\cal O}(G^4),
\end{alignat}
where the $I$ functions are given in Eqs.~(\ref{JradEM}) and (\ref{JradGR}).  We confirm below that this result for $\chi_\mr{rad,EM}$, coming via the relation (\ref{chiradfromJ}) from $J_\mr{rad,EM}$ (computed from the second-order trajectories), matches the result we obtain directly from the third-order trajectories.  The result of Ref.~\cite{Damour:2020tta} for $\chi_\mr{rad,GR}$ from $J_\mr{rad,GR}$ via (\ref{chiradfromJ}) has also been independently confirmed by the direct calculation of the third-order impulse in Ref.~\cite{Herrmann:2021tct}.

The final ingredients in the expression for radiative impulse in Eq.~(\ref{dissp}) is the radiated momentum $K^\mu$.  We compute $K^{\mu}_\mr{EM}$ below from the full third-order impulses, using $K^\mu=-\Delta p_1^\mu-\Delta p_2^\mu$, obtaining
\begin{alignat}{3}
\label{KradEM}
& K^{\mu}_\mr{EM}  = \frac{ \pi e^6 q_1^2 q_2^2}{4  |b|^3} \bigg\{\left (\frac{q_1^2}{m_1^2} u_1^{\mu} + \frac{q_2^2}{m_2^2} u_2^{\mu} \right ) \frac{3 \gamma^2 +1}{3 \gamma v} \nonumber \\& \qquad\qquad - \frac{q_1 q_2}{m_1 m_2} \frac{u_1^{\mu} + u_2^{\mu}}{\gamma +1} \mc F(v) \bigg\}+\mc {\cal O}(e^8),
\\&
 \mc F(v) = \frac{1}{(\gamma v)^3} \bigg\{(3\gamma^2+1)\Big(\gamma - \frac{\text{arctanh}\,v}{\gamma v} \Big) - 4 (\gamma -1)^2\bigg\}.
\end{alignat}
By performing the momentum-space integral given in Eq.~(6.32) of Ref.~\cite{Kosower:2018adc}, we find that Eq.~(\ref{KradEM}) precisely matches the direct calculation of the momentum radiated to infinity by the EM field. The radiated momentum for the GR case	 was first computed in Ref.~\cite{Herrmann:2021lqe} using the KMOC formalism \cite{Kosower:2018adc}, with the result
\begin{alignat}{3}
\label{KradGR}
K_\mr{GR}^\mu&=\frac{G^3m_1^2m_2^2}{|b|^3}\frac{u_1^{\mu} + u_2^{\mu}}{\gamma +1}\mc E(v)+\mc {\cal O}(G^4),
\\
\frac{\mc E(v)}{\pi}&=f_1+f_2\log\frac{\gamma+1}{2}+f_3\frac{\mr{arctanh}\,v}{2v},
\end{alignat}
where the $f_n$ are polynomials in $\gamma$ divided by powers of $\gamma v$ given in Eq.~(9) of Ref.~\cite{Herrmann:2021lqe}.  This, along with the full (third-order) impulse structure Eqs.~(\ref{cons}, \ref{dissp}) and all of the GR results above, has also been confirmed by the calculation in Ref.~\cite{Herrmann:2021tct} of the complete $\Delta p_{1,\mr{GR}}^\mu$ using the KMOC formalism~\cite{Kosower:2018adc} applied to the two-loop amplitude.

\subsection{Unbound to bound continuation}
\label{sec:cont}

An ultimate objective is to use the results obtained from analyzing scattering problems to learn more about the bound-orbit problem.
One way to achieve this is by mapping corresponding observables from unbound to bound orbits, as was exemplified in Refs.~\cite{Kalin:2019rwq,Kalin:2019inp}, by relating the scattering angle $\chi$ for unbound orbits to the periastron advance $\Delta\phi$ for bound orbits,
\begin{equation}
\Delta \phi(\mathcal{E},J) = \chi(\mathcal{E},J) + \chi(\mathcal{E},-J).
\end{equation}
In Sec.~\ref{utob}, we show how this relation can be extended to other observables, provided certain conditions are satisfied.
We show this explicitly by relating energy and angular momentum losses between unbound and bound orbits.
The relations for the energy loss reads
\begin{alignat}{3}
\label{powerrel}
E_{\mr{rad}}^{\mr{bound}}(\mathcal{E},J) = E_{\mr{rad}}^{\mr{unbound}}(\mathcal{E},J) - E_{\mr{rad}}^{\mr{unbound}}(\mathcal{E},-J),  \nonumber
\\&
\end{alignat}
as previously noted in Refs.~\cite{Bini:2020hmy}.
Here we find an analogous relation for the angular monentum loss,
\begin{equation}
\label{amomrel}
J_{\mr{rad}}^{\mr{bound}}(\mathcal{E},J) = J_{\mr{rad}}^{\mr{unbound}}(\mathcal{E},J) + J_{\mr{rad}}^{\mr{unbound}}(\mathcal{E},-J).
\end{equation}

These maps can be used to recover resummed relativistic expressions for radiative losses in bound orbits. However, since the PM/weak field expansion is also an expansion in $1/J$ (see expressions of scattering angles in Eq.~(\ref{chiradEM}) for relevant dimensionless quantities, the initial angular momentum is related to impact parameter as $J = |p| |b|$), we only recover the leading order $1/J^3$ ($1/J$) part of the energy (angular momentum) losses, respectively, from the leading order radiative losses. For gravity, the 0PN energy loss for generic bound orbits also contains terms that scale as $1/J^5$ and $1/J^7$, and the coefficient of these terms cannot be recovered directly through the map. The situation is worse for angular-momentum loss where the 0PN angular-momentum loss only contains even powers of $1/J$ (as would be expected from Eq.~(\ref{amomrel})) and thus leading 2PM angular momentum loss does not provide any contribution. 

Naively, this would lead to the discouraging conclusion that one needs to solve till 7PM for gravity to recover even the 0PN energy loss. However, as we shall discuss in detail below, we can circumvent this by using an alternate method of fixing radiative losses for generic orbits in the PN expansion. We do it by constructing an ans\"{a}tz for radiative losses in generic orbits, parametrized by a finite number of unknown coefficients which are then fixed by evaluating the energy losses in the weak field expansion where it should match with the PM results. Further constraints on the coefficients can be obtained by using the relation between energy and angular-momentum losses for circular orbits and some coefficients can be discarded by adding schott/total time derivative terms to the radiative losses. It can be shown that this method allows us to fix the 0PN energy losses from just 4PM results.

\subsection{Outline}

The paper is organized as follows. In Sec.~\ref{sec:EMcals}, we explain in detail the process of solving for the worldline corrections iteratively via the classical method to $3^{\mr{rd}}$ order. In Sec.~\ref{sec:kchi}, we write down the impulse up to $3^{\mr{rd}}$ order and express it in terms of the conservative and radiative parts of the scattering angle and radiated momentum. In Sec.~\ref{angmomsec}, we derive the leading order angular momentum loss and show that it is correctly related to the leading order radiative correction to the scattering angle. We then investigate the high-energy and nonrelativistic limit of relevant observables in Sec.~\ref{hel} and Sec.~\ref{nrelsec}, respectively. In Sec.~\ref{utob}, we show how the method of analytic continuation shown in Refs.~\cite{Kalin:2019rwq,Kalin:2019inp} can be extended to other observables if certain conditions are satisfied, and explicitly show how these mappings can be used to extract partial results for bound orbits, and their scope. We then conclude in Sec.~\ref{sum} with a short summary. In Appendix.~\ref{2oapp}, we explain in detail how we compute the $2^{\mr{nd}}$ order correction worldlines and explicitly write down the $2^{\mr{nd}}$ order worldline correction for particle 1. In Appendix~\ref{3oimpulse}, we translate the $3^{\mr{rd}}$ order force correction diagrams to expressions and give brief comments regarding the source of each diagram.

\section{Relativistic scattering of two point charges}\label{sec:EMcals}

We consider here the motion of two charged point particles in classical relativistic electromagnetism in flat spacetime.  The particles have charges $e q_1$ and $e q_2$, masses $m_1$ and $m_2$, and proper-time--parametrized worldlines $x^\mu=z_1^\mu(\tau_1)$ and $x^\mu=z_2^\mu(\tau_2)$.  The Lorenz-gauge field equation for the gauge field $e A_\mu(x)$, with $\doe_\mu A^\mu=0$, is
\be
\doe^2( e A^\mu)=4\pi  J^\mu.
\ee
We take out a factor of the order-counting parameter $e$ from the gauge field (and the field strength below) for later convenience.  The current density is
\be\label{Jmu}
J^\mu(x)=e\sum_aq_a \int d\tau_a \, \dot z_a^\mu \delta^4(x-z_a),
\ee
with $a=1,2$.  The equation of motion for each worldline can be taken as the Lorentz-Dirac equation,
\be\label{LDEgen}
m_a\ddot z_a^\mu=e^2 q_a  F^\mu{}_\nu(z_a)\dot z^\nu_a+\frac{2e^2 q_a^2}{3}(\dddot z_a^{\,\mu}+\dot z_a^\mu\ddot z^2_a),
\ee
where $eF_{\mu\nu}=2e\doe_{[\mu}A_{\nu]}$ is the \emph{external} field strength (due to the field of the other particle), while the second term accounts for the action of the self field~\cite{Poisson:1999tv}.  Our task is to construct an iterative solution to these equations working perturbatively in the coupling strength, measured by $e^2$.

At zeroth order in $e^2$, the particles follow inertial trajectories, and their worldlines can be written as
\begin{equation}
\label{zeroth}
z_{1}^{(0)\mu} = b^{\mu} + u_1^{\mu} \tau_1, \quad z_2^{(0)\mu} = u_2^{\mu} \tau_2,
\end{equation}
where $u_1^\mu$ and $u_2^\mu$ are the zeroth-order 4-velocities, with $ u_1\cdot u_2\equiv\gamma$, and $b^\mu$ is the impact parameter vector with $b\cdot u_1=0=b\cdot u_2$, so that the minimum separation between the zeroth-order trajectories is given by $|b|$. 

The field sourced by particle $a$, on a general trajectory $x^{\mu}=z_a^{\mu}(\tau_a)$, obeys
\begin{equation}
 \doe^2 ( A_a^{\mu})  = 4 \pi  q_{a} \int d\tau_a\,  \dot z^{\mu}_a \, \delta^{4}(x-z_a(\tau_a)),
\end{equation}
which has multiple solutions, corresponding to different boundary conditions. Choosing the retarded (no-incoming radiation) boundary conditions, the solution is
\begin{equation}
 A_a^{\mu}(x) =  \frac{  q_a \dot{z}^\mu_{a,\mr{ret}}}{ \dot{z}_{a,\mr{ret}}\cdot(x-z_{a,\mr{ret}})},
\label{ret}
\end{equation}
where we use the shorthand $f_{\mr{ret}} = f(\tau_{\mr{ret}})$. The retarded proper time $\tau_{a,\mr{ret}}$ is the proper time at which the trajectory of the source particle $a$ intersects the past light cone of the field point $x$. It is obtained by solving $|x-z_{a}(\tau_a)|^2=0$, and choosing the solution that makes $x-z_{a}(\tau_a)$ a future-directed null vector. 
Then, the field strength sourced by particle $a$ is given by
\begin{alignat}{4}
F_a^{\mu\nu}(x)&= 2\partial^{[\mu}A_a^{\nu]}\label{field}\\  
&= \frac{2  q }{ r_a^3}\rho_a^{[\mu}\Big( r \ddot{z}_{a,\mr{ret}}^{\nu]}-\dot{z}_{a,\mr{ret}}^{\nu]}(\ddot{z}_{a,\mr{ret}}\cdot \rho_a-1) \Big)  , \nonumber 
\end{alignat}
where we have defined $\rho_a^{\mu} = (x-z_{a,\mr{ret}})^{\mu}$, $r_a = \dot{z}_{a,\mr{ret}}\cdot \rho_a$ and used  $\partial^{\mu} \tau_{a,\mr{ret}} = \rho^{\mu}_a/r_a$. 

Taking the field-sourcing particle above to be particle $a=2$, the equation of motion for the other particle, 1, reads
\begin{equation}
m_1 \ddot{z}^{\mu}_1 = e^2 q_1 F_2^{\mu \nu}(z_1) \dot{z}_{1,\nu} + \frac{2 e^2q_1^2}{3} \Big(\dddot{z}_1^{\mu}+\ddot{z}_1^2 \dot{z}_1^{\mu}\Big).
\label{aldf}
\end{equation}
The first term in the RHS of Eq.~(\ref{aldf}) is the well known Lorentz force, giving the influence of the field sourced by particle 2 on particle 1. The second term is the Abraham-Lorentz-Dirac (ALD) self-force which accounts for the particle's influence on itself, which can be derived from the traditional Lorentz force law in various ways.  The simplest method for our purposes is to use a regularization procedure to deal with the divergence when evaluating particle's field on its own worldline, as was done in Ref.~\cite{Westpfahl:1985tsl}. In the case of EM, this can be done once, and the divergence-free expression in Eq.~(\ref{aldf}) can be used at all orders in the perturbative expansion in $e^2$ (see however the following paragraph). In contrast, in the GR case, one must repeat the regularization procedure at each order to compute the self-influence due to non-linearities in the theory.
 
  Dissipative effects are included in the time-asymmetric part of the force which come from the choice of retarded propagator as well as from the ALD term (which is odd in $\tau_1\rightarrow -\tau_1$). The validity of the self-force expression in EM is a nontrivial issue, associated with runaway solutions and pre-acceleration. In general, one expects the ALD force to be valid as long as the particle is not too large or too small. In Ref.~\cite{Poisson:1999tv}, this was defined more precisely by requiring that the particle be small enough to enable fast communication between portions of its body ($l \ll t_c$, where $l$ is the size of the particle and $t_c$ is the time scale of changes in any external force field) but not so compact that the electromagnetic self-energy is larger than the mass (i.e., the positive bare mass condition $0 \leq m_0 \equiv m - e^2 q^2/l$, or $e^2 q^2/ l \leq m$). The latter of these constraints in particular turns out to be important for making sense of divergences in the high-energy limit for some radiative quantities, as will be shown later in Sec.~\ref{hel}.

To iteratively solve Eq.~(\ref{aldf}), we expand the worldlines in powers of $e^2$, with the zeroth orders given by Eq.~(\ref{zeroth}),
\begin{alignat}{3}
z_1^{\mu}(\tau_1) &= b^{\mu} + u_1^{\mu} \tau_1 + e^2 z_1^{(1),\mu}(\tau_1)+e^4 z_1^{(2),\mu}(\tau_1)+\ldots , \nonumber \\&\\ \nonumber
z_2^{\mu}(\tau_2) &= u_2^{\mu} \tau_2 + e^2 z_2^{(1),\mu}(\tau_2)+e^4 z_2^{(2),\mu}(\tau_2)+\ldots.
\\&
\end{alignat}
We use $|z_1(\tau_1)-z_2(\tau_{2,\mr{ret}})|^2=0$ to solve for the retarded time $\tau_{2,\mr{ret}}$ in terms of the worldline corrections ($z^{(n),\mu}$).
\begin{alignat}{3}
\label{tauexp}
\tau_{2,\mr{ret}} = \tau_{2,\mr{ret}}^{(0)} + e^2 \tau_{2,\mr{ret}}^{(1)} + e^4 \tau_{2,\mr{ret}}^{(2)}+ \ldots ,
\end{alignat}
which we then substitute into Eq.~(\ref{aldf}) to get a similar expansion for the force,
\begin{alignat}{3}
\nonumber & e^2 m_1  \ddot{z}_1^{(1)} + e^4 m_1 \ddot{z}_1^{(2)} + e^6 m_1 \ddot{z}_1^{(3)} + \ldots\\&= e^2 f^{(1)} + e^4 f^{(2)} + e^6 f^{(3)}+\ldots .\label{expansion}
\end{alignat}
Now collecting terms by powers of $e^2$, we get the correction to the force ($e^{2n}f^{(n)}$) at each order as a function of lower-order worldline corrections.
 These $2^{\mr{nd}}$ order differential equations can be then iteratively solved starting from the lowest order to acquire the worldline corrections at each order. 
 
 It is sufficient to solve for the worldline corrections of particle 1 since the analogous quantities for the second particle can be then obtained by simply swapping $1\leftrightarrow 2$ and $b^{\mu} \rightarrow -b^{\mu}$.

\subsubsection{Diagrams and rules}

\label{diag}
The correction to the force at each order $f^{(n)}$ [see the RHS of Eq.~(\ref{expansion})] depends on the corrections to worldlines and retarded time at lower orders (along with zeroth-order quantities). For instance, $f^{(1)}$ only depends on the zeroth-order quantities ($z_1^{(0)}$, $z_2^{(0)}$, $\tau_{2,\mr{ret}}^{(0)}$), whereas $f^{(2)}$ depends on $z^{(1)}_{a=1,2}$ and $\tau_{2,\mr{ret}}^{(1)}$, as well. Put simply, $f^{(n)}$ is the $n^{\rm{th}}$ term in the Taylor expansion of the Lorentz and ALD forces w.r.t.\ the variables $z_1$, $z_2$, and $\tau_{2,\mr{ret}}$. Thus, it is a linear combination of derivatives of the Lorentz (and ALD) force multipled by the corrections to these quantities. At higher orders, we find it convenient to split the force correction $f^{(n)}$ into various diagrams, as a way to illustrate the method of calculation and its increasing complexity with each order. It is important to note that these will \textit{not} be Feynman or Feynman-like diagrams. We illustrate the rules for understanding the diagrams with an example below.

Diagrams with two worldlines represent terms in the Taylor expansion of the Lorentz force term i.e.,~\ $e^2 q_1 \dot{z}_{1,\nu} F_2^{\mu\nu}[z_1,z_2(\tau_{2,\mr{ret}}),\dot{z}_2(\tau_{2,\mr{ret}}), \ddot{z}_2(\tau_{2,\mr{ret}})] $, via corrections from $z_1$, $z_2$ (and derivatives), and $\tau_{2,\mr{ret}}$, for example: 
\begin{center}
\begin{equation}
\label{examplediag}
\begin{tikzpicture}
\begin{scope}[very thick, every node/.style={sloped,allow upside down}]
\node[text width=0.25cm] at (-0.5,0) {2};
\node[text width=0.25cm] at (-0.5,1.5) {1};
\node[text width=0.25cm] at (2.5,-0.25) {$\tau_{2,\mr{ret}}^{(1)}$};
\draw[dashed](0,0)--(2.5,0);
\draw[very thick](0,1.5)--(3.0,1.5);
\draw[very thick, double](3.0,1.5)--(5,1.5);
\draw[very thick](3.0,1.6)--(5,1.6);
\draw[snake it](2.5,0)--node {\midarrow}(3.0,1.5);
\end{scope}
\end{tikzpicture}
\end{equation}
\end{center}
 The directed photon (wavy) line conveys that the field is sourced by 2 to affect 1's worldline. 
\begin{itemize}[leftmargin=*]
\item[1.] The part of the diagram to the \textbf{right} of the photon line tells the order at which this diagram contributes, according to the conventions:

\begin{tikzpicture} 
\draw[very thick](0,0)--(2,0); 
\end{tikzpicture}
$1^{\mr{st}}$ order correction\vspace{\baselineskip} \newline
\begin{tikzpicture} 
\draw[very thick, double](0,0)--(2,0); 
\end{tikzpicture}
$2^{\mr{nd}}$order corrections
\vspace{\baselineskip} \newline
and so on.
\item[2.] The part of the diagram to the \textbf{left} of the photon line tells us which derivatives to act on $e q_1 F_2^{\mu\nu}\dot{z}_{1,\nu}$ and which corrections to multiply. The derivatives are to be evaluated at zeroth order. \newline
\emph{a.}~ Dashed lines represent contributions solely from unperturbed zeroth-order worldlines. They merely contribute a factor of 1 to the diagram.

\begin{tikzpicture} 
\draw[dashed](0,0)--(2,0); 
\end{tikzpicture}
$\quad$ no derivative (zeroth order)

\emph{b.}~Linear corrections (of any order) are represented using single/closely-spaced lines,

\begin{tikzpicture} 
\draw[very thick](0,0)--(2,0); 
\end{tikzpicture}
$\quad$ $e^2\sum_{n=0}^3 d^{n}_{\tau_i} z^{(1)}_i\frac{\partial}{\partial d_{\tau_i}^n z_i}$ ($1^{\mr{st}}$ order)\vspace{\baselineskip} \newline
\begin{tikzpicture} 
\draw[very thick, double](0,0)--(2,0); 
\end{tikzpicture}
$\quad$ $e^4\sum_{n=0}^3 d^{n}_{\tau_i} z^{(2)}_i\frac{\partial}{\partial d_{\tau_i}^n z_i}$ ($2^{\mr{nd}}$ order)\vspace{\baselineskip} \newline
and so on. Here we are using the shorthand notation $d^{n}_{\tau_i} z_i = (d/d\tau_i)^{n}z_i(\tau_i)$

\emph{c.}~Quadratic corrections are represented using wedges as follows, \vspace{\baselineskip} \newline
\begin{tikzpicture} 
\draw[very thick](0,-0.125)--(2,0); 
\draw[very thick](0,+0.125)--(2,0); 
\end{tikzpicture}
$
\frac{e^4}{2}\sum_{m,n=0}^{m,n=3} (d_{\tau_1}^m z_{i}^{(1)})(d_{\tau_1}^n z_{i}^{(1)})\frac{\partial^2}{\doe d_{\tau_1}^m z_{i} \doe d_{\tau_1}^n z_{i}} 
$ \vspace{\baselineskip}\newline
\begin{tikzpicture} 
\draw[very thick,double](0,-0.125)--(2,0); 
\draw[very thick](0,+0.125)--(2,0); 
\end{tikzpicture}
$
\frac{e^6}{2}\sum_{m,n=0}^{m,n=3} (d_{\tau_i}^m z_{i}^{(1)})(d_{\tau_i}^n z_{i}^{(2)})\frac{\partial^2}{\doe d_{\tau_i}^m z_{i} \doe d_{\tau_i}^n z_{i}} 
$ \vspace{\baselineskip}\newline
and so on. All the derivatives are to be evaluated at zeroth-order retarded time ($\tau_2 = \tau_{2,\mr{ret}}^{(0)}$).

\item[4.]The label at the intersection of particle 2's worldline with the photon (wavy line) $(\tau_{2,\mr{ret}}^{(m)})^{n}$ contributes $$
e^{2nm}(\tau_{2,\mr{ret}}^{(m)})^n\times(1/n!)(d/d\tau_{2,\mr{ret}})^{n},
$$
where $\tau_{2,\mr{ret}}^{(m)}$ is the $m^{\rm th}$ order correction to retarded time. Once again, the derivatives are evaluated at zeroth-order retarded time $\tau_{2,\mr{ret}}^{(0)}$. Note that we are separating the dependence on retarded time purely for convenience, since $\tau_{2,\mr{ret}}^{(n)}$ can be written in terms of worldline corrections $z_a^{(m)}$ via the relation $|z_1(\tau_1)-z_2(\tau_{2,\mr{ret}})|^2=0$.\newline
\end{itemize}
Thus, based on these rules, the above diagram (\ref{examplediag}) gives a $3^{\mr{rd}}$ order contribution and is equal to 
\begin{alignat}{3}
& e^{6}\sum_{n=0}^{3} (d_{\tau_1}^n z_{1}^{(1)}) \frac{\partial }{\partial(d_{\tau_1}^n z_{1})}  (\tau_{2,\mr{ret}}^{(1)})(d/d\tau_{2,\mr{ret}})  ( q_1 F_2^{\mu\nu} \dot{z}_{1,\nu})|_{(0)}\nonumber.
\end{alignat}

Contributions from the ALD (self-)force are constructed using the same rules, but only have one particle's worldline. Also, the derivatives act on the ALD force term i.e.,~ $e^2 (2 q_1^2/3) (\dddot{z}_1+\ddot{z}_1^2 \dot{z}_1^{\mu})$, and there is no contribution to ALD force from $z_2^{(m)}$ (particle 2's worldline corrections) or retarded time corrections $\tau_{2,\mr{ret}}^{(n)}$. For instance,
\begin{center}
\begin{equation}
  \begin{tikzpicture}
    \draw[very thick](4.0,0.75) -- ( 5.0,0.75);
    \draw[very thick, double](5.0, 0.75) -- (6.0,0.75);
  \end{tikzpicture}
\label{aldfig}
\end{equation}
\end{center}
is a $2^{\mr{nd}}$ self force (ALD) contribution which gives
\begin{alignat}{3}
& \nonumber e^2\sum_{n=0}^{3} (d_{\tau_1}^n z_{1}^{(1)}) \frac{2 e^2 q_1^2\partial }{3\partial(d_{\tau_1}^n z_{1})}(\dddot{z}_1+\ddot{z}_1^2 \dot{z}_1^{\mu})  =e^4 (2 q_1^2/3) (\dddot{z}^{(1)}_1).
 \end{alignat}
Note that there are no photon lines in diagram (\ref{aldfig}) since ALD force is not due to the interaction between two particles. We thus simply have a second order worldline correction (the double line on the right) produced by first order worldline correction (single line on the left) through the ALD force term.
\subsection{$1^{\mr{st}}$ order}
At leading order, we can substitute the zeroth-order worldlines (\ref{zeroth}) into the formula for retarded field tensor (\ref{field}), the self-force can be neglected. The only relevant diagram is given here.
\begin{equation}
\begin{tikzpicture}
\begin{scope}[very thick, every node/.style={sloped,allow upside down}]
\node[text width=0.25cm] at (-0.5,0) {2};
\node[text width=0.25cm] at (-0.5,1.5) {1};
\draw[dashed](0,0)--(2.5,0);
\draw[dashed](0,1.5)--(3.0,1.5);
\draw[very thick](3.0,1.5)--(5,1.5);
\draw[snake it](2.5,0)--node {\midarrow}(3.0,1.5);
\end{scope}
\end{tikzpicture}
\label{fig1st}
\end{equation}
 The corresponding equations are
\begin{alignat}{2}
& e^2 m_1 \frac{d^2 z_1^{(1)\mu}}{d\tau^2} = e^2 q_1  u_{1,\nu}  F_2^{\mu \nu}(z_1^{(0)})[z_2^{(0)}(\tau_{2,\mr{ret}}^{(0)})], \\&
 F_2^{(0)\mu \nu} = \frac{2  q_2 \rho_2^{(0)[\mu} u_{2}^{\nu]}}{ (r_2^{(0)})^3},\quad r_2^{(0)} =u_2\cdot \rho_2^{(0)},
\label{basic1st} \\&
\rho_2^{(0)\mu} = [z_1^{(0)}(\tau_1)-z_2^{(0)}(\tau_{2,\mr{ret}}^{(0)})]^{\mu}.
\end{alignat}
The retarded proper-time at zeroth order $\tau_{2,\mr{ret}}^{(0)}$ is obtained by solving
\begin{alignat}{3}
& |\rho^{(0)}|^2=(|b|^2+\tau_1^2+\tau_{2,\mr{ret}}^{(0),2}-2 \tau_1 \tau_{2,\mr{ret}}^{(0)} \gamma) = 0, \nonumber  \\&
\Rightarrow \tau_{2,\mr{ret}}^{(0)} =  \gamma \tau_1 - \sqrt{|b|^2 + (\gamma v)^2 \tau_1^2}.
\end{alignat}
Defining
 \be
  r_{21} = r^{(0)}_2 = \sqrt{|b|^2+(\gamma v)^2 \tau_1^2 },
 \ee
  we can now write the field tensor explicitly as
\begin{equation}
 F^{\mu \nu}_2=\frac{2  q_2}{ r_{21}^3}(b^{[\mu}u_2^{\nu]}+\tau_1 u_1^{[\mu} u_2^{\nu]}),
\end{equation}
which we substitute into Eq.~(\ref{basic1st}) to get the 4-acceleration $a_a^{\mu}=\ddot{z}^{\mu}(\tau_a)$ for particle 1,
\begin{alignat}{3} 
&  a_1^{(1)\mu}  = \frac{ q_1 q_2[\gamma b^\mu + \tau_1 (\gamma u_1 - u_2)^\mu ]}{ m_1  r_{21}^3}.
\\ & \text{This can be integrated twice w.r.t.~$\tau_1$ to get} \nonumber
\\ \label{P11o}
&  v_1^{(1)\mu}  = \frac{ q_1 q_2[\gamma^2 v s_1 b^{\mu}  +|b|^2 (-\gamma u_1^{\mu}+u_2^{\mu})]}{ \gamma v^2  m_1 |b|^2 r_{21} },
\\ \nonumber
&  z_1^{(1)\mu}  = \frac{ q_1 q_2[\gamma v s_1 b^{\mu}  +|b|^2 \log(\gamma v s_1)(-\gamma u_1^{\mu} + u_2^{\mu})]}{ m_1 |b|^2  (\gamma v)^3}, \label{1wd} \\&
\end{alignat}
as the $1^{\mr{st}}$ order velocity [$v_a^{\mu} = \dot{z}^{\mu}_a(\tau_a)$] and worldline corrections, where 
\be
s_1 = \gamma v \tau_1 + r_{21}(\tau_1)=\gamma v \tau_1 + \sqrt{|b|^2+(\gamma v)^2 \tau_1^2}.
\ee
 The constants of integration have been chosen such that initial velocity and impact parameter remain unchanged, i.e.~$v_{1}^{(1)\mu}\rightarrow 0$ and $z_1^{(1)}\cdot b = 0$  for $\tau_{1} \rightarrow -\infty$.  
The corrections for particle 2 can be obtained by sending $(1\leftrightarrow 2)$ and $b^\mu\rightarrow -b^\mu$ to get
\begin{alignat}{2}
&  z_2^{(1),\mu} =  \frac{ q_1 q_2[-\gamma v s_2 b^{\mu}  + |b|^2 \log(\gamma v s_2)(-\gamma u_2^{\mu} + u_1^{\mu})]}{4 m_2 |b|^2 \pi (\gamma v)^3},\nonumber
\\&  s_2 = \gamma v \tau_2 + r_{12}, \quad r_{12}(\tau_2) = \sqrt{|b|^2 + (\gamma v \tau_2)^2}.
\label{P21o}
\end{alignat}

We will find that the complexity of expressions (for force correction) greatly increases at $2^{\mr{nd}}$ order and beyond. These expressions can be rather simplified by the use of certain substitutions and variables. Thus, at $2^{\mr{nd}}$ and $3^{\mr{rd}}$ order, we use the variables $s_1 = \gamma v \tau_1 + r_{21} $ as the worldline variable (in place of $\tau_1$) to express all integrands. This helps get rid of all square roots involving expressions of $\tau_1$ while also making the integrals analytically tractable. In some places, we also use the rapidity parameter $\cosh(\phi) = \gamma$ which simplifies terms involving $\mr{arcsinh}(\gamma v)=\mr{arcosh}(\gamma)$, etc.

\subsection{$2^{\mr{nd}}$ order}
We can now use the $1^{\mr{st}}$ order worldline corrections ($z_1^{(1)}$ and $z_2^{(1)}$) to compute the worldlines at $2^{\mr{nd}}$ order. We need to evaluate the $2^{\mr{nd}}$ order force correction, $f^{(2)}=m_1 \ddot{z}_1^{(2)}$, that is
\begin{alignat}{3}
\label{2ndorderlaw}&  m_1 \ddot{z}_1^{(2),\mu} = [e^4] \left[ e^2 q_1 F_2^{\mu\nu}\dot{z}_{1,\nu} +   \frac{2e^2 q_1^2}{ 3} \Big(\dddot{z}_1^\mu+\ddot{z}_1^2 \dot{z}_1^{\mu}\Big) \right],\nonumber \\& \\&
 z_1\rightarrow z_1^{(0)}(\tau_1) + e^2z_1^{(1)}(\tau_1),\quad z_{2,\mr{ret}} \rightarrow z_{2,\mr{ret}}^{(0)}+ e^2 z_{2,\mr{ret}}^{(1)},\nonumber \\& \tau_{2,\mr{ret}} = \gamma \tau_1 - r_{21} +e^2 \tau_{2,\mr{ret}}^{(1)}.
	\end{alignat}
Here $[x^n]f(x)$ is the coefficient of $x^n$ in $f(x)$. To evaluate this, we split the contributions from corrections to $z_1$, $z_2$, and $\tau_{2,\mr{ret}}$ as mentioned before. The $1^{\mr{st}}$ order worldline corrections $z_1^{(1)}$ and $z_2^{(1)}$, were derived in the last subsection (see Eq.~(\ref{1wd}) and Eq.(\ref{P21o})), the $1^{\mr{st}}$ order retarded time correction $\tau_{2,\mr{ret}}^{(1)}$, in terms of $z_a^{(1)}$ is given in Eq.~(\ref{1ret}). The relevant diagrams are shown and labeled here,
\begin{equation*}
\begin{tikzpicture}
\begin{scope}[very thick, every node/.style={sloped,allow upside down}]
\node[text width=0.25cm] at (-1.5,0.75) {$e^4 f^{(2)}=$};
\node[text width=0.25cm] at (1.1,2.5) {$\mr{I}$};
\node[text width=0.25cm] at (5.1,2.5) {$\mr{II}$};
\draw[dashed](0,0)--(1.1,0);
\draw[very thick](0,1.5)--(1.4,1.5);
\draw[very thick, double](1.4,1.5)--(2.2,1.5);
\draw[snake it](1.1,0)--node{\midarrow}(1.4,1.5);
\node[text width=0.25cm] at (3.0,0.75) {+}; 
\draw[dashed](4.0,1.5)--(5.4,1.5);
\draw[very thick, double](5.4,1.5)--(6.2,1.5);
\draw[snake it](5.1,0)--node{\midarrow}(5.4,1.5);
\draw[very thick](4,0)--(5.1,0.0);
\end{scope}
\end{tikzpicture}
\end{equation*}

\begin{equation}
\label{fig2nd}
\begin{tikzpicture}
\begin{scope}[very thick, every node/.style={sloped,allow upside down}]
\node[text width=0.25cm] at (1.1,2.5) {$\mr{III}$};
\node[text width=0.25cm] at (5.1,2.5) {$\mr{IV}$};
\node[text width=0.25cm] at (-0.5,0.75){+};
\node[text width=0.25cm] at (1.1,-0.25){$\tau^{(1)}_{2,\mr{ret}}(z_a^{(1)})$};
\draw[dashed](0,0)--(1.1,0);
\draw[dashed](0,1.5)--(1.4,1.5);
\draw[very thick, double](1.4,1.5)--(2.2,1.5);
\draw[snake it](1.1,0)--node{\midarrow}(1.4,1.5);
\node[text width=0.25cm] at (3.0,0.75){+};
\draw[very thick](4.0,0.75) -- ( 5.4,0.75);
\draw[very thick, double](5.4, 0.75) -- (6.2,0.75);
\end{scope}
\end{tikzpicture}
\end{equation}

The diagrams in Fig.~(\ref{fig2nd}) can be used to compute the second-order correction to the force ($ f^{(2)} =  m_1 \ddot{z}_1^{(2)}$) using the rules given in Sec.~(\ref{diag}). This has been done explicitly in Appendix.~\ref{2oapp}. One needs to then integrate each term twice w.r.t $\tau_1$ to get the $2^{\mr{nd}}$ order worldline corrections, $z_1^{(2)}(\tau_1)$. We impose the same boundary conditions as in first order worldline corrections (see the paragraph below Eq.~(\ref{P11o})). We briefly discuss the integration process below.

 Diagrams $\mr{I}$, $\mr{II}$, and $\mr{III}$ are due to corrections to the Lorentz force term ($eq_1F_2^{\mu\nu}\dot{z}_{1,\nu}$) from $1^{\mr{st}}$ order worldline corrections. Their contribution to second-order force correction consists of various terms such as
\begin{alignat}{3}
\frac{s_1}{r_{21}^4}, \frac{s_1^3}{r_{21}^3 r_{12}^3}, \frac{s_1^2}{r_{21}^4}, \frac{\log(s_1)}{r_{21}^3}, \frac{s_1 \log(s_1)}{r_{21}^4}\ldots 
\end{alignat}
where $\tau_2$ and all quantities that depend on it \begin{equation}r_{12} = \sqrt{|b|^2+\gamma^2 v^2 \tau_2^2}, \quad  s_2=\gamma v \tau_2 + r_{12},\end{equation} are to be evaluated at zeroth order retarded time \begin{equation}
\tau_{2,\mr{ret}}^{(0)}=\gamma \tau_1 - \sqrt{|b|^2+\gamma^2 v^2 \tau_1^2}.
\end{equation}
 Thus, there are many square roots and nested square roots in the expression, which makes analytical integration complicated. The square roots can be eliminated by rewriting them as functions of $s_1=\gamma v \tau_1 + r_{21}$ along with the relations given in Eq.~(\ref{conv1}).
With these simplifications, we get expressions in terms of $s_1$ that only contain rational functions and logarithms as the ingredients, i.e. terms of the form
\begin{alignat}{3}
\nonumber &\frac{s_1^3\mr{Poly}(s_1) \log(s_1)}{(m_1 \text{ or } m_2)(1+s_1^2)^5}, \quad \frac{s_1^3\mr{Poly}(s_1)}{(m_1 \text{ or } m_2)(1+s_1^2)^5},\\& \frac{s_1^3\mr{Poly}(s_1)}{m_2(1+s_1^2)^4(e^{2\phi}+s_1^2)^3},
\end{alignat}
where $\mr{Poly}(s_1)$ stands for polynomial function of $s_1$.
 Such terms can be easily integrated twice w.r.t.~$\tau_1$ via the relation $d\tau_1 = (d\tau_1/ds_1)ds_1 = \{r_{21}/(s_1\sinh(\phi))\}ds_1$ to obtain their contribution to the $2^{\rm nd}$ order worldline correction ($z^{(2)}_1$).
 The reader is reffered to Appendix.~\ref{2oapp} for a detailed discussion of the contributions from various diagrams and the process of integration.

Diagram $\mr{IV}$ comes from the ALD force term [$(2e^2 q_1^2/3) \Big(\dddot{z}_1+\ddot{z}_1^2 \dot{z}_1^{\mu}\Big)$] and only its first term contributes at $2^{\mr{nd}}$ order, which is proportional to the $1^{\mr{st}}$ order jerk $\dddot{z}_1^{(1)}$. Thus, its contribution to 2nd order worldline correction is simply
\begin{alignat}{3}
\label{2orc}
\nnm z^{(2)}_1|_{\mr{IV}}&=\frac{2 e^4 q_1^2}{3 m_1} \dot{z}_1^{(1),\mu} \\&= \frac{2 e^4 q_1^3 q_2[\gamma^2 v s_1 b^{\mu}  + |b|^2 (-\gamma u_1^{\mu}+u_2^{\mu})]}{3 \gamma v^2  m_1^2 |b|^2 r_{21} }. \nonumber \\&
\end{alignat}
After integration, collecting the contribution of all diagrams gives us the the complete 2nd order worldline correction $z_1^{(2)}(\tau_1)$. The explicit expression for $z^{(2)}_1(\tau_1)$ is given in Eq.~(\ref{2ndworld}). We can now use this to evaluate the force correction and subsequently impulse at $3^{\mr{rd}}$ order where dissipative effects appear for the first time.
\subsection{$3^{\mr{rd}}$ order}
\label{3c}
At $3^{\mr{rd}}$ the force correction given by $ m_1 \ddot{z}^{(3)}_1= f^{(3)}$, gains contributions from both $1^{\mr{st}}$ and $2^{\mr{nd}}$ order worldline corrections. The $1^{\mr{st}}$ order corrections contribute quadratically as $e^4 (z^{(1)}_a)^2$, whereas the $2^{\mr{nd}}$ order corrections contribute linearly via $e^4 z^{(2)}_a$. Further, their contributions are independent, in the sense that there are no cross terms since $e^2 z^{(1)}_a \times e^4 z^{(2)}_b \sim e^6$. It is thus possible to separate their contributions. As before, there are also corrections to retarded time $\tau_{2,\mr{ret}}$ to take into account. Since the retarded time is obtained by solving a quadratic relation $|z_1-z_2|^2=0$, the $2^{\mr{nd}}$ order retarded time correction $\tau_{2,\mr{ret}}^{(2)}$ gains contribution from both $1^{\mr{st}}$ and $2^{\mr{nd}}$ order worldline corrections, once again quadratic in $e^2 z^{(1)}_a$ and linear in $e^4 z^{(2)}_a$. Since there are no cross contributions ($z_a^{(1)}\times z_b^{(2)}$), we can also separate the $2^{\mr{nd}}$ order retarded time correction $\tau_{2,\mr{ret}} = \tau_{2,\mr{ret}}^{(1)}(z_a^{(1)}) +  \tau_{2,\mr{ret}}^{(1)}(z_a^{(2)})$. 

Thus, we can divide the contributions into quadratic (in $z^{(1)}$) and linear (in $z^{(2)}$) types. The various diagrams corresponding to these contributions are given in Fig.~(\ref{fig3rdq}) and Fig.~(\ref{fig3rdl}) respectively, where we have accordingly separated the force correction, $f^{(3)}=f^{(3)}(z_a^{(1)})+f^{(3)}(z_b^{(2)})$ .

\begin{widetext}
\begin{equation}
\begin{tikzpicture}
\begin{scope}[very thick, every node/.style={sloped,allow upside down}]
\node[text width=0.25cm] at 	(-2.5,0.75){$e^6f^{(3)}(z_a^{(1)})=$};
\node[text width=0.25cm] at 	(1.25,2.0){$\mr{I}$};
\node[text width=0.25cm] at 	(4.75,2.0){$\mr{II}$};
\node[text width=0.25cm] at 	(8.25,2.0){$\mr{III}$};
\node[text width=0.25cm] at 	(11.75,2.0){$\mr{IV}$};
\node[text width=0.25cm] at 	(1.25,-1.5){$\mr{V}$};
\node[text width=0.25cm] at 	(4.75,-1.5){$\mr{VI}$};
\node[text width=0.25cm] at 	(8.25,-1.5){$\mr{VII}$};
\node[text width=0.25cm] at 	(11.75,-1.5){$\mr{VIII}$};
\node[text width=0.25cm] at (3.0,0.75) {+};
\node[text width=0.25cm] at (3,-2.75) {+};
\node[text width=0.25cm] at (6.5,0.75) {+};
\node[text width=0.25cm] at (6.5,-2.75) {+};
\node[text width=0.25cm] at (10,0.75) {+};
\node[text width=0.25cm] at (10,-2.75) {+};
\node[text width=0.25cm] at (13.5,0.75) {+};
\draw[very thick,double](1.5,1.5)--(2.5,1.5);
\draw[very thick](1.5,1.6)--(2.5,1.6);
\draw[very thick](0,1.25)--(1.5,1.5);
\draw[very thick](0,1.5)--(1.5,1.5);
\draw[snake it](1.25,0)--node{\midarrow}(1.50,1.5);
\draw[dashed](0,0)--(1.25,0);
\draw[very thick](5,1.6)--(6,1.6);
\draw[very thick,double](5,1.5)--(6,1.5);
\draw[very thick](3.5,0)--(4.75,0);
\draw[very thick](3.5,0.25)--(4.75,0);
\draw[snake it](4.75,0)--node{\midarrow}(5.0,1.5);
\draw[dashed](3.5,1.5)--(5,1.5);
\draw[very thick](7.0,0)--(8.25,0);
\draw[very thick](7.0,1.5)--(8.5,1.5);
\draw[snake it](8.25,0)--node{\midarrow}(8.5,1.5);
\draw[very thick](8.5,1.6)--(9.5,1.6);
\draw[very thick,double](8.5,1.5)--(9.5,1.5);
\draw[dashed](10.5,0)--(11.75,0);
\draw[snake it](11.75,0)--node{\midarrow}(12.0,1.5);
\draw[very thick](10.5,1.5)--(12.0,1.5);
\draw[very thick](12.0,1.6)--(13.0,1.6);
\draw[very thick,double](12.0,1.5)--(13.0,1.5);
\node[text width=0.25cm]at (11.75,-0.25) {$\tau^{(1)}_{2,\mr{ret}}(z_a^{(1)})$};
\draw[dashed](0,-2)--(1.5,-2);
\draw[snake it](1.25,-3.5)--node{\midarrow}(1.5,-2);
\draw[very thick](0,-3.5)--(1.25,-3.5);
\draw[very thick](1.5,-1.9)--(2.5,-1.9);
\draw[very thick,double](1.5,-2.0)--(2.5,-2.0);
\node[text width=0.25cm]at (1.25,-3.75) {$\tau^{(1)}_{2,\mr{ret}}(z_a^{(1)})$};
\draw[dashed](3.5,-2)--(5.0,-2);
\draw[snake it](4.75,-3.5)--node{\midarrow}(5.0,-2);
\draw[dashed](3.5,-3.5)--(4.75,-3.5);
\draw[very thick](5.0,-1.9)--(6.0,-1.9);
\draw[very thick,double](5.0,-2.0)--(6.0,-2.0);
\node[text width=0.25cm]at (4.75,-3.75) {$(\tau^{(1)}_{2,\mr{ret}}(z_a^{(1)}))^2$};
\draw[dashed](7,-2)--(8.5,-2);
\draw[snake it](8.25,-3.5)--node{\midarrow}(8.5,-2);
\draw[dashed](7,-3.5)--(8.25,-3.5);
\draw[very thick](8.5,-1.9)--(9.5,-1.9);
\draw[very thick,double](8.5,-2.0)--(9.5,-2.0);
\node[text width=0.25cm]at (8.25,-3.75) {$\tau^{(2)}_{2,\mr{ret}}(z_a^{(1)})$};
\draw[very thick](10.5,-2.75)--(11.5,-2.75);
\draw[very thick](10.5,-3.00)--(11.5,-2.75);
\draw[very thick](11.5,-2.65)--(12.5,-2.65);
\draw[very thick, double](11.5,-2.75)--(12.5,-2.75);
\end{scope}
\end{tikzpicture}
\label{fig3rdq}
\end{equation}

\begin{equation}
\begin{tikzpicture}
\begin{scope}[very thick, every node/.style={sloped,allow upside down}]	
\node[text width=0.25cm] at 	(-2.5,0.75){$e^6 f^{(3)}(z_a^{(2)})=$};
\node[text width=0.25cm] at 	(1.25,2.0){$\mr{IX}$};
\node[text width=0.25cm] at 	(5.5,2.0){$\mr{X}$};
\node[text width=0.25cm] at 	(9.75,2.0){$\mr{XI}$};
\node[text width=0.25cm] at (3.25,0.75) {+};
\node[text width=0.25cm] at (7.5,0.75) {+};
\draw[very thick](1.5,1.6)--(2.5,1.6);
\draw[very thick,double](1.5,1.5)--(2.5,1.5);
\draw[very thick,double](0,1.5)--(1.5,1.5);
\draw[snake it](1.25,0)--node{\midarrow}(1.50,1.5);
\draw[dashed](0,0)--(1.25,0);
\draw[very thick](5.75,1.6)--(6.75,1.6);
\draw[very thick,double](5.75,1.5)--(6.75,1.5);
\draw[dashed](4.25,1.5)--(5.75,1.5);
\draw[snake it](5.5,0)--node{\midarrow}(5.75,1.5);
\draw[very thick, double](4.25,0)--(5.50,0);
\draw[dashed](8.5,0)--(9.75,0);
\draw[dashed](8.5,1.5)--(10.0,1.5);
\draw[snake it](9.75,0)--node{\midarrow}(10.0,1.5);
\draw[very thick](10.0,1.6)--(11.0,1.6);
\draw[very thick,double](10.0,1.5)--(11.0,1.5);
\node[text width=0.25cm] at (9.75,-0.25) {$\tau^{(2)}_{2,\mr{ret}}(z_a^{(2)})$};
\end{scope}
\end{tikzpicture}
\label{fig3rdl}
\end{equation}
\end{widetext}
\paragraph{Quadratic contributions from $z^{(1)}_a$ and derivatives:} 
Diagrams in Fig.~(\ref{fig3rdq}) are contributions to $3^{\mr{rd}}$ force corrections ($f^{(3)}$) that are quadratic in $1^{\mr{st}}$ order worldline corrections (and derivatives).

 Among these, diagrams $\mr{I}$ through $\mr{VII}$ are corrections to the Lorentz force term ($e^2 q F_{2}^{\mu\nu}\dot{z}_{1,\nu}$) term and give expressions composed of similar terms. Explicit calculation gives a linear combination with terms having $\log(b s_1)$, $\log(b s_1)^2$, and polynomial functions of $(s_1)$ (written as Poly($s_1$) from now on) as numerators, and $(1+s_1^2)^n\times(e^{2\phi}+s_1^2)^m$ as denominators. This is in accordance with what we expect from the form of $1^{\mr{st}}$ order worldline corrections, see Eqs.~(\ref{P11o}) and (\ref{P21o}), and the expression for the field tensor Eq.~(\ref{field}), after writing everything in terms of $s_1$ and using the relations in Eq.~(\ref{conv1}). To evaluate their contribution to the impulse, we multiply by the Jacobian factor and integrate over the entire worldline $\tau_1 \in (-\infty,\infty)$, or $s_1 \in (0,\infty)$ using the relation $d\tau_1 = (d\tau_1/ds_1)\times ds_1 = \{r_{21}/[s_1\sinh(\phi)]\}ds_1$. The integrals to be evaluated are a linear combination of the following types of terms,
\begin{alignat}{3}
&\frac{\mr{Poly}(s_1)}{(1+s_1^2)^4 (e^{2\phi}+s_1^2)^3},\quad \nonumber \frac{\log(s_1)\mr{Poly}(s_1)}{(1+s_1^2)^4(e^{2\phi}+s_1^2)^3},\\&
\frac{\log(s_1)^2}{(1+s_1^2)^4}.
\end{alignat}
{\tt Mathematica} is once again able to evaluate them in a relatively short time once they have been separated into these types. Although we only require the definite integral, it is convenient for some terms to evaluate the indefinite integrals and then take the limits.

Diagram $\mr{VIII}$ comes from the second term in the ALD force [$(2e^2 q_1^2/3) \Big(\dddot{z}_1+\ddot{z}_1^2 \dot{z}_1^{\mu}\Big)$] (the first term does not contribute to the impulse). It's contribution at $3^{\mr{rd}}$ order is given by $(2/3) \, e^6 q_1^2 (\ddot{z}_1^{(1)})^2 u_1^{\mu}$ and it contributes to the radiative part of the impulse.  It is worth noting here that this term is proportional to the initial momentum $p_1 = m_1 u_1$ and thus it cannot be the sole contribution to the radiative part of the impulse, even in the test-body limit, since that would change the rest mass. Additional contributions to the radiative part of the impulse come from corrections to the Lorentz-force term ($e^2q_1F_2^{\mu\nu}\dot{z}_{1\nu}$) (included in other diagrams). Together, they lead to a radiative contribution to the impulse that conserves the rest mass. Specifically in the test-body limit, the only other contribution to radiative impulse comes from the contribution of $z^{(2)}|_{\mr{IV}}$, see Eq.~(\ref{2orc}) to the Lorentz force term ($e^2q_1F_2^{\mu\nu}\dot{z}_{1\nu}$). This additional contribution was ignored in Ref.~\cite{Kosower:2018adc} in their calculation of net impulse in the test-body limit, leading to an incorrect result for the net radiative impulse in the test-body case that was proportional to $p_1$ .

\paragraph{Linear contributions from $z^{(2)}$:}  
Diagrams in Fig.~(\ref{fig3rdl}) are contributions to $3^{\mr{rd}}$ force corrections ($f^{(3)}$) that are linear in $2^{\mr{nd}}$ order worldline corrections (and derivatives).

All three diagrams are due to corrections to the Lorentz-force term ($e^2 q_1 F^{\mu\nu}_2 \dot{z}_{1,\nu}$). Only the first term in the ALD force [$(2e^2 q_1^2/3) \Big(\dddot{z}_1+\ddot{z}_1^2 \dot{z}_1^{\mu}\Big)$] contributes linearly in $2^{\mr{nd}}$ order worldline corrections ($z^{(2)}$) at $3^{\mr{rd}}$ order and we have not included it's contribution since it is a total time derivative of a quantity that vanishes at infinite past/future (acceleration), and we are only interested in the net impulse.

Thus, all three diagrams give expressions composed of similar terms. We get a linear combination of terms with $\arctan(\gamma v s_1/|b|)$, $\arctan(s_1)$, $\log(|b|\gamma v s_1)$, $\arctan(r_{21}/|b|)$, and $\mr{Poly}(s_1)$ in numerators, and $(1+s_1^2)^m \times (e^{2\phi}+s_1^2)^n$ in denominators. This is in accordance with the expression for $2^{\mr{nd}}$ order worldline correction given in Eq.~(\ref{2ndworld}) and the expression for the field tensor Eq.~(\ref{field}), after expressing everything in terms of $s_1$ using the relations in Eq.~(\ref{conv1}).

 Once again, to get the impulse, we multiply with the Jacobian factor and integrate over the whole worldline $s_1\in(0,\infty)$, where 
{\tt Mathematica} has no trouble evaluating the integrals. Indefinite integrals sometimes give non-elementary polylogarithms, but they do not pose a challenge as far as computing the impulse is concerned. This however indicates that one might have to deal with non-elementary functions starting from $4^{\rm th}$ order (i.e.~$e^8$).
\comm{there are theorems about which integrands lead to elementary integrals; is there more that can be said here about why nonelementary functions appear?}
 
Finally, performing the integration and adding the quadratic and linear contributions to the impulse, we obtain the complete expression for the $3^{\mr{rd}}$ correction to the impulse $\Delta p_1$ ($\Delta p_1^{(3)}$). The explicit expression for the same is given below in Eq.~(\ref{Dp13}).

Now, we will use the key results of this section (i.e. $1^{\mr{st}}$ and $2^{\mr{nd}}$ order worldline corrections), and the $3^{\mr{rd}}$ order correction to the impulse $\Delta p^{(3)}_1$, to compute observables associated with the scattering process (e.g., the impulses).

\section{Observables in the scattering process}

\subsection{Net impulses}\label{sec:kchi}

The net impulse (defined as change in momentum) of particle 1 to $3^{\mr{rd}}$ order can be written as
 \begin{equation}
 \Delta p_1 = 
e^2 \Delta p_1^{(1)} + e^4 \Delta p_1^{(2)} + e^6 \Delta p_1^{(3)}.
 \end{equation}
 We can derive $\Delta p_1^{(1)}$ and $\Delta p_1^{(2)}$ from the time-dependent worldlines upto $2^{\rm nd}$ order given in Eq.~(\ref{1wd}) and Eq.~(\ref{2ndworld}). With  $z_1(\tau_1) = z_1^{(0)} + e^2 z_1^{(1)} + e^4 z_1^{(2)}$, we have
\begin{alignat}{3}
& e^2 \Delta p_1^{(1)} + e^4 \Delta p_2^{(2)} = \lim_{\tau_1\rightarrow \infty} m_1 \dot{z}_1(\tau_1) - \lim_{\tau_1\rightarrow -\infty} m_1 \dot{z}_1(\tau_1), \nnm
\\
&= \frac{2 e^2q_1 q_2 }{ v |b|^2 } b^{\mu} -  \frac{e^4q_1^2 q_2^2 E (  v^2 |p| M \pi b^{\mu} + 4 E |b| p^{\mu})}{2 m_1^2 m_2^2 \gamma^2 v^4 |b|^3 }.  \nnm \\& \label{imp2m}
\end{alignat}
We can now add to Eq.~(\ref{imp2m}) the $3^{\mr{rd}}$ order correction to the impulse. We described the process of computing $\Delta  p^{(3)}_1$ in Sec.~\ref{3c}, and the result is 
\begin{widetext}
\begin{alignat}{3}\label{Dp13}
&\Delta p_{1}^{(3)}  = - \frac{2 q_1^3 q_2^3 [\gamma m_1^2 +\gamma m_2^2 + 2 m_1 m_2 (1+\gamma^4 v^2)]}{ m_1^2 m_2^2 |b|^4 \gamma^5 v^5 }b^{\mu} +  \frac{\pi M E^2 q_1^3 q_2^3}{ m_1^3 m_2^3 |b|^3  \gamma^3 v^4}p^{\mu} \nnm \\& + \frac{4 q_1^2 q_2^2}{3m_1^2 m_2^2 |b|^4  \gamma^2 v^4}b^{\mu}\bigg[(m_2^2 q_1^2 +  m_1^2 q_2^2) \gamma^2 v^2 - 3m_1m_2q_1q_2 \Big(\gamma-\frac{\mr{arctanh}\, v}{\gamma v}\Big) \bigg] \nnm \\& + \frac{\pi q_1^2 q_2^2}{4 \gamma^2 v^2 |b|^3}\bigg[\Big(\frac{q_1^2}{m_1^2} \gamma + \frac{q_2^2}{m_2^2} \Big) \frac{3 \gamma^2 +1}{3 \gamma v} - \frac{q_1 q_2}{m_1 m_2} \mathcal{F}(\gamma)\bigg](u_2^{\mu}-\gamma u_1^{\mu}),
\end{alignat}
\end{widetext}
where
\begin{alignat}{3}
\mathcal{F}(\gamma) &= \frac{1}{(\gamma v)^3}  \bigg[(3\gamma^2+1)\Big(\gamma - \frac{\text{arctanh}\,v}{\gamma v} \Big) - 4 (\gamma -1)^2\bigg].\label{helpg}
\end{alignat}
We can simplify this somewhat bulky expression by splitting the result into conservative and radiative parts and writing these in terms of the scattering angles and the total radiated momentum. To do so, we first evaluate the total scattering angle as
\begin{alignat}{3}
\sin \chi = \frac{-\Delta p_1 \cdot b}{|p||b|}+O(e^8),
\end{alignat}
which yields
\begin{alignat}{3}
& \chi = \chi_{\mr{cons}}+\chi_{\mr{rad}}, \\
&\chi_{\text{cons}}= \frac{ 2 e^2 E q_1 q_2  }{ m_1 m_2 \gamma |b| v^2 } - \frac{\pi e^4  M E q_1^2 q_2^2}{2 m_1^2 m_2^2 \gamma^2 |b|^2 v^2 } 
\nnm
\\ & +\frac{e^6 q_1^3 q_2^3 E  [(m_1^2+m_2^2)(4 \gamma^2 - 6)- 4 m_1 m_2 \gamma ( \gamma^2 -  3 v^2 )]}{3 m_1^3 m_2^3 |b|^3 \gamma^5 v^6 }\nonumber, \\ \\&\label{chirad3}
\chi_{\mr{rad}}=
 -\frac{4 e^6 q_1^2 q_2^2 E }{3 m_1^3 m_2^3 |b|^3 \gamma^3 v^5} \Big[(\gamma v)^2(m_2^2 q_1^2 + m_1^2 q_2^2) \nonumber \\& - 3 m_1 m_2 q_1 q_2  \Big(\gamma - \frac{\text{arctanh}\,v}{\gamma v}\Big)\Big]. 
\end{alignat}
where the conservative (radiative) scattering angle has been defined to be the part that is even (odd) under $v \rightarrow -v$.
We can now define the conservative part of the impulse as the result of a simple rotation in the scattering plane by the angle $\chi_\mr{cons}$, and define the radiative part of the impulse to be the remainder,
\begin{alignat}{3}
&\Delta p_{1,\mr{cons}}^\mu=|p|\sin \chi_\mr{cons}\frac{b^\mu}{|b|}+(\cos\chi_\mr{cons}-1)p^\mu, \\
&\Delta p_{1,\mr{rad}} = \Delta p_1 - \Delta p_{1,\mr{cons}}
\end{alignat}
As we will see below, $\Delta p_{1,\mr{rad}}$ is fully determined by the radiative contribution to the scattering angle $\chi_\mr{rad}$ together with the total radiated momentum $K^\mu$.  The latter is found from summing the impulse (\ref{Dp13}) on particle 1 and its $(1\leftrightarrow2)$ version, yielding
 \begin{alignat}{3}
  K^{\mu} &= -\Delta p_1^{\mu} - \Delta p_2^{\mu}=-\Delta p_{1,\mr{rad}}^{\mu}-\Delta p_{2,\mr{rad}}^{\mu} \phantom{\bigg|}\nnm\\
\label{power}
  &= \frac{\pi e^6 q_1^2 q_2^2}{4 |b|^3} \bigg[\Big(\frac{q_1^2}{m_1^2} u_1^{\mu} + \frac{q_2^2}{m_2^2} u_2^{\mu} \Big) \frac{3 \gamma^2 +1}{3 \gamma v} \nonumber
\\&\quad- \frac{q_1 q_2}{m_1 m_2} \frac{u_1^{\mu} + u_2^{\mu}}{\gamma +1} \mathcal{F}(\gamma) \bigg],
\end{alignat}
where $\mathcal{F}(\gamma)$ was defined in Eq.~(\ref{helpg}).
Now, the conservative part of the impulse of each particle separately conserves the particle's rest mass ( i.e.~$(p_{1,i} + \Delta p_{1,\mr{cons}})^2 = m_1^2$). Thus, the leading-order radiative effect must satisfy $\Delta p_{1,\mr{rad}} \cdot u_1 =0 $, so that the total impulse conserves the rest mass to $3^{\mr{rd}}$ order. Furthermore, $\Delta p_{1,\mr{rad}}$ is  solely responsible for both radiated momentum and radiative part of the scattering angle. Thus, we obtain the following expression for $\Delta p_{1,\mr{rad}}$ 
 \begin{equation}
 \Delta p_{1,\text{rad}}^{\mu} = \frac{K\cdot u_2}{(\gamma v)^2}(u_2^{\mu} - \gamma u_{1}^{\mu} ) + |p| \chi_{\text{rad}} \frac{b^{\mu}}{|b|},
 \end{equation}
  as anticipated in Eq.~(\ref{dissp}). 

  \subsection{Radiation of angular momentum}
  \label{angmomsec}
  The emitted radiation also carries away angular momentum. Unlike radiated momentum, angular momentum loss can be seen already at $2^{\mr{nd}}$ order. Radiated angular momentum can be evaluated using the relations
  \begin{alignat}{3}
& J_{\mr{rad}}^{\mu} = -J^{\mu}_{\mr{f}}+	J^{\mu}_{\mr{i}},  
\\
& J^{\mu}_{\mr{f/i}} = \lim_{\tau_1,\tau_2\rightarrow \pm\infty}-\epsilon^{\mu}{}_{\nu \rho\sigma} \frac{p_1^{\nu} p_2^{\sigma}}{E} (z_1 - z_2)^{\sigma}, \\&
p_a^{\mu} = m_a \dot{z}_a^{\mu}.
\end{alignat}
Note that the evaluation of this quantity requires the complete time-dependent worldlines and thus we cannot evaluate it to $3^{\mr{rd}}$ order. However, we can evaluate the radiated angular momentum at $2^{\mr{nd}}$ order by substituting $z_a^{\mu}=z^{(0)\mu}_a+e^2 z_a^{(1)\mu}+e^4 z_a^{(2)\mu}$ to get
\begin{alignat}{3}
\label{angmomloss}
 J_{\mr{rad}}^{\mu}  &= -2\frac{e^4 q_1^2 q_2^2  }{   E |b|^2 \gamma v }I(v)\epsilon^{\mu}{}_{\nu \rho \sigma}b^{\nu} u_1^{\rho} u_2^{\sigma} + {\cal O}(e^6),
\\
I(v)&=  -\frac{2}{3}\gamma(\frac{q_1/m_1}{q_2/m_2} +\frac{q_2/m_2}{q_1/m_1}) +\frac{2}{ v^2} - \frac{2\,\text{arctanh}\,v}{\gamma^2 v^3}. 
\end{alignat}
As discussed in Sec.~\ref{sec:exec}, this leading ($2^\mr{nd}$) order change in angular momentum determines the leading ($3^\mr{rd}$) order radiative contribution to the scattering angle (\ref{chirad3}), via the relation (\ref{chiradfromJ}) above, as derived in Ref.~\cite{Bini:2012ji}.

 \subsection{High-energy limits of observables}
 \label{hel}

We define the high-energy (HE) limit by requiring that the system of particles have energies much higher than their total rest mass in the COM reference frame. The energy of the system in this frame is given by $E=\sqrt{m_1^2+m_2^2+2m_1m_2\gamma}$ and we choose it to be much larger than $m_1+m_2$ by setting $\gamma \gg 1$, which is the high-energy limit. Note that we are not sending $m_1$, $m_2$ to $0$, which is the massless limit. The high-energy/ultra-relativistic limit is not equivalent to the massless limit in EM. We will see the need for this distinction soon. We further require that $q_1 \sim q_2$ and $m_1 \sim m_2$ and they will be treated as fixed when taking the high-energy limit.
 
 The conservative scattering angle to $3^{\mr{rd}}$ order for EM in the HE limit is given by
 \begin{equation}
 {\chi_{\mr{cons},\mr{EM}}}_|{_{\mr{HE}}} \rightarrow \frac{4 e^2 q_1 q_2}{|b| E} -\frac{16 e^6 q_1^3 q_2^3}{3 |b|^3 E^3}.
\end{equation}
It is finite and well behaved unlike GR where the conservative part of the scattering angle exhibits a logarithmic divergence~$\propto \log(\gamma)$ at $3^{\mr{rd}}$ order that cancels against contributions from the radiative part.
The radiative part of the scattering angle in the HE limit is given by
\begin{alignat}{3}
\label{chidiv}
{\chi_{\mr{rad}}}_|{_{\mr{HE}}} \rightarrow -\frac{8 e^6 q_1 q_2}{3 E |b|}\Bigg(\frac{q_1^3 q_2}{m_1^2 |b|^2}+\frac{q_2^3 q_1}{m_2^2 |b|^2}-\overbrace{\frac{6 q_1^2 q_2^2}{E^2 |b|^2}}^{\mr{finite}}\Bigg).\nonumber \\&
\end{alignat}
Here, we see the importance of distinguishing between the HE limit and the massless limit.  The above expression is linearly divergent in $\gamma$ in the massless limit ($m_1,m_2\rightarrow 0$, $\gamma\rightarrow \infty$ while fixing $E$, the overbrace highlights the part that is finite in this limit).
  We encounter a similar situation for the radiated angular momentum, as well, which is given by
\begin{alignat}{3}
\label{angdiv}
& \frac{|J_{\mr{rad},\mr{EM}}|}{J}{|_{\mr{HE}}} \rightarrow -\frac{4 e^4}{3 }\Bigg(\frac{q_1^3q_2}{m_1^2 |b|^2}+\frac{q_2^3 q_1}{m_2^2 |b|^2}-\overbrace{\frac{6 q_1^2 q_2^2}{E^2 |b|^2}}^{\mr{finite}}\Bigg). \nonumber \\&
\end{alignat}
This kind of divergent behaviour for radiative scattering angle and radiated angular momentum is absent for gravity. This can be understood by noting that the divergence in these quantities comes from the ALD force, which is asymmetric in charges $\sim q_1^3 q_2$, $\sim q_1 q_2^3$. The expression for classical ALD (self-)force should be regarded as valid only when $e^2 q_a^2/ l \leq m_a$ (i.e.~the bare mass is positive, $m \sim m_0 + e^2 q_a^2/ l$) as argued in Ref.~\cite{Poisson:1999tv}, where $l$ is the size of the charged particle. In our setup, the size of the charged particles must be much smaller than the impact parameter $|b|$, and thus we must have $e^2 q_a^2 / m_a \ll |b|$.  It is easy to see that the above expressions diverge precisely when this condition is violated.
\comm{so the ALD force is not valid for massless particles, and/or are massless charged particles pathological? the latter might be true, since photons are rather unique in QFT and you can't have them charged, I think.}

However, the fraction of energy radiated in the COM~frame per mass-energy, $K \cdot u_{\mr{com}}/E$, seems to diverge in the HE limit regardless of masses. In the HE limit, it is given by
\begin{alignat}{3}
\label{ediv}
\nonumber \frac{K\cdot u_{\mr{com}}}{E}|_{\mr{HE}} \rightarrow \gamma \frac{\pi e^6 q_1 q_2}{2 |b|^3}\Bigg[&  \Bigg(\frac{q_1^3 q_2}{m_1^3} + \frac{q_2^3 q_1}{m_2^3}\Bigg) \\& - \frac{3  q_1^2  q_2^2}{ E^3 } \Bigg].
\end{alignat}
This diverges for $E\gg M$ ($\gamma \gg 1$), once again due to the self-force (asymmetric in charges) terms. It is important to note however that the divergence in fraction of energy radiated is also present in gravity for which the analogous result was recently obtained in Ref.~\cite{Herrmann:2021lqe}. Either way, this signals that the $e^2$- or $G$-expansion is not uniformly valid at arbitrarily high energies.

\subsection{Nonrelativistic limit}
\label{nrelsec}
 We now consider both particles be moving at nonrelativistic (NR) speeds in the COM~frame. This is achieved by simply sending $\gamma \rightarrow 1$, and we recover (to leading order) the Newtonian result $E=m_1+m_2$ (where $E$ is the total energy in the COM~frame). The radiated energy (in the COM~frame) is then given by
\begin{alignat}{3}
\label{nrlim}
K_{\mr{EM}}\cdot u_{\mr{com}}|_{\mr{NR}} \rightarrow \frac{\pi e^6 q_1^2 q_2^2}{3 |b|^3 v}\Bigg(\frac{q_1}{m_1}-\frac{q_2}{m_2}\Bigg)^2.
\end{alignat}
The dependence on $(q_1/m_1 - q_2/m_2)^2$ is consistent with the expectation that the dipole approximation should suffice for computing the radiated energy in the NR limit via the term $2\ddot{p}^2/3$. We can see this explicitly by considering a hyperbolic orbit (an exact trajectory in the NR limit) and evaluating the dipolar energy loss along it. We work in the centre-of-mass frame (equivalent to the centre-of-momentum frame in the NR limit) where the expression for the dipolar energy loss is given by
\begin{alignat}{3}
P = -\frac{dE}{dt}|_{\mr{NR}} = \frac{2|\ddot{\bs {p}}|^2}{3},\quad\bs{p} = \Bigg(\frac{q_1}{m_1} - \frac{q_2}{m_2} \Bigg) \mu \bs{r} .
\end{alignat}
We denote the NR energy per reduced mass and angular momentum with the symbols $\mathcal{E} = (E - M)/\mu$ and $J$ respectively. The orbit is then given in the NR limit as
\begin{alignat}{3}
& \mu \ddot{\bs{r}} = \frac{e^2 q_1 q_2 \bs{r}}{r^3},\quad r=\frac{R}{1+ e\cos\phi},\quad R = \frac{J^2}{e^2 q_1 q_2 \mu} \nonumber \\&
e^2 = 1+ \frac{2 \mathcal{E} J^2}{e^4  q_1^2 q_2^2}>1,\quad J = \mu r^2 \frac{d\phi}{dt}.
\end{alignat}
We can now integrate $P$ over the entire trajectory, $\phi \in (-\arccos(-1/e),\arccos(-1/e))$ and obtain the total radiated energy 
\begin{alignat}{3}
\label{Eradub}
-\mu\Delta \mathcal{E} = \int d\phi \frac{2e^4 q_1^2 q_2^2}{3 R^4}\Bigg(\frac{q_1}{m_1} - \frac{q_2}{m_2}\Bigg)^2 (1+ e\cos\phi)^4.
\end{alignat}
 To make contact with Eq.~(\ref{nrlim}), we take the limit of high angular momentum (low ${e^2 q^2}/{J}$, weak-field limit) where a perturbative analysis of the orbit is valid and get
\begin{equation}
-\mu\Delta {\mathcal{E}}_{|_{HE}} = \frac{2 \pi e^6 q_1^2 q_2^2 \mu^3 \mathcal{E}}{3  J^3}\Big(\frac{q_1}{m_1}-\frac{q_2}{m_2}\Big)^2.
\end{equation}
This is equal to Eq.~(\ref{nrlim}) in the NR limit once we identify $\mathcal{E} = v^2/2 + {\cal O}(v^4)$ and $J = \mu v |b|(1+{\cal O}(v^2))$ thus confirming our expectations.

\section{From scattering to bound orbits}
\label{utob}

We now consider the case of generic classical scattering of point-charges to study general relations between bound and unbound
motion~\cite{Kalin:2019rwq, Kalin:2019inp}. This is of particular
interest for gravitational interactions and gravitational-wave
physics. Indeed, mergers of bound compact binaries are much more likely 
to be observed than scattering encounters through gravitational-wave 
radiation. We are then motivated to investigate the methods by
which knowledge of bound orbits can be obtained by looking at
scattering events. One such method is via the maps for certain
observables between bound and unbound orbits based on analytic
continuation, as shown in Ref.~\cite{Kalin:2019inp}, which related the
unbound scattering angle to the bound periastron-advance angle. These
maps can be extended or motivated for some other observables as well,
as shown below.

Let us briefly explain the basis of the mapping procedure given in Refs.~\cite{Kalin:2019rwq, Kalin:2019inp} before extending it to other observables.
Consider a system of two nonspinning relativistic compact bodies whose interaction can be effectively described by conservative local dynamics. We can write down an effective Hamiltonian for the system in the COM\ frame (in an isotropic gauge \cite{Cheung:2018wkq,Vines:2018gqi,Bern:2019nnu,Bern:2019crd,Kalin:2019rwq, Kalin:2019inp}) as 
\begin{alignat}{3}
H(\bs{p},r)=\sqrt{\bs{p}^2+m_1^2}+\sqrt{\bs{p}^2+m_2^2}+V(r,\bs{p}^2),
\end{alignat}
where $\bs{p}^2(E,J,r)=p_{r}^2(E,J,r)+J^2/r^2$. The position coordinates are $q=(r,\phi)$, in polar coordinates in the plane of the motion. The Hamiltonian has no explicit dependence on $t$, $\phi$, and thus we have conserved quantities $E=H(\bs{p},r)$ (energy) and $J=p_{\phi}$ (angular momentum). We interpret $\pm\bs{p}$ to be the physical spatial momentum of either particle when they are far apart (which happens when $r\rightarrow \infty$, $V(r,\bs{p}^2)\rightarrow 0$). 

When the system is bound ($E<M=m_1+m_2$), there are two turning points $r_{\mr{min}}$ and $r_{\mr{max}}$ where $\dot{r}=0$.  Since $\dot{r} = \partial_{p_{r}} H \propto p_r$, the turning points occur at $p_{r}(E,J,r)=0$. When the system is unbound ($E>M$), one of the turning points $r_{\mr{max}}$ becomes negative, and thus non-physical. However, the roots can be generally related to each other via the relation $r_{\mr{min}}(\mathcal{E},-J) = r_{\mr{max}}(\mathcal{E},J)$ (where $\mathcal{E} = (E - M)/\mu$) for both bound and unbound orbits. Since $r_{\mr{max}}$ tends to $\pm\infty$ as $\mathcal{-E}\rightarrow \pm 0$, we work instead with the variable $u=1/r$. We define $u_{+} = 1/r_{\mr{min}}$ and $u_{-}=1/r_{\mr{max}}$ which are continuous if we vary $\mathcal{E}$ for fixed $J$ with $u_{\mr{max}}\rightarrow0$ if $\mathcal{E}\rightarrow0$. The relation between the roots is the same as before, $u_{-}(\mathcal{E},J) = u_{+}(\mathcal{E},-J)$.

This relation between the roots was used to relate the scattering angle for unbound orbits to periastron advance for bound orbits in Ref.~\cite{Kalin:2019inp} as follows. The scattering angle for an unbound system is given by 
\begin{alignat}{3}
& \chi(\mathcal{E},J) = -\pi + \Delta_{\mr{tot}}\phi=-\pi+\int_{\mr{tot}} \frac{d\phi/dt}{du/dt}du, \\& 
= -\pi - 2 \int_{u_{+}(\mathcal{E},J)}^{0} \frac{J}{p_{r}(u,\mathcal{E},J)} du,
\end{alignat}
where we use $\dot{\phi}/\dot{r} = J/(r^2 p_r)$ and the fact that the contribution to $\Delta_{\mr{tot}}\phi$ from $r=\infty$   to $r_{\mr{min}}$ (or $u=0$ to $u_{+}$) is the same as that from the subsequent journey from $r_{\mr{min}}$ to $r=\infty$, hence the factor of 2.

Now, consider the quantity 
\begin{alignat}{3}
&\chi(\mathcal{E},J) + \chi(\mathcal{E},-J) = -2\pi - 2\int^0_{u_{+}}\frac{J}{p_{r}}du-2\int^0_{u_{-}}\frac{-J}{p_{r}}du, \nonumber  \\&
= -2\pi + 2\int_{u_{-}(\mathcal{E},J)}^{u_{+}(\mathcal{E},J)} \frac{J}{ p_r(\mathcal{E},J)}du, \label{intper}
\end{alignat}
where we use $u_{+}(\mathcal{E},J)=u_{-}(\mathcal{E},-J)$. It is not difficult to recognize the second term in Eq.~(\ref{intper}) as the expression for the total angle subtended by a bound binary in one radial orbit (analytically continued to $\mathcal{E}>0$). Thus, we have for Eq.~(\ref{intper})
\begin{alignat}{3}
-2 \pi + 2\pi(K+1) = 2\pi K(\mathcal{E}>0,J),
\end{alignat}
where $2\pi K$ is the angle of periastron advance. This gives us the relation
\begin{equation}
\chi(\mathcal{E},J)+\chi(\mathcal{E},-J) = 2\pi K(\mathcal{E},J),
\end{equation}
as obtained in Ref~\cite{Kalin:2019inp}. Note that the LHS of the relation is only valid for $\mathcal{E}>0$ and vice-versa for the RHS; this relation is thus based on analytical continuation of the expressions for scattering angle and periastron advance.

\subsection{Analytic continuation of general observables}
\label{sec:acgo}
Assume that an observable $\mr{O}_{\mr{bound}}(\mathcal{E},J)$ associated with a nonspinning bound system can be expressed as an integral over one radial period (of the conservative dynamics) as 
\begin{equation}
\label{ansatz}
2 \int_{u_{-}}^{u_{+}} f(u,\mathcal{E},J) du = \mr{O}_{\mr{bound}}(\mathcal{E},J),
\end{equation}
Here, we are assuming that the function $f$ only depends on $u$, not on $\phi$ (reflecting rotational invariance). Now, the corresponding observable for an unbound orbit $\mr{O}_{\mr{unbound}}(\mathcal{E},J)$ can be written as 
\begin{equation}
2 \int_{0}^{u_{-}}  f(u,\mathcal{E},J) du = \mr{O}_{\mr{unbound}}(\mathcal{E},J).
\end{equation}
Again, with this integral evaluated along the conservative dynamics, we have the mapping given in Ref.~\cite{Kalin:2019rwq}, $u_{-}(\mathcal{E},-J) = u_{+}(\mathcal{E},J)$. 

Further assuming that $f(u,\mathcal{E},J)$ is either odd or even in $J$ (reflecting a certain behavior under time reversal), we have
\begin{alignat}{3}
\label{main}
\mr{O}_{\mr{bound}}(\mathcal{E},J) = \mr{O}_{\mr{unbound}}(\mathcal{E},J) + \theta(f)\mr{O}_{\mr{unbound}}(\mathcal{E},-J), \nonumber \\&
\end{alignat}
with $\theta(f) = \pm 1$ if $f$ is odd/even, respectively.
Note that the LHS is not really an observable for bound orbits unless $\mathcal{E}<0$. This is a formal relation between the functions based on analytical continuation of the expressions.

\subsection{Radiated energy}
\label{sec:emap}
It is reasonable to assume that the rate of energy loss (power) for generic orbits can be expressed as $dE/dt = P(u=1/r, \mathcal{E}, J^2)$ 
(i.e.\ as an even function of $J$) --- for example see Sec.~\ref{sec:Fluxes} for a motivation of this property in the PN context. Thus, we can write the energy radiated per radial orbit in the bound case as \begin{alignat}{3}
E_{\mr{rad}}^{\mr{bound}} =  \oint P(u,\mathcal{E},J^2) dt =2 \int_{u_{-}}^{u_{+}} P(u,\mathcal{E},J^2) \frac{-1}{u^2 p_{r}} du, \nonumber \\&
 \end{alignat}
and for the unbound case
\begin{equation}
E_{\mr{unbound}}^{\mr{rad}}=2 \int_{0}^{u_{-}} P(u,\mathcal{E},J^2) \frac{-1}{u^2 p_{r}} du,
\end{equation}
where we have defined $E_{\mr{bound}}^{\mr{rad}}$ as the energy loss per orbit for bound systems and $E_{\mr{unbound}}^{\mr{rad}}$ as the total energy loss for an unbound trajectory.
Then, using the general relation derived in Eq.~(\ref{main}), we get the relation between unbound and bound energy losses as
\begin{equation}
\label{immedate}
E_{\mr{rad}}^{\mr{bound}}(\mathcal{E},J) = E_{\mr{rad}}^{\mr{unbound}}(\mathcal{E},J) - E_{\mr{rad}}^{\mr{unbound}}(\mathcal{E},-J),
\end{equation}
which was given earlier in Ref.~\cite{Bini:2020hmy}.
We now use this relation to compute partial results for bound orbits in EM and verify it with explicit calculations in the NR limit.

To compute the RHS of Eq.~(\ref{immedate}), we use the expression for the energy radiated in the COM frame in a scattering event in the NR limit given in Eq.~(\ref{nrlim}). We write it in terms of $\mathcal{E}$ and $J$ using the relations $\mathcal{E} = v^2/2$, and $J=\mu v |b|$,  obtaining
\begin{equation}
E_{\mr{rad}}^{\mr{unbound}}(\mathcal{E},J)= \frac{2 \pi e^6 q_1^2 q_2^2 \mu^3 \mathcal{E}}{3 J^3}\Big(\frac{q_1}{m_1}-\frac{q_2}{m_2}\Big)^2,
\end{equation}
which is odd in $J$. Thus, in this case the RHS in Eq.~(\ref{immedate}) is 
\begin{equation}
\textrm{RHS of \eqref{immedate}}=\frac{4 \pi e^6 q_1^2 q_2^2 \mathcal{E} \mu^3}{ 3 J^3}\Big(\frac{q_1}{m_1}-\frac{q_2}{m_2}\Big)^2.
\label{rhsenmap}
\end{equation}
We should acquire the same result to the highest power in $1/J$ if we compute the radiated energy per orbit for bound orbits in the NR limit. We use the fact that, in this limit, the orbits are conic sections and the power can be obtained from the dipole approximations. We computed the dipole-power loss for unbound orbits in Eq.~(\ref{Eradub}), we now do the same for bound orbits.

Thus, once again using the dipole formula for the power $P = 2\ddot{\bs{p}}^2/3$, with $\bs{p} = \Big(q_1/m_1 - q_2/m_2 \Big) \mu \bs{r}$, where $\bs{r}$ is the separation vector from particle 2 to particle 1, $|\bs r| = r = R/(1+ e \cos \phi)$, $\mu r^2 d\phi/dt = J$, with $R=J^2/(e^2 q_1q_2\mu), \text{ } e^2 = 1 +2 \mathcal{E} J^2/(e^4  q_1^2q_2^2)$, we integrate over one period and obtain
 \begin{alignat}{3}
 \label{embound}
& E^{\mr{bound}}_{\mr{rad}} = \frac{2 \pi \mu^3 e^{6} q_1^2 q_2^2(3 e^4 q_1^2 q_2^2 +2 \mathcal{E} J^2)}{3 J^5}\Big(\frac{q_1}{m_1}-\frac{q_2}{m_2}\Big)^2. \nonumber \\&
 \end{alignat} 
 We then take the limit $J \rightarrow \infty$ and get
 \begin{equation}
 E^{\mr{bound}}_{\mr{rad}} \to \frac{4 \pi e^6 q_1^2 q_2^2 \mathcal{E} \mu^3}{ J^3}\Big(\frac{q_1}{m_1}-\frac{q_2}{m_2}\Big)^2,
\end{equation}
 which is identical to Eq.~(\ref{rhsenmap}) as expected.

\subsection{Radiated angular momentum}

The angular momentum is a vector, and symmetry requires that the rate of angular-momentum loss be in the same direction as the angular momentum. Thus, we expect a relation of the form 
\begin{equation}
\frac{d\bs{J}}{dt} = \bs{J}\,P_{J}(u,\mathcal{E},J^2),
\end{equation}
which gives the following rate of change for the magnitude of the angular-momentum loss 
\begin{equation}
\frac{dJ}{dt} = J \, P_{J}(u,\mathcal{E},J^2),
\end{equation}
which is odd in $J$. Now, defining $J_{\mr{unbound}}^{\mr{rad}}$ as the total angular-momentum loss for an unbound trajectory and $J_{\mr{bound}}^{\mr{rad}}$ as the angular-momentum loss per orbit for bound systems, we derive the relation
\begin{equation}
J_{\mr{rad}}^{\mr{bound}}(\mathcal{E},J) = J_{\mr{rad}}^{\mr{unbound}}(\mathcal{E},J) + J_{\mr{rad}}^{\mr{unbound}}(\mathcal{E},-J),
\label{angmom2}
\end{equation}
which has the opposite sign in the RHS compared to Eq.~(\ref{immedate}). 

The above relation, unfortunately, cannot be used to obtain partial result for the angular-momentum loss for bound orbits using only the leading-order result we have derived in this paper, since the angular-momentum loss at leading order is odd in $J$, as seen in Eqs.~(\ref{JradEM}) and (\ref{JradGR}). We need to evaluate it to at least $3^{\mr{rd}}$ order in the weak-field expansion. Nevertheless, as an illustration, we verify this relation at 1PN order in gravity. We compute the 1PN angular momentum loss for unbound orbits following the method in Ref.~\cite{Junker:1992kle} to obtain
\begin{widetext}
\begin{alignat}{3}
&\Delta J_{\mr{rad}}^{\mr{unbound}}= \frac{-8  G m_1 m_2 \nu}{5c^5 h^4}\Bigg\{\arccos\Bigg(\frac{-1}{h\sqrt{1/h^2+2\mathcal{E}}}\Bigg)(15+14\mathcal{E} h^2) + \sqrt{2 \mathcal{E}} h (15 + 4 \mathcal{E} h^2)\nonumber \\&\nonumber +\frac{1}{1008 h^2 c^2}\Bigg[\arccos\Bigg(\frac{-1}{h\sqrt{1/h^2+2\mathcal{E}}}\Bigg)[105(1077-940 \nu)+252(535-748 \nu)\mathcal{E}h^2  + 12(4283 - 3976 \nu)\mathcal{E}^2h^4] \\& + \frac{\sqrt{2\mathcal{E} h^2}}{1 + 2 \mathcal{E} h^2}\{105(1077-940 \nu) + 224(1275 - 1429 \nu) \mathcal{E} h^2 + 4(42711 - 61600 \nu)\mathcal{E}^2h^4 + 288(109-35 \nu)\mathcal{E}^2h^6\}\Bigg]\Bigg\}.  \nonumber \\&
\label{Jub}
\end{alignat}
Substituting this in the RHS of Eq.~(\ref{angmom2}) and using $\arccos(-x) = \pi - \arccos(x)$, we get
\begin{alignat}{3}
&\Delta J_{\mr{rad}}^{\mr{bound}}=\frac{-8 \pi G m_1 m_2 \nu}{5 c^5 h^4}\Bigg[(15+14\mathcal{E} h^2)+\frac{105(1077-940 \nu)+252(535-748 \nu)\mathcal{E}h^2  + 12(4283 - 3976 \nu)\mathcal{E}^2h^4}{1008 h^2 c^2}\Bigg] \nonumber, \\&
\end{alignat}
\end{widetext}
where $h= J/(G M \mu)$ and thus we recover the correct expression for the angular-momentum radiated per bound orbit at 1PN order, obtained by multiplying Eq.~(30) in Ref.~\cite{Junker:1992kle} for the average angular momentum flux by the orbital period $P = 2 \pi/n$, with $n$ given in Eq.~(26) of Ref.~\cite{Junker:1992kle}. We also verified this expression by doing the explicit calculation for angular momentum loss in case of bound orbits (at 1PN). Note that the expression for angular momentum losses in unbound orbit in Eq.~(\ref{Jub}) does not completely match that given in Ref.~\cite{Junker:1992kle}, there is a minor computational error in their result. 

\subsection{Total radiative losses from instantaneous fluxes}\label{sec:Fluxes}

Given the maps provided in the last two subsections, one can find resummed relativistic expressions for the energy and angular momentum losses for bound orbits via the following simple algorithm:   
(i) solve the scattering problem in the weak-field expansion to compute energy- and angular-momentum losses, and
(ii) use the maps provided earlier to find partial expressions for the energy- and angular-momentum losses of bound orbits.

However, this method is limited by the order to which the scattering problem can be solved in the weak-field regime. Since this is an expansion in large impact parameters $|b|$, the observables are obtained in powers of $1/|b|$ --- for example see Eq.~(\ref{JradGR}) and Eq.~(\ref{KradGR}). Using the relation between $|b|$ and initial angular momentum, $J = (\mu M/E) \gamma v |b|$, we see that this is also an expansion in $1/J$. 

In particular, the leading-order energy loss in the scattering case goes as $1/J^3$ [see Eq.~(\ref{KradGR})]. Thus, the map only gives us the $1/J^3$ part of the energy loss for bound orbits. However, the expression for 0PN energy loss per orbit for bound systems is given (from Einstein's quadrupole formula) by
\begin{alignat}{3}
\nnm&\Delta E_\mr{GR}=
\\\nnm
&\quad\frac{2\pi\mu^2}{Mc^5}\bigg(\frac{148}{15}\mc E^2\frac{G^3M^3}{J^3}+\frac{244}{5}\mc E\frac{G^5M^5}{J^5}+\frac{85}{3} \frac{G^7M^7}{J^7}\bigg) \\&
\end{alignat}
$+\mc {\cal O}({1/c^7})$, and we do not recover the $1/J^5$ and $1/J^7$ terms. Each higher order in the PM expansion adds a power of $1/J$, and thus naively, one needs to solve to $7^{\rm th}$ order (7PM, $G^7$) to recover the complete expressions for even the 0PN energy loss via the maps. This leads to a discouraging conclusion regarding the possibility of recovering expressions for bound-orbit radiative losses via results for the scattering encounter.

However, an alternative way of recovering bound-orbit observables is to fix the form of (gauge-dependent) instantaneous fluxes of energy and angular momentum that are directly applicable to both bound and unbound orbits (at least for local-in-time contributions).  Following the known forms of PN expansions of these fluxes, one can write down general ans\"atze parametrized by unknown coefficients, which can then be fixed by computing the energy and angular momentum losses along near-straight-line (small-deflection-regime) trajectories, which should equal the results obtained from the weak-field expansion.

For example, in the gravitational case, the expressions for energy and angular momentum fluxes through 1PN order~\cite{1989MNRAS239845B,Junker:1992kle} suggest that suitable ans\"atze can be written as
\begin{alignat}{3}
\Phi_{E} &= \frac{G^3 M^2 \mu^2}{c^5 r^4}\Bigg(\sum_{i=1}^{3}\alpha_{i} \mathcal{X}_{i} + \frac{1}{c^2}\sum_{i,j=1}^{3}\alpha_{ij} \mathcal{X}_{i} \mathcal{X}_{j}+\ldots\Bigg), \nnm\\&\\
\Phi_{J} &= \frac{G^2 M \mu J}{r^3 c^5}\Bigg(\sum_{i=1}^{3}\beta_{i}\mathcal{X}_{i}+ \frac{1}{c^2}\sum_{i,j=1}^{3} \beta_{ij} \mathcal{X}_{i} \mathcal{X}_{j}+\ldots\Bigg), \nnm\\& 	\\ \mathcal{X}_{i} &=  \Bigg\{\bs{v}^2, \left(\frac{\bs{v}\cdot \bs{r}}{r}\right)^2, \frac{GM}{r}\Bigg\}, \quad \mu = \frac{m_1 m_2}{m_1+m_2},
\end{alignat}
where $\bs{v}$ is the velocity and $\bs{r}$ is the relative position, in the PN context.
For the purpose of demonstration, consider the leading-PN-order (0PN) fluxes,
\begin{alignat}{3}
\Phi_{E} &= \frac{G^3 M^2 \mu^2}{c^5 r^4} \Bigg [\alpha_1 \bs{v}^2 + \alpha_2 \left(\frac{\bs{v}\cdot \bs{r}}{r}\right)^2 + \alpha_3 \frac{G M }{r} \Bigg],& \label{0PNfluxese} \\
\Phi_{J} &= \frac{G^2 M \mu J}{r^3 c^5} \Bigg[\beta_1 \bs{v}^2+ \beta_2 \left(\frac{\bs{v}\cdot \bs{r}}{r}\right)^2 + \beta_3 \frac{GM}{r}\Bigg].
\label{0PNfluxesj}
\end{alignat}
 The above expressions are subject to a gauge freedom in that we can add terms that are total time derivatives, the so-called Schott terms $\dot{E}_{\mr{Schott}}$ and $\dot{J}_{\mr{Schott}}$, which are functions of $\bs r$ and $\bs v$ (under the Newtonian equations of motion $\dot{\bs r}=\bs v$ and $\dot{\bs v}=-GM\bs r/r^3$) and vanish at infinity. This does not change the total energy and angular momentum losses, obtained by integrating over one radial period for the bound case or the entire orbit for the unbound case.  The relevant Schott terms at 0PN order are
\begin{alignat}{3}
E_{\mr{Schott}} &= a \frac{G^3 M^2 \mu^2}{c^5 r^4} (\bs{v}\cdot \bs{r}), \\
J_{\mr{Schott}} &= b \frac{G^2  M \mu J}{c^5 r^3} (\bs{v}\cdot\bs{r}).
\end{alignat}
We find that we can choose $a=\alpha_2/4$ and $b=\beta_2/3$ and set the coefficients of the $(\bs{v}\cdot\bs{r})^2$ terms in Eqs.~(\ref{0PNfluxese}) and (\ref{0PNfluxesj}) to zero.  This leaves us with an isotropic gauge for the fluxes, in which they depend only on $\bs v^2$ and $r$; we can then also express the fluxes as functions only of $r$ and $\mc E=\bs v^2/2-GM/r+\mc {\cal O}(1/c^2)$.  With $\tilde\Phi_E=\Phi_E+\dot E_\mr{Schott}$ and similarly for $J$, we are left with
\begin{alignat}{3}\label{isotropicfluxes}
\tilde\Phi_{E} &= \frac{G^3 M^2 \mu^2}{c^5 r^4} \Bigg(\tilde\alpha_1 \mc E + \tilde\alpha_2  \frac{G M }{r} \Bigg)+\mc {\cal O}\left (\frac{1}{c^7}\right ),  \\
\tilde\Phi_{J} &= \frac{G^2 M \mu J}{r^3 c^5} \Bigg(\tilde\beta_1 \mc E+ \tilde\beta_2  \frac{GM}{r}\Bigg)+\mc {\cal O}\left (\frac{1}{c^7} \right ),
\end{alignat}
with $\tilde\alpha_1=2\alpha_1+\alpha_2/2$, $\tilde\alpha_2=2\alpha_1+\alpha_2/4+\alpha_3$, $\tilde\beta_1=2\beta_1+2\beta_2/3$ and $\tilde\beta_2=2\beta_1+\beta_2/3+\beta_3$.  There is now a direct correspondence (via integration over the Newtonian orbit) between these coefficients $\tilde\alpha_{1,2}$ and $\tilde\beta_{1,2}$ and the coefficients in the PN-PM expansion of the total radiative losses for a scattering orbit; the former can be determined from the latter.  We see that the orders in the weak-field-scattering expansion needed to recover the Newtonian fluxes (and thus also the bound-orbit losses) are considerably less than those needed in the direct use of the analytic-continuation maps.  For example, to determine the (effective) 0PN energy flux $\tilde\Phi_E$, instead of 7PM, we need $\Delta E_\mr{rad}^\mr{unbound}$ only to 4PM order, to $\mc {\cal O}(G^4)$.  Still we cannot recover the complete 0PN fluxes from the leading orders in the weak-field expansions of $\Delta E_\mr{rad}^\mr{unbound}$ (leading $G^3$) and $\Delta J_\mr{rad}^\mr{unbound}$ (leading $G^2$).\footnote{Note however that at least one further constraint on the four coefficients in Eq.~(\ref{isotropicfluxes}) is provided by the condition $\tilde\Phi_{E} - \omega\tilde\Phi_{J}=0$ for circular orbits, where $\omega=\dot\phi$ is the orbital frequency; this ensures that adiabatic radiation reaction takes circular orbits to circular orbits (adiabatically).

It is interesting to note in this context that Ref.~\cite{Dlapa:2021npj} has derived a relativistic expression for the instantaneous flux at 3PM order for the gravity case, noting its relation to the coefficient of the dimensional-regularization pole in the effective action at 4PM order.
}
\section{Summary}
\label{sum}

We have considered the relativistic scattering of two charged point
particles in classical electrodynamics and have calculated, via direct
iteration of the classical equations of motion, the impulse on each
particle through $3^{\mr{rd}}$ order in the weak-field expansion
(through $6^{\rm th}$ order in the charges). This is the order at which
radiative effects first appear in the impulse, and we have   
consistently included them by using retarded boundary conditions and by
accounting for each particle's influence on itself by using the ALD
force (see Sec.~\ref{sec:EMcals}). We have related the impulse up to $3^{\mr{rd}}$ order to the
conservative scattering angle, the radiated momentum and the radiative
correction to the scattering angle (see Sec.~\ref{sec:kchi}). We have completely or partially verified 
the latter quantities by comparisons with other results and consistency tests 
in the literature.  In particular, we have verified the general
relationship derived in Ref.~\cite{Bini:2012ji} between the radiated
angular momentum (see Eq.~(\ref{angmomloss})) at $2^{\mr{nd}}$ order and the radiative contribution
to the scattering angle (see Eq.~(\ref{chirad3})) at $3^{\mr{rd}}$ order, by separately
computing these quantities within our setup.  We have also verified that the conservative scattering angle matches the result of Ref.~\cite{Bern:2019nnu}, and that the total radiated momentum matches the result of an integral given in Ref.~\cite{Kosower:2018adc}.  We have considered both the
nonrelativistic ($v\ll c$) and high-energy ($v\to c$) limits of
observables such as the scattering angle and the radiated energy (see Sec.~\ref{hel} and Sec.~\ref{nrelsec}). We  
found consistency with known (well-behaved) results in the
nonrelativistic limit, but encountered certain divergences in the
high-energy limit (See Eq.~(\ref{chidiv}), Eq.~(\ref{angdiv}) and Eq.~(\ref{ediv})).  Whilst some of these divergences seem to arise
from limitations of the validity of ALD self-force (or more generally
of the zero-size point-particle idealization), the divergence in the
fraction of energy radiated signals a breakdown of weak-field
perturbation theory for arbitrarily high energies, a feature in common
with the gravitational case, as discussed in
Ref.~\cite{Herrmann:2021lqe}.  It would be highly instructive to see if our results for the complete 3$^\mr{rd}$ order impulse match those which would be produced by applying the KMOC formalism \cite{Kosower:2018adc} to the relevant amplitudes up to 2-loop order in (scalar) quantum electrodynamics, and to explore how the methodologies compare.

We have also investigated the scope of relating observables via
analytic continuation between unbound and bound orbits, following
Ref.~\cite{Kalin:2019rwq} (see Sec.~\ref{utob}).  We have derived a general map  between generic unbound and bound observables that satisfy certain reasonable
requirements (see Sec.~\ref{sec:acgo}).  We have shown that the different maps known so far are special
cases of this general map (see Eq.~(\ref{main})), and we have also derived a new map between loss of angular
momentum (see Eq.~(\ref{angmom2})).  We have explicitly verified this new map for the
gravitational case through the next-to-leading order in the
PN expansion, and we have uncovered an error in the expression
for the loss of angular momentum in hyperbolic encounters obtained 
in Ref.~\cite{1989MNRAS239845B} (See Eq.~(\ref{Jub}) for corrected expression). The previously known map between
energy losses was further verified for the electromagnetic case by
comparing to the leading order in the nonrelativistic limit (see Sec.~\ref{sec:emap}). We have found  that directly using the analytic-continuation maps, to obtain
complete expressions for bound-orbit radiative observables in the
nonrelativistic limit, requires going to quite high orders in the
weak-field expansion in the scattering regime. Conversely, we saw 
that lower orders for scattering are required to fix the coefficients in
general ans\"atze for (gauge-dependent) instantaneous fluxes (see Eq.~(\ref{0PNfluxese}) and Eq.~(\ref{0PNfluxesj})), from which one
can derive the total radiated energy and angular momentum for both
unbound and bound orbits (for local-in-time contributions) (see Sec.~\ref{sec:Fluxes}). These
investigations are valuable for understanding and modeling the
relativistic binary problem in EM and GR (its dynamics and radiation)
over the full range of eccentricities.

\py

\acknowledgements

We thank Zvi Bern, Juan Pablo Gatica, Enrico Herrmann, David Ksosower, Andr\'es Luna, Ben Maybee, Donal O'Connell, Julio Parra-Martinez, Radu Roiban, Michael S. Ruf, Chia-Hsien Shen, Mikhail P. Solon, and Mao Zeng for useful discussions.  

{\tt Mathematica} expressions of the third-order contributions to the acceleration are available upon request.
\appendix
\begin{widetext}

\section{The $2^{\rm nd}$ order worldline corrections}
\label{2oapp}

\subsection{Evaluating the force}
Here, we show the steps leading to the explicit computation of the $2^{\mr{nd}}$ order worldline corrections. We first need to find the partial derivatives of the field tensor ($F^{\mu\nu}$) w.r.t.~the coordinates of the particles' worldlines, velocities and accelerations. We list them below and then use them to compute the $2^{\mr{nd}}$ order force in terms of $1^{\mr{st}}$ order worldline corrections $z^{(1)}$. We have
\begin{alignat}{4}
&  F^{\mu\nu}(z_1(\tau_1))[z_2(\tau_{2,\mr{ret}})=z_{2,\mr{ret}}] = \frac{2 q_2 \rho_2^{[\mu}[r_2 \ddot{z}_{2,\mr{ret}}^{\nu]}-\dot{z}_{2,\mr{ret}}^{\nu]}(\ddot{z}_{2,\mr{ret}}\cdot \rho_2-1) ]}{ r_2^3} ,\quad \rho_{2}^{\mu} = z_1^{\mu} - z^{\mu}_{2,\mr{ret}},\quad r_2 = \dot{z}_{2,\mr{ret}}\cdot\rho_2.  \nonumber\\&
\end{alignat}
This is the field sourced by particle 2 at particle 1's position ($x=z(\tau_1)$). It depends on the worldlines directly via the expression shown above and indirectly through the retarded time $\tau_{2,\mr{ret}}$. It is convenient to separately deal with the dependence on retarded time. The required partial derivaties are
\begin{alignat}{3}
& \frac{\partial F_{\mu\nu}(z_1)[z_{2,\mr{ret}}]}{\partial z_{1,\alpha}} =\frac{2 q_2 [ (r_2\delta^{\alpha}_{[\mu}-3\dot{z}_{2,\mr{ret}}^{\alpha}\rho_{2,[\mu})(r_2 \ddot{z}_{2,\mr{ret},\nu]}-\dot{z}_{2,\mr{ret},\nu]}(\ddot{z}_{2,\mr{ret}}\cdot \rho_2-1) )+r_2 \rho_{2,[\mu} (\dot{z}_{2,\mr{ret}}^{\alpha}\ddot{z}_{2,\mr{ret},\nu]}-\dot{z}_{2,\mr{ret},\nu]}\ddot{z}_{2,\mr{ret}}^{\alpha})]}{r_2^4},  \\& \frac{\partial F_{\mu\nu}(z_1)[z_{2,\mr{ret}}]}{\partial z_{2,\mr{ret},\alpha}} = -\frac{\partial F_{\mu\nu}(z_1)[z_{2,\mr{ret}})]}{\partial z_{1,\alpha}},\\&
\frac{\partial F_{\mu\nu}(z_1)[z_{2,\mr{ret}}]}{\partial \dot{z}_{2,\mr{ret},\alpha}} = \frac{ -2  q_2  \rho_{2,[\mu}[ 2\rho^{\alpha}_2  \ddot{z}_{2,\mr{ret},\nu]}-(3\rho^{\alpha}_2 \dot{z}_{2,\mr{ret},\nu]}-\delta_{\nu]}^{\alpha})(\ddot{z}_{2,\mr{ret}}\cdot \rho_2-1) ]}{ r_2^4},\\& \frac{\partial F_{\mu\nu}(z_1)[z_{2,\mr{ret}}]}{\partial \ddot{z}_{2,\mr{ret},\alpha}} = \frac{ 2 q_2 \rho_{2,[\mu}( r_2 \delta^{\alpha}_{\nu]}-\dot{z}_{2,\mr{ret},\nu]}\rho_2^{\alpha} )}{r_2^3},\\&\nonumber\\&\nonumber\text{where we have kept the retarded time fixed while varying the coordinates. We now quantify the dependence on}\\&\text{retarded time by the total derivative} \nonumber\\&\nonumber\\&
\frac{d F_{\mu\nu}(z_1)[z_2(\tau_{2,\mr{ret}})]}{d\tau_{2,\mr{ret}}} = \frac{\partial F_{\mu\nu}(z_1)[z_2(\tau_{2,\mr{ret}})]}{\partial z_{2,\mr{ret},\alpha}}\dot{z}_{2,\mr{ret},\alpha} + \frac{\partial F_{\mu\nu}(z_1)[z_2(\tau_{2,\mr{ret}})]}{\partial \dot{z}_{2,\mr{ret},\alpha}}\ddot{z}_{2,\mr{ret},\alpha} + \frac{\partial F_{\mu\nu}(z_1)[z_2(\tau_{2,\mr{ret}})]}{\partial \ddot{z}_{2,\mr{ret},\alpha}} \dddot{z}_{2,\mr{ret},\alpha}.
\end{alignat}
We can now compute the $2^{\mr{nd}}$ order corrections to the force. The $2^{\mr{nd}}$ order correction to the force is obtained by substituting $1^{\mr{st}}$ order worldlines (and retarded time) in the force and taking the coefficient of $e^4$, i.e.,
\begin{alignat}{3}
&  m_1 \ddot{z}_1^{(2)\mu} = \Big[e^4\Big] \left[ e^2 q_1 F^{\mu\nu}\dot{z}_{1,\nu} +  e^2 \frac{2 q^2}{ 3} \Big(\dddot{z}_1^{\mu}+\ddot{z}_1^2 \dot{z}_1^{\mu}\Big)\right ] = f^{(2)\mu} 
\nonumber\,\text{where $[x^2]f(x)=\text{Coefficient of $x^2$ in } f(x)$}, \\& \text{and in RHS, we have } z_1^{\mu}\rightarrow z_1^{(0)\mu} + e^2 z_1^{(1)\mu}, \, z_2^{\mu} \rightarrow z_2^{(0)\mu} + e^2 z_2^{(1)\mu},\, \tau_{2,\mr{ret}} = \tau_{2,\mr{ret}}^{(0)}+e^2 \tau_{2,\mr{ret}}^{(1)},  \label{2ofapp}
\end{alignat}
where $f^{(2)\mu}$ is the total force correction at $2^{\mr{nd}}$ order. We expand the RHS of Eq.~(\ref{2ndorderlaw}) into four parts using Taylor series, as was shown in the diagrams in figure~\ref{fig2nd} in main text. Here, we write those contributions explicitly in terms of $1^{\mr{st}}$ order worldline corrections using the partial derivatives of the field tensor derived above.
\begin{wrapfigure}{l}{0.15\textwidth}
\begin{tikzpicture}
\begin{scope}[very thick, every node/.style={sloped,allow upside down}]
\node[text width=0.25cm] at (1.1,2.5) {$\mr{I}$};
\draw[dashed](0,0)--(1.1,0);
\draw[very thick](0,1.5)--(1.4,1.5);
\draw[very thick, double](1.4,1.5)--(2.2,1.5);
\draw[snake it](1.1,0)--node{\midarrow}(1.4,1.5);
\end{scope}
\end{tikzpicture}
\end{wrapfigure}
\textbf{a) Correction to $e^2 q_1 F^{\mu\nu} \dot{z}_{1,\nu}$ due to $e^2 z_1^{(1)}$.} \newline
Diagram $\mr{I}$ is the force correction due to the $1^{\mr{st}}$ order worldline corrections to particle 1 in the zeroth order field of particle 2, via the explicit dependence of $z_1(\tau_1)$, $ \dot{z}_1(\tau_1)$ in the Lorentz force (first term in RHS of Eq.~(\ref{2ofapp})). It thus scales as $1/m_1$. It is given by
\begin{alignat}{3}
e^4 f_{\mr{I}}^{(2)\mu} = e^4 \frac{\partial ( q_1 F^{\mu\nu}\dot{z}_{1,\nu})}{\partial z_1^{\alpha}}|_{(0)}z_1^{(1),\alpha} +  e^4 \frac{\partial(  q_1 F^{\mu\nu}\dot{z}_{1,\nu})}{\partial \dot{z}_1^{\alpha}}|_{(0)}\dot{z}_1^{(1),\alpha}.
\end{alignat}
At zeroth order, we have $\ddot{z}^{(0)}=0$, $\dot{z}_i=u_i$, $\rho_2^{(0),\mu}=b^{\mu}+u_1^{\mu}\tau_1-u_2^{\mu} \tau^{(0)}_{2,\mr{ret}}$, $r_2= -\tau^{(0)}_{2,\mr{ret}}+\gamma \tau_1$. We define $r_2(\tau_1,\tau_{2,\mr{ret}}^{(0)})=r_{21}=\sqrt{|b|^2+(\gamma v)^2\tau_1^2}$ and  $s_1 = \gamma v \tau_1 + r_{21}$. Thus, the contribution to force is given by
\begin{alignat}{3}
f_{\mr{I}}^{(2)\mu}=& \frac{ q_1 q_2 [\gamma(r_{21} z_1^{(1).\mu}-3\rho_2^{(0),\mu}u_2\cdot z_1^{(1)})-u_{2}^{\mu}(r_{21} u_1\cdot z_1^{(1)} - 3  u_1 \cdot\rho_2^{(0)} u_2\cdot z_1^{(1)})  ]}{ r_{21}^4}\nonumber \\&+\frac{ q_1 q_2[ \rho_2^{(0),\mu} (u_2\cdot \dot{z}_1^{(1)})  - u_2^{\mu} (\rho_2^{(0)}\cdot \dot{z}_1^{(1)}) ]}{ r_{21}^3},
\end{alignat}
where the $1/m_1$ dependence comes from $1^{\mr{st}}$ order worldline corrections $z^{(1)}_1$. Using this expression and the $1^{\mr{st}}$ order corrections (Eq.~(\ref{P11o})), it is easy to see that this is composed of a linear combination of terms such as
\begin{equation}
\label{setI}
\frac{1}{m_1}\Bigg\{\frac{(r_{21} \text{ or }s_1\text{ or }\tau_1)\times\log(s_1)}{r_{21}^4}, \frac{(r_{21} \text{ or }s_1\text{ or }\tau_1)\times s_1}{r_{21}^4},\frac{(\tau_1\text{ or }s_1)}{r_{21}^4}\Bigg\}
\end{equation}\newline

\begin{wrapfigure}{l}{0.15\textwidth}
\begin{tikzpicture}
\begin{scope}[very thick, every node/.style={sloped,allow upside down}]
\node[text width=0.25cm] at (1.1,2.5) {$\mr{II}$};
\draw[very thick](0,0)--(1.1,0);
\draw[dashed](0,1.5)--(1.4,1.5);
\draw[very thick, double](1.4,1.5)--(2.2,1.5);
\draw[snake it](1.1,0)--node{\midarrow}(1.4,1.5);
\end{scope}
\end{tikzpicture}
\end{wrapfigure}
\textbf{b) Correction to $e^2 q_1 F^{\mu\nu} \dot{z}_{1,\nu}$ due to $e^2 z_2^{(1)}$.} \newline
Diagram $\mr{II}$ is the force correction due to $1^{\mr{st}}$ order worldline correction of particle 2 via explicit dependence of the Lorentz force term on $[z_2]$. It thus scales as $1/m_2$. It is given by,
\begin{alignat}{2}
e^4 f_{\mr{II}}^{(2)\mu} = &  e^4\frac{\partial ( q_1 F^{\mu\nu}u_{1,\nu})}{\partial z_2^{\alpha}}|_{(0)}z_2^{(1),\alpha} + e^4 \frac{\partial(q_1 F^{\mu\nu}u_{1,\nu})}{\partial \dot{z}_2^{\alpha}}|_{(0)}\dot{z}_2^{(1),\alpha} \nonumber \\& +  e^4 \frac{\partial( q_1 F^{\mu\nu}u_{1,\nu})}{\partial \ddot{z}_2^{\alpha}}|_{(0)}\ddot{z}_2^{(1),\alpha} 
\end{alignat} 
Evaluating the partial derivatives with zeroth order worldlines gives us
\begin{alignat}{5}
& f_{\mr{II}}^{(2)\mu}= \frac{- q_1 q_2 [\gamma(r_{21} z_2^{(1).\mu}-3\rho_2^{(0),\mu}u_2\cdot z_2^{(1)})-u_{2}^{\mu}(r_{21} u_1\cdot z_2^{(1)} - 3  u_1 \cdot\rho^{(0)}_2 u_2\cdot z_2^{(1)})  ]}{ r_{21}^4}\nonumber\\&-\frac{ q_1 q_2  [\rho_2^{(0),\mu}(3\rho_2^{(0)}\cdot \dot{z}_2^{(1)} \gamma-u_1\cdot \dot{z}_2^{(1)} )-(\rho_2^{(0)}\cdot u_1)(3 u_{2}^{\mu}\rho_2^{(0)} \cdot \dot{z}_{2}^{(1)}-\dot{z}_2^{(1),\mu}) ]}{ r_{21}^4}\nonumber\\&+\frac{ q_1 q_2[ \rho_2^{(0),\mu}( r_{21} (u_1 \cdot \ddot{z}_2^{(1)})-\gamma\rho_2^{(0)}\cdot \ddot{z}_2^{(1)}) -(u_1\cdot \rho_2^{(0)})(\ddot{z}_2^{(1),\mu}-u_2^{\mu}(\rho_2^{(0)} \cdot \ddot{z}_2^{(1)}))]}{ r_{21}^3},
\end{alignat}
where the $1/m_2$ dependence comes from the first order worldline corrections $z_2^{(1)}$, and we can use this expression and the $1^{\mr{st}}$ order corrections (Eq.~(\ref{P21o})) to see that it is a linear combination of terms like
\begin{alignat}{3}
\label{setII}
\nonumber\frac{1}{m_2}\Bigg\{&\frac{(r_{21}\text{ or 	}\tau_1\text{ or }s_1)\times s_2}{r_{21}^4},\frac{(r_{21}\text{ or }\tau_1\text{ or }s_1)\times \log(s_2)}{r_{21}^4},\frac{(s_1 \tau_1\text{ or }\tau_1^2\text{ or }s_1^2\text{ or }s_1\text{ or }\tau_1)\times(1 \text{ or }s_2)}{r_{21}^4 r_{12}},\\& \frac{r_{21}\text{ or } s_1 \text{ or } \tau_1\text{ or }s_1^2\text{ or }\tau_1s_1 \text{ or }\tau_1^2 \text{ or }s_1^3 \text{ or }s_1^2\tau_1 \text{ or }\tau_1^3}{r_{21}^3 r_{12}^3}\Bigg\}, \nonumber\\& 
r_{12}=\sqrt{|b|^2+(\gamma v)^2\tau_{2}^2},\quad s_2=\gamma v\tau_2+r_{12}.  \\& 
\text{In the above expressions, all functions of $\tau_2$ should be evaluated at $\tau_{2,\mr{ret}}^{(0)}$.} \nonumber
\end{alignat}

\begin{wrapfigure}{l}{0.15\textwidth}
\begin{tikzpicture}
\begin{scope}[very thick, every node/.style={sloped,allow upside down}]
\node[text width=0.25cm] at (1.1,2.5) {$\mr{III}$};
\node[text width=0.25cm] at (1.1,-0.25){$\tau^{(1)}_{2,\mr{ret}}(z_a^{(1)})$};
\node[text width=0.25cm] at (1.4,1.75){$\tau_{1}$};
\draw[dashed](0,0)--(1.1,0);
\draw[dashed](0,1.5)--(1.4,1.5);
\draw[very thick, double](1.4,1.5)--(2.2,1.5);
\draw[snake it](1.1,0)--node{\midarrow}(1.4,1.5);
\end{scope}
\end{tikzpicture}
\end{wrapfigure}
\textbf{c) Correction to $e^2 q_1 F^{\mu\nu} \dot{z}_{1,\nu}$ due to $e^2 \tau_{2,\mr{ret}}^{(1)}$.} \newline
Diagram $\mr{III}$ is the force correction due to first order corrections to retarded time which in turn comes from 1st order correction to both particles' worldlines. It gains contributions linear in both $z_1^{(1)}$ and $z_2^{(1)}$ (and their derivatives) and thus has terms with both kinds of scaling ($1/m_1$, $1/m_2$).
It is given by, 
\begin{alignat}{3}
e^2 q_1 \frac{d F^{\mu \nu} u_{1,\nu}}{d \tau_{2,\mr{ret}}} \tau_{2,\mr{ret}}^{(1)} = e^4 \frac{\partial( q_1 F_{\mu\nu}(z_1,z_{2,\mr{ret}})u_{1,\nu})}{\partial z_{2,\mr{ret}}^{\alpha}}u_{2}^{\alpha} \tau_{2,\mr{ret}}^{(1)}.
\end{alignat}
The $1^{\mr{st}}$ order correction to retarded time is obtained by solving $|\rho|^2=0$ with first order corrected worldlines to order $e^2$. Thus, we have 
\begin{equation}
|z_1 - z_2(\tau_{2,\mr{ret}})|^2 \approx (b^2 + \tau_1^2 + \tau_{2,\mr{ret}}^2 - 2 \tau_1 \tau_{2,\mr{ret}} \gamma) + 2e^2(b+ u_1 \tau_1 - u_2 \tau_{2,\mr{ret}})\cdot(z_1^{(1)}-z_2^{(1)})=0.	 
\end{equation}
To solve iteratively, we substitute $\tau_{2,\mr{ret}} = \tau_{2,\mr{ret}}^{(0)}+e^2 \tau_{2,\mr{ret}}^{(1)}$, $\tau_{2,\mr{ret}}^{(0)} = \gamma \tau_1 - r_{21}$. We get
\begin{alignat}{2}
& e^2 \tau_{2,\mr{ret}}^{(1)}(\tau^{(0)}_{2,\mr{ret}} -  \gamma\tau_{1}) =e^2 \tau_{2,\mr{ret}}^{(1)}r_{21}=- e^2 (b + u_1 \tau_1 - u_2 \tau_{2,\mr{ret}}^{(0)})\cdot(z_1^{(1)}-z_2^{(2)}),\nonumber \\&
e^2 \tau_{2,\mr{ret}}^{(1)}= \frac{e^2 \rho_2^{(0)}\cdot(z_1^{(1)}-z_2^{(1)})}{r_{21}}. \label{1ret}
\end{alignat}
As expected, the retarded time is linear in both $z_1^{(1)}$ and $z_2^{(1)}$ and thus has terms with both types of mass dependencies $m_1^{-1}, m_2^{-1}$. We can now evaluate its contribution to force to be
\begin{alignat}{3}
f_{\mr{III}}^{(2)\mu} = \frac{- q_1 q_2(\rho_2^{(0)}\cdot(z_1^{(1)}-z_2^{(1)}))	 [ (r_{21}u_{2}^{\mu}-3\rho_2^{\mu}\gamma)\gamma-(\gamma r_{21} - 3 (\rho_2^{(0)}\cdot u_1))u_{2}^{\mu}]}{ r_{21}^5} .
\end{alignat}
It contains new types of terms such as
\begin{alignat}{3}
\label{setIII}
\frac{(s_1 \text{ or } \tau_1 \text{ or }s_1^2 \text{ or } \tau_1 s_1 \text{ or } \tau_1^2)\times (s_1 \text{ or }\log(s_1)\text{ or }s_2 \text{ or }\log(s_2))}{r_{21}^5}.
\end{alignat}
   We omit discussion of Diagram $\mr{IV}$ in the appendix since it is much simpler and was explicitly treated in the main text. The total force is the sum of all four contributions,
\begin{equation}
e^4 m_1 z_1^{(2)\mu} = e^4(f^{(2)\mu}_{\mr{I}}+f^{(2)\mu}_{\mr{II}}+f^{(2)\mu}_{\mr{III}}+f^{(2)\mu}_{\mr{IV}} ).
\end{equation}
\subsection{Performing the integrals}
All three sets of terms in Eq.~(\ref{setI}), Eq.~(\ref{setII}), and Eq.~(\ref{setIII}) contain many square roots coming from $r_{21} = \sqrt{|b|^2 + (\gamma v)^2 \tau_1^2}$, $r_{12}(\tau_{2,\mr{ret}}^{(0)}) =\sqrt{|b|^2 + (\gamma v)^2 (\gamma \tau_1 - r_{21})^2}$. The latter even appears to contribute nested square roots. This however can be resolved by using the relation 
\begin{alignat}{3}
&r^{12}_2(\tau_{2,\mr{ret}}^{(0)})= |b|^2 + (\gamma v)^2 (\gamma \tau_1 - r_{21})^2 = |b|^2 + (\gamma^2-1)\gamma^2 \tau_1^2 + (\gamma^2-1)r_{21}^2 -2 (\gamma^2-1)\gamma \tau_1 r_{21}\nonumber\\& = \gamma^2( |b|^2 + (\gamma^2-1) \tau_1^2) + (\gamma^2-1)^2 \tau_1^2 + 2 (\gamma^2-1)\gamma \tau_1 r_{21}\nonumber\\& = (\gamma r_{21} + (\gamma v) \tau_1)^2,
\end{alignat}
which removes the nested square roots but the square roots still complicate analytical integration. We can remove all the square roots by using the very convenient variable $s_1 = \gamma v \tau_1 + r_{21}$ along with the relations
 \begin{alignat}{3}
\nonumber
\tau_1 = \frac{|b|(s_1^2-1)}{2 s_1 \sinh(\phi)}&, \quad\tau_2|_{\tau_2=\tau_{2,\mr{ret}}^{(0)}} = \frac{|b|(s_1^2-e^{2\phi})}{2 s_1 e^{\phi}},\quad
r_{21} = \frac{|b|(1+s_1^2)}{2 s_1},\\& \quad
r_{12}|_{\tau_2=\tau_{2,\mr{ret}}^{(0)}} = \frac{|b|(e^{2\phi}+s_1^2)}{2 e^{\phi} s_1},\quad
s_2|_{\tau_2=\tau_{2,\mr{ret}}^{(0)}}=e^{-\phi} s_1,\label{conv1}
\end{alignat}
where $\phi = \mr{arccosh}(\gamma)$. This reduces the many terms in the acceleration to these simpler classes of terms
\begin{equation}
 \frac{s_1^3\mr{Poly}(s_1) \log(s_1)}{(m_1 \text{ or } m_2)(1+s_1^2)^5}, \quad \frac{s_1^3\mr{Poly}(s_1)}{(m_1 \text{ or } m_2)(1+s_1^2)^5}, \quad\frac{s_1^3\mr{Poly}(s_1)}{m_2(1+s_1^2)^4(e^{2\phi}+s_1^2)^3}.
\end{equation}
We continue to use $s_1$ as our main variable during integration, we multiply the acceleration (force/$m_1$) with the Jacobian $(d\tau_1/ds_1) = r_{21}/(s_1 \sinh(\phi))) = (1+s_1^2)/(2 s_1^2 \sinh(\phi))$. We can now integrate w.r.t.~$s_1$ to get the correction to the velocity ($\dot{z}^{(2)}_1$) up to a constant. The integrals are of the type(s)
\begin{alignat}{3}
\int d s_1 \frac{s_1 \times \mr{Poly}(s_1) \log(s_1)}{(1+s_1^2)^4}, \quad \int d s_1 \frac{s_1 \times \mr{Poly}(s_1)}{(1+s_1^2)^4}, \quad\int d s_1 \frac{s_1 \times \mr{Poly}(s_1)}{(1+s_1^2)^3(e^{2\phi}+s_1^2)^3}.
\end{alignat} 
{\tt Mathematica} has no trouble evaluating these integrals in this form, no futher simplification is required from a practical point of view. It may be tempting to divide them further into basis integrals (say of the form $\int ds_1 s_1^{n} \log(s_1)/(1+s_1^2)^3$) but this does not provide any insight or lead to further simplification. In fact, dividing in such a way can lead to non-elementary functions (PolyLogs) upon integration which cancel out in the overall expression. A better way to divide them further (if one wishes to) is to divide the terms in the force in the forms given in Eq.~(\ref{setI}),  Eq.~(\ref{setII}), and Eq.~(\ref{setIII}) except expressed as functions of $s_1$. Then each individual integral is made only of elementary functions. Regardless, performing all the integrals and putting them together gives an expression for $2^{\mr{nd}}$ order velocity ($\dot{z}^{(2)}_1$) made of a linear combination of terms of the form
\begin{alignat}{3} 
\nonumber &\frac{\mr{Poly}(s_1)}{(m_1m_2\text{ or }m_1^2)(1+s_1^2)^3},\quad \frac{1}{(m_1m_2\text{ or }m_1^2)}\arctan(s_1),  \frac{\mr{Poly}(s_1)\log(|b| s_1 \sinh(\phi))}{(m_1m_2\text{ or }m_1^2)(1+s_1^2)^3},\quad\frac{1}{m_1m_2}\arctan(e^{-\phi} s_1) \nonumber,\quad \\& \frac{\mr{Poly}(s_1)}{m_1m_2 (1+s_1^2)^3(e^{2\phi}+s_1^2)^2}.
\end{alignat}
We get the worldline correction ($z_1^{(2)}$) by repeating the integration process one more time. Thus, we once again multiply the Jacobian $(d\tau_1/ds_1) = r_{21}/(s_1 \sinh(\phi))) = (1+s_1^2)/(2 s_1^2 \sinh(\phi))$ integrate w.r.t.~$s_1$. {\tt Mathematica} can do these integrals with relative ease and we get the $2^{\mr{nd}}$ order corrections to the worldlines. Explicit calculation gives the following expression after imposing the boundary conditions $\lim_{\tau_1\rightarrow -\infty} z_1^{(2)} \cdot b =0$ and $\lim_{\tau_1\rightarrow -\infty} \dot{z}_1^{(2)} = 0$:
\begin{alignat}{3}
&  z_1^{(2)\mu} =  \nonumber \frac{ q_1^2 q_2^2}{m_1^2}\Bigg\{\arctan\Bigg(\frac{\tau_1 \gamma v}{|b|}\Bigg)\frac{(-\cosh(2\phi) \tau_1 \hat{b}^{\mu}+|b| (u_1^{\mu}- \gamma u_2^{\mu}))}{2 |b|^2  (\gamma v)^3}+\frac{\tau_1   \arctan(s_1/|b|) \hat{b}^{\mu}}{|b|^2 \gamma v^3}\\& - \frac{\log( s_1 \gamma v)(s_1 \gamma^3 v \hat{b}^{\mu} +\gamma |b|(u_2^{\mu} - \gamma u_1^{\mu}))}{ |b| r_{21} (\gamma v)^5}+\frac{-2(3|b|^2+s_1^2)\gamma u_1^{\mu}+[5|b|^2 +s_1^2 +(|b|^2+s_1^2)\cosh(2\phi)]u_2^{\mu}}{4 |b|^2 r_{21} \gamma^4 v^5}
   \nonumber \\&
   +\frac{2 |b|(|b|^2 s_1 + 5 s_1^3)+[\pi (|b|^4-s_1^4) - 2 s_1 |b|^3 + 6 s_1^3 |b|]\cosh(2\phi)}{16 |b|^3 s_1^2 r_{21} (\gamma v)^4}b^{\mu}\Bigg\}
   \nonumber \\&
   \nonumber + \frac{ q_1^2 q_2^2}{m_1m_2}\Bigg\{\frac{\tau_1 [2 \arctan(s_{1}/|b|) - \arctan(\tau_1 \gamma v/|b|)] \hat{b}^{\mu} }{2 |b|^2 \gamma^2 v^3 }-\frac{\log(e^{-\phi} s_1 \gamma v)(s_1  \gamma^2 v \hat{b}^{\mu} - (\gamma u_1^{\mu} - u_2^{\mu})) }{ r_{21} (\gamma v)^5}\nonumber \\&+\frac{[\arctan(r_{21}/(|b|\gamma v))+\arctan(v \tau_1/|b|)](-\tau_1  \gamma v^2 \hat{b}^{\mu}+  |b| (\gamma u_1^{\mu}  - u_2^{\mu}))}{2 |b|^2 \gamma^2 v^3}+\frac{e^{-3\phi}(u_2^{\mu}-\gamma u_1^{\mu})}{32 r_{12} r_{21} s_1 |b|^2 (\gamma v)^5}[|b|^4+11 s_1^2 |b|^2 + 2 s_1^4 \nnm \\&+e^{4\phi}(15|b|^4+9s_1^2 |b|^2+2s_1^4)+e^{2\phi}(7|b|^4+11s_1^2|b|^2+4s_1^4)+2e^{6\phi}r_{21} s_1 |b|^2]+\frac{\hat{b}^{\mu}(1-v)s_1}{|b| r_{21} \gamma^3 v^4}-\frac{\pi \tau_1 \hat{b}^{\mu}}{4 |b|^2 \gamma^2 v^3}\nnm\\&+\frac{\hat{b}^{\mu}  (s_1^2-e^{2\phi}\gamma^2 v^2 \tau_1^2)}{2 e^{2\phi}\gamma^2 v^2  r_{21}r_{12} |b|}\Bigg\}    + \frac{2 e^4 q_1^3 q_2[\gamma^2 v s_1 b^{\mu}  + |b|^2 (-\gamma u_1^{\mu}+u_2^{\mu})]}{3 \gamma v^2  m_1^2 |b|^2 r_{21} }.\label{2ndworld}
\end{alignat}

\section{Contributions to the $3^{\mr{rd}}$ order impulse}
\label{3oimpulse}

The force correction at $3^{\mr{rd}}$ order is obtained similarly by expanding the expression for force using $1^{\mr{st}}$ and $2^{\mr{nd}}$ order trajectories. We also need to find the retarded time corrections at next order, i.e., we need to compute, 
\begin{alignat}{3}
& m_1 \ddot{z}_1^{(3)\mu} = \Big[e^6\Big] \left[e^2 q_1 F^{\mu\nu}\dot{z}_{1,\nu} +  e^2 \frac{q^2}{ 6 \pi} \Big(\dddot{z}_1+\ddot{z}_1^2 \dot{z}_1^{\mu}\Big)\right] = f^{(3)\mu},
\\& \nonumber (z_1\rightarrow b+ u_1 \tau_1 + e^2 z_1^{(1)}+e^4 z_1^{(2)},\, z_2 \rightarrow u_2 \tau_2 + e^2 z_2^{(1)}+e^4 z_2^{(2)}, \,\tau_{2,\mr{ret}} = \tau_{2,\mr{ret}}^{(0)}+e^2 \tau_{2,\mr{ret}}^{(1)}+e^4 \tau_{2,\mr{ret}}^{(2)}).
\end{alignat}
In the main text, the various contributions to force correction at $3^{\mr{rd}}$ order were given in diagrammatic form in figures~(\ref{fig3rdl}) and (\ref{fig3rdq}). Here, we write them down in terms of the partial derivatives of the field tensor explicitly and lower order worldline corrections. We also provide the formulae for retarded time corrections, and elaborate on the kind of terms that appear at this order. As mentioned in the main text, contributions at this order come from quadratic in $1^{\mr{st}}$ order worldline corrections ($\sim e^4 (z^{(1)})^2$) and linear in $2^{\mr{nd}}$ order worldline corrections ($\sim e^4z^{(2)}$). It is convenient to deal with them separately.  
\subsection{Quadratic Contributions}
\begin{wrapfigure}{l}{0.15\textwidth}
\begin{tikzpicture}
\begin{scope}[very thick, every node/.style={sloped,allow upside down}]
\node[text width=0.25cm] at (1.1,2.5) {$\mr{I}$};
\draw[dashed](0,0)--(1.1,0);
\draw[very thick](0,1.5)--(1.4,1.5);
\draw[very thick](0,1.25)--(1.4,1.5);
\draw[very thick](1.4,1.6)--(2.2,1.6);
\draw[very thick, double](1.4,1.5)--(2.2,1.5);
\draw[snake it](1.1,0)--node{\midarrow}(1.4,1.5);
\end{scope}
\end{tikzpicture}
\end{wrapfigure}
\textbf{I) Correction to $e^2 q_1 F^{\mu\nu} \dot{z}_{1,\nu}$ from $e^4(z_1^{(1)})^2$.} \newline
Diagram $\mr{I}$ comes from quadratic contribution of $1^{\mr{st}}$ order worldlines for particle 1 (i.e.,~$e^4 (z_1^{(1)})^2$), thus it scales as $m_1^{-2}$. We can write this down either by using the rules for diagrams or Taylor series as
\begin{alignat}{2}
 f^{(3)\mu}_{\mr{I}}  = &  \frac{e^6}{2}\frac{\partial^2 ( q_1 F^{\mu\nu} \dot{z}_{1,\nu})}{\partial z_1^{\alpha}\partial z_1^{\beta}} z_1^{(1),\alpha} z_1^{(1),\beta} + e^6\frac{\partial^2 ( q_1 F^{\mu\nu} \dot{z}_{1,\nu})}{\partial \dot{z}_1^{\alpha}\partial z_1^{\beta}} \dot{z}_1^{(1),\mu} z_1^{(1),\beta} \nonumber \\& + e^6 \frac{1}{2}\frac{\partial^2 ( q_1 F^{\mu\nu} \dot{z}_{1,\nu})}{\partial \dot{z}_1^{\alpha}\partial \dot{z}_1^{\beta}} \dot{z}_1^{(1),\alpha} \dot{z}_1^{(1),\beta},
 \end{alignat} 
 and since the field tensor does not depend on particle 1's velocity, we can simplify this to 
\begin{alignat}{2}
e^6 f^{(3)\mu}_{\mr{I}} = & \frac{1}{2}\frac{\partial^2 ( q_1 F^{\mu\nu} \dot{z}_{1,\nu})}{\partial z_1^{\alpha}\partial z_1^{\beta}} z_1^{(1),\alpha} z_1^{(1),\beta} + \frac{\partial ( q_1 F^{\mu\alpha} )}{\partial z_1^{\beta}} \dot{z}_1^{(1),\alpha} z_1^{(1),\beta}.
 \end{alignat} 
 \end{widetext}
 \begin{widetext}
\begin{wrapfigure}{l}{0.15\textwidth}
\begin{tikzpicture}
\begin{scope}[very thick, every node/.style={sloped,allow upside down}]
\node[text width=0.25cm] at (1.1,2.5) {$\mr{II}$};
\draw[very thick](0,.25)--(1.1,0);
\draw[very thick](0,0)--(1.1,0);
\draw[dashed](0,1.5)--(1.4,1.5);
\draw[very thick](1.4,1.6)--(2.2,1.6);
\draw[very thick, double](1.4,1.5)--(2.2,1.5);
\draw[snake it](1.1,0)--node{\midarrow}(1.4,1.5);
\end{scope}
\end{tikzpicture}
\end{wrapfigure}
\textbf{II) Correction to $e^2 q_1 F^{\mu\nu} \dot{z}_{1,\nu}$ from $e^4(z_2^{(2)})^2$.} \newline
Diagram $\mr{II}$ comes from quadratic contribution of $1^{\mr{st}}$ order worldlines for particle 2 (i.e.,~$e^4(z_1^{(2)})^2$), thus it scales as $m_1^{-2}$. This is more complicated since the field tensor depends on position, velocity and acceleration of particle 2,
\begin{alignat}{3}
 & e^6 f^{(3)\mu}_{\mr{II}}  =  \frac{e^6}{2}\frac{\partial^2 ( q_1 F^{\mu\nu} \dot{z}_{1,\nu})}{\partial z_1^{\alpha}\partial z_1^{\beta}} z_1^{(1),\alpha} z_1^{(1),\beta} + \frac{e^6}{2}\frac{\partial^2 ( q_1 F^{\mu\nu} \dot{z}_{1,\nu})}{\partial \dot{z}_1^{\alpha}\partial \dot{z}_1^{\beta}} \dot{z}_1^{(1),\alpha} \dot{z}_1^{(1),\beta} \nonumber \\&+ \frac{e^6}{2}\frac{\partial^2 ( q_1 F^{\mu\nu} \dot{z}_{1,\nu})}{\partial \ddot{z}_1^{\alpha}\partial \ddot{z}_1^{\beta}} \ddot{z}_1^{(1),\alpha} \ddot{z}_1^{(1),\beta} + e^6 \frac{\partial^2 ( q_1 F^{\mu\nu} \dot{z}_{1,\nu})}{\partial \dot{z}_1^{\alpha}\partial z_1^{\beta}} \dot{z}_2^{(1),\mu} z_2^{(1),\beta}+ e^6\frac{\partial^2 ( q_1 F^{\mu\nu} \dot{z}_{1,\nu})}{\partial \dot{z}_1^{\alpha}\partial \ddot{z}_1^{\beta}} \dot{z}_2^{(1),\mu} \ddot{z}_2^{(1),\beta} \nonumber \\& + e^6\frac{\partial^2 ( q_1 F^{\mu\nu} \dot{z}_{1,\nu})}{\partial \ddot{z}_1^{\alpha}\partial z_1^{\beta}} \ddot{z}_2^{(1),\mu} z_2^{(1),\beta}.
\end{alignat}
\end{widetext}
\begin{widetext}
\begin{wrapfigure}{l}{0.15\textwidth}
\begin{tikzpicture}
\begin{scope}[very thick, every node/.style={sloped,allow upside down}]
\node[text width=0.25cm] at (1.1,2.5) {$\mr{III}$};
\draw[very thick](0,0)--(1.1,0);
\draw[very thick](0,1.5)--(1.4,1.5);
\draw[very thick](1.4,1.6)--(2.2,1.6);
\draw[very thick, double](1.4,1.5)--(2.2,1.5);
\draw[snake it](1.1,0)--node{\midarrow}(1.4,1.5);
\end{scope}
\end{tikzpicture}
\end{wrapfigure}
\textbf{III) Correction to $e^2 q_1 F^{\mu\nu} \dot{z}_{1,\nu}$ from cross terms $e^4(z_1^{(1)} \times z_2^{(1)})$.}\newline
Diagram $\mr{III}$ ise due to quadratic contribution of cross terms from the $1^{\mr{st}}$ order worldline corrections of both particles, thus it scales as $(m_1m_2)^{-1}$,
\begin{alignat}{3}
& e^6 f^{(3)\mu}_{\mr{III}} =e^6\frac{\partial^2 ( q_1 F^{\mu\nu} \dot{z}_{1,\nu})}{\partial z_1^{\alpha}\partial z_2^{\beta}} z_1^{(1),\alpha} z_2^{(1),\beta} + e^6 \frac{\partial^2 ( q_1 F^{\mu\nu} \dot{z}_{1,\nu})}{\partial \dot{z}_1^{\alpha}\partial z_2^{\beta}} \dot{z}_2^{(1),\alpha} z_2^{(1),\beta} \nonumber \\& + e^6 \frac{\partial^2 ( q_1 F^{\mu\nu} \dot{z}_{1,\nu})}{\partial z_1^{\alpha}\partial \dot{z}_2^{\beta}} z_1^{(1),\alpha} \dot{z}_2^{(1),\beta}+ e^6\frac{\partial^2 ( q_1 F^{\mu\nu} \dot{z}_{1,\nu})}{\partial \dot{z}_1^{\alpha}\partial \dot{z}_2^{\beta}} \dot{z}_1^{(1),\alpha} \dot{z}_2^{(1),\beta}+e^6\frac{\partial^2 ( q_1 F^{\mu\nu} \dot{z}_{1,\nu})}{\partial z_1^{\alpha}\partial \ddot{z}_2^{\beta}} z_1^{(1),\alpha} \ddot{z}_2^{(1),\beta} \nonumber \\& +e^6 \frac{\partial^2 ( q_1 F^{\mu\nu} \dot{z}_{1,\nu})}{\partial \dot{z}_1^{\alpha}\partial \ddot{z}_2^{\beta}} \dot{z}_1^{(1),\alpha} \ddot{z}_2^{(1),\beta}.
\end{alignat}
\end{widetext}
\begin{widetext}
\begin{wrapfigure}{l}{0.15\textwidth}
\begin{tikzpicture}
\begin{scope}[very thick, every node/.style={sloped,allow upside down}]
\node[text width=0.25cm] at (1.1,2.5) {$\mr{IV}$};
\node[text width=0.25cm] at (1.1,-0.25){$\tau^{(1)}_{2,\mr{ret}}(z_a^{(1)})$};
\draw[dashed](0,0)--(1.1,0);
\draw[very thick](0,1.5)--(1.4,1.5);
\draw[very thick, double](1.4,1.5)--(2.2,1.5);
\draw[very thick](1.4,1.6)--(2.2,1.6);
\draw[snake it](1.1,0)--node{\midarrow}(1.4,1.5);
\end{scope}
\end{tikzpicture}
\end{wrapfigure}
\textbf{IV) Correction to $e^2 q_1 F^{\mu\nu} \dot{z}_{1,\nu}$ linear in $e^2 z_1^{(1)}$ and $1^{\mr{st}}$ order retarded time correction ($e^2 \tau_{2,\mr{ret}}^{(1)}$)}.\newline
Diagram $\mr{IV}$ is due to the combined contribution of first order worldline correction of particle 1 ($e^2 z_1^{(1)}$) and 1st order retarded time correction ($e^2 \tau_{2,\mr{ret}}^{(1)}$). This has terms that scale as $(m_1m_2)^{-1}$ or $m_1^{-2}$ (see expression of $\tau_{2,\mr{ret}}^{(1)}$ in Eq.~(\ref{1ret})),
\begin{alignat}{3}
e^6 f^{(3)\mu}_{\mr{IV}}  = e^6\frac{d}{d\tau_{2,\mr{ret}}}\Bigg[\frac{\partial ( q_1 F^{\mu\nu}\dot{z}_{1,\nu})}{\partial z_1^{\alpha}}|_{(0)}z_1^{(1),\alpha} + e^6\frac{\partial(  q_1 F^{\mu\nu}\dot{z}_{1,\nu})}{\partial \dot{z}_1^{\alpha}}|_{(0)}\dot{z}_1^{(1),\alpha} \Bigg]\tau_{2,\mr{ret}}^{(1)}.
\end{alignat}
\end{widetext}
\begin{widetext}
\begin{wrapfigure}{l}{0.15\textwidth}
\begin{tikzpicture}
\begin{scope}[very thick, every node/.style={sloped,allow upside down}]
\node[text width=0.25cm] at (1.1,2.5) {$\mr{V}$};
\node[text width=0.25cm] at (1.1,-0.25){$\tau^{(1)}_{2,\mr{ret}}(z_a^{(1)})$};
\draw[very thick](0,0)--(1.1,0);
\draw[dashed](0,1.5)--(1.4,1.5);
\draw[very thick](1.4,1.6)--(2.2,1.6);
\draw[very thick, double](1.4,1.5)--(2.2,1.5);
\draw[snake it](1.1,0)--node{\midarrow}(1.4,1.5);
\end{scope}
\end{tikzpicture}
\end{wrapfigure}
\textbf{V) Correction to $e^2 q_1 F^{\mu\nu} \dot{z}_{1,\nu}$ linear in $z_2^{(1)}$ and $1^{\mr{st}}$ order retarded time correction ($e^2 \tau_{2,\mr{ret}}^{(1)}$)}.\newline
This is the counterpart of $\mr{IV}$. This has terms that scale as $(m_1m_2)^{-1}$ or $m_2^{-2}$,
\begin{alignat}{3}
e^6 f^{(3)\mu}_{\mr{V}} =&e^6 \frac{d}{d\tau_{2,\mr{ret}}}\Bigg[ \frac{\partial ( q_1 F^{\mu\nu}u_{1,\nu})}{\partial z_2^{\alpha}}|_{(0)}z_2^{(1),\alpha} + \frac{\partial(  q_1 F^{\mu\nu}u_{1,\nu})}{\partial \dot{z}_2^{\alpha}}|_{(0)}\dot{z}_2^{(1),\alpha} \nonumber \\& + \frac{\partial(  q_1 F^{\mu\nu}u_{1,\nu})}{\partial \ddot{z}_2^{\alpha}}|_{(0)}\ddot{z}_2^{(1),\alpha} \Bigg]\tau_{2,\mr{ret}}^{(1)}.
\end{alignat}
\end{widetext}
\begin{widetext}
\begin{wrapfigure}{l}{0.15\textwidth}
\begin{tikzpicture}
\begin{scope}[very thick, every node/.style={sloped,allow upside down}]
\node[text width=0.25cm] at (1.1,2.5) {$\mr{VI}$};
\node[text width=0.25cm] at (1.1,-0.25){$(\tau^{(1)}_{2,\mr{ret}}(z_a^{(1)}))^2$};
\draw[dashed](0,0)--(1.1,0);
\draw[dashed](0,1.5)--(1.4,1.5);
\draw[very thick](1.4,1.6)--(2.2,1.6);
\draw[very thick, double](1.4,1.5)--(2.2,1.5);
\draw[snake it](1.1,0)--node{\midarrow}(1.4,1.5);
\end{scope}
\end{tikzpicture}
\end{wrapfigure}
\textbf{VI) Correction to $e^2 q_1 F^{\mu\nu} \dot{z}_{1,\nu}$ quadratic in $1^{\mr{st}}$ order retarded time correction ($e^4(\tau_{2,\mr{ret}}^{(1)})^2$)}.\newline
Diagram $\mr{VI}$ should be self explanatory. This scales as $(m_1 m_2)^{-1}, m_1^{-2}$ or $m_2^{-2}$,
\begin{alignat}{3}
e^6 f^{(3)\mu}_{\mr{VI}}  = e^6\frac{(\tau_{2,\mr{ret}}^{(1)})^2}{2}\frac{d^2}{d\tau_{2,\mr{ret}}^2}(e^2 q_1 F^{\mu\nu}\dot{z}_{1,\nu}).
\end{alignat}
\end{widetext}
\begin{widetext}
\begin{wrapfigure}{l}{0.15\textwidth}
\begin{tikzpicture}
\begin{scope}[very thick, every node/.style={sloped,allow upside down}]
\node[text width=0.25cm] at (1.1,2.5) {$\mr{VII}$};
\node[text width=0.25cm] at (1.1,-0.25){$\tau^{(2)}_{2,\mr{ret}}(z^{(1)})$};
\draw[dashed](0,0)--(1.1,0);
\draw[dashed](0,1.5)--(1.4,1.5);
\draw[very thick](1.4,1.6)--(2.2,1.6);
\draw[very thick, double](1.4,1.5)--(2.2,1.5);\draw[snake it](1.1,0)--node{\midarrow}(1.4,1.5);
\end{scope}
\end{tikzpicture}
\end{wrapfigure}
\textbf{VII) Correction to $e^2 q_1 F^{\mu\nu} \dot{z}_{1,\nu}$ linear in $2^{\mr{nd}}$ order retarded time correction ($e^4 \tau_{2,\mr{ret}}^{(2)}$) (due to 1st order worldline corrections , $e^2 z_1^{(1)}$ and $e^2 z_2^{(1)}$)}.\newline
$1^{\mr{st}}$ order worldline corrections also produce a $2^{\mr{nd}}$ order correction to retarded time since the relation $|z_1 - z_2|^2 =  0 $ is quadratic. To find the $2^{\mr{nd}}$ order retarded time correction (due to $1^{\mr{st}}$ order worldline corrections) $e^4 \tau_{2,\mr{ret}}^{(2)}(z^{(1)})$, we need to solve the relation $|z_1-z_2|^2=0$ to NNLO ($e^4$). Substituting worldlines with $1^{\mr{st}}$ order corrections, we have
\begin{alignat}{3}
&|z_1 - z_2|^2=b^2 + \tau_1^2 + \tau_{2,\mr{ret}}^2 - 2 \tau_1 \tau_{2,\mr{ret}} \gamma 
\nnm\\
&+ 2e^2(b+ u_1 \tau_1 - u_2 \tau_{2,\mr{ret}})\cdot(z_1^{(1)}-z_2^{(1)})+e^4(z_1^{(1)}-z_2^{(1)})^2 = 0 ,
\end{alignat}
where we substitute $\tau_{2,\mr{ret}} = \tau_{2,\mr{ret}}^{(0)}+e^2 \tau_{2,\mr{ret}}^{(1)}+e^4 \tau_{2,\mr{ret}}^{(2)}$, and then solve for $\tau_{2,\mr{ret}}^{(2)}$. We get
\begin{alignat}{3}
&
 (\tau_{2,\mr{ret}}^{(1)})^2+2 r_1\tau_{2,\mr{ret}}^{(2)}-2  u_2\cdot (z_1^{(1)}-z_2^{(1)}(\tau_{2,\mr{ret}}^{(0)})) \tau_{2,\mr{ret}}^{(1)} - 2 \rho^{(0)} \dot{z}_2^{(1)}(\tau_{2,\mr{ret}}^{(0)}) \tau_{2,\mr{ret}}^{(1)}+(z_1^{(1)}-z_2^{(2)}(\tau_{2,\mr{ret}}^{(0)}))^2=0, \nonumber \\&
 \tau_{2,\mr{ret}}^{(2)} = \frac{-(z_1^{(1)}-z_2^{(2)}(\tau_{2,\mr{ret}}^{(0)}))^2+2 \rho^{(0)} \dot{z}_2^{(1)}(\tau_{2,\mr{ret}}^{(0)}) \tau_{2,\mr{ret}}^{(1)}+2  u_2\cdot (z_1^{(1)}-z_2^{(1)}(\tau_{2,\mr{ret}}^{(0)})) \tau_{2,\mr{ret}}^{(1)}-(\tau_{2,\mr{ret}}^{(1)})^2}{2 r_1}.
\end{alignat}
Once we have $\tau_{2,\mr{ret}}^{(2)}$, we can simply substitute this in
\begin{alignat}{3}
e^6 f^{(3)\mu}_{\mr{VII}}  = e^6 q_1 \frac{d F^{\mu \nu} u_{1,\nu}}{d \tau_{2,\mr{ret}}} \tau_{2,\mr{ret}}^{(2)} = e^6 \frac{\partial( q_1 F_{\mu\nu}(z_1,z_{2,\mr{ret}})u_{1,\nu})}{\partial z_{2,\mr{ret}}^{\alpha}}u_{2}^{\alpha} \tau_{2,\mr{ret}}^{(2)}.
\end{alignat}
These are all the quadratic corrections to Lorentz force at $3^{\mr{rd}}$ order ($e^6$). \newline
\textbf {IX) Self-force contribution} -
In addition, there is a quadratic contribution to the ALD force
$(2 e^2q_1^2/3)(\dddot{z}_1^{\mu}+\ddot{z}_1^2\dot{z}_1^{\mu}) $.
We only need to include the second term at this order since we are interested in the impulse, and the first term is a total derivative of acceleration (which vanishes at boundaries). Thus, it gains no relevant contribution from $2^{\mr{nd}}$ order worldline corrections at $3^{\mr{rd}}$ order. We omit the diagram here. Thus, the relevant contribution is
\begin{alignat}{3}
e^6 f^{(3)\mu}_{\mr{IX}}  = (2 e^6q_1^2/3)\times (\ddot{z}_1^{(1)})^2 u_1^{\mu} = \frac{2 e^6 q_1^4 q_2^2[\gamma^2 b^2 + \tau_1^2(\gamma-1)^2]}{3 m_1^2 \pi^3 r_1^6}u_1^{\mu}.
\end{alignat}
\end{widetext}

\begin{widetext}
\subsection{Linear contributions}
The next set of contributions are corrections to the Lorentz force term that are linear in $2^{\mr{nd}}$ order worldline corrections. These are similar in form to the diagrams at $2^{\mr{nd}}$ order. We thus have three diagrams again, 
 \begin{wrapfigure}{l}{0.15\textwidth}
\begin{tikzpicture}
\begin{scope}[very thick, every node/.style={sloped,allow upside down}]
\node[text width=0.25cm] at (1.1,2.5) {$\mr{IX}$};
\draw[dashed](0,0)--(1.1,0);
\draw[very thick,double](0,1.5)--(1.4,1.5);
\draw[very thick](1.4,1.6)--(2.2,1.6);
\draw[very thick, double](1.4,1.5)--(2.2,1.5);
\draw[snake it](1.1,0)--node{\midarrow}(1.4,1.5);
\end{scope}
\end{tikzpicture}
\end{wrapfigure}
\textbf{IX) Correction to $e^2 q_1 F^{\mu\nu} \dot{z}_{1,\nu}$ linear in $e^4 z_1^{(2)}$}.\newline
Diagram $\mr{IX}$ is due to the $2^{\mr{nd}}$ order world line correction to particle 1 to the Lorentz force, while fixing retarded time and particle 2's worldline at zeroth order. It is linear in $e^4 z_1^{(2)}$, thus scales as $1/m_1m_2$ or $1/m_1^2$. It is given by
\begin{alignat}{3}
e^6 f^{(3)\mu}_{\mr{IX}}  = e^6\frac{\partial ( q_1 F^{\mu\nu}\dot{z}_{1,\nu})}{\partial z_1^{\alpha}}|_{(0)}z_1^{(2),\alpha} + e^6 \frac{\partial( q_1 F^{\mu\nu}\dot{z}_{1,\nu})}{\partial \dot{z}_1^{\alpha}}|_{(0)}\dot{z}_1^{(2),\alpha}.
\end{alignat}
\end{widetext}
\begin{widetext}
\begin{wrapfigure}{l}{0.15\textwidth}
\begin{tikzpicture}
\begin{scope}[very thick, every node/.style={sloped,allow upside down}]
\node[text width=0.25cm] at (1.1,2.5) {$\mr{X}$};
\draw[very thick, double](0,0)--(1.1,0);
\draw[dashed](0,1.5)--(1.4,1.5);
\draw[very thick](1.4,1.6)--(2.2,1.6);
\draw[very thick, double](1.4,1.5)--(2.2,1.5);
\draw[snake it](1.1,0)--node{\midarrow}(1.4,1.5);
\end{scope}
\end{tikzpicture}
\end{wrapfigure}
\textbf{X) Correction to $e^2 q_1 F^{\mu\nu} \dot{z}_{1,\nu}$ linear in $e^4 z_2^{(2)}$}.\newline
Diagram $\mr{X}$ is due to $2^{\mr{nd}}$ order worldline correction to particle 2 while fixing retarded time and particle 1's worldline at zeroth order. It is linear in $e^4 z_2^{(2)}$, thus scales as $1/m_1m_2$ or $1/m_2^2$. It is given by
\begin{alignat}{2}
e^6 f^{(3)\mu}_{\mr{X}}  = & e^6 \frac{\partial ( q_1 F^{\mu\nu}u_{1,\nu})}{\partial z_2^{\alpha}}|_{(0)}z_2^{(2),\alpha} + e^6 \frac{\partial(  q_1 F^{\mu\nu}u_{1,\nu})}{\partial \dot{z}_2^{\alpha}}|_{(0)}\dot{z}_2^{(2),\alpha} \nonumber \\& + e^6 \frac{\partial(  q_1 F^{\mu\nu}u_{1,\nu})}{\partial \ddot{z}_2^{\alpha}}|_{(0)}\ddot{z}_2^{(2),\alpha}.
\end{alignat}
\end{widetext}
\begin{widetext}
\begin{wrapfigure}{l}{0.15\textwidth}
\begin{tikzpicture}
\begin{scope}[very thick, every node/.style={sloped,allow upside down}]
\node[text width=0.25cm] at (1.1,2.5) {$\mr{XI}$};
\node[text width=0.25cm] at (1.1,-0.25){$\tau^{(2)}_{2,\mr{ret}}(z^{(2)})$};
\draw[dashed](0,0)--(1.1,0);
\draw[dashed](0,1.5)--(1.4,1.5);
\draw[very thick](1.4,1.6)--(2.2,1.6);
\draw[very thick, double](1.4,1.5)--(2.2,1.5);
\draw[snake it](1.1,0)--node{\midarrow}(1.4,1.5);
\end{scope}
\end{tikzpicture}
\end{wrapfigure}
\textbf{XI) Correction to $e^2 q_1 F^{\mu\nu} \dot{z}_{1,\nu}$ from $2^{\mr{nd}}$ order retarded time correction due to $2^{\mr{nd}}$ order worldline corrections ($e^4 \tau_{2,\mr{ret}}^{(2)}(z^{(2)}_a)$)}.\newline
The second order correction to retarded time also gets contribution from $z^{(2)}$ as one would expect. We can find it in the same manner we found the first order retarded time correction (see derivation in Eq.~\ref{1ret}), to get  $\tau_{2,\mr{ret}}^{(2)}= \frac{ \rho^{(0)}\cdot(z_1^{(2)}-z_2^{(2)})}{r_1}$. Thus, we have
\begin{alignat}{3}
e^6 f^{(3)\mu}_{\mr{XI}}  = e^6\frac{d ( q_1 F^{\mu\nu}\dot{z}_{1,\nu})}{d \tau_{2,\mr{ret}}}|_{(0)}\tau_{2,\mr{ret}}^{(2)} =e^6 \frac{\partial( q_1 F_{\mu\nu}(z_1,z_{2,\mr{ret}})u_{1,\nu})}{\partial z_{2,\mr{ret}}^{\alpha}}u_{2}^{\alpha} \tau_{2,\mr{ret}}^{(2)}.
\end{alignat}
These are all the contributions to the acceleration and subsequently impulse at $3^{\mr{rd}}$ order. 
\end{widetext}


\begin{thebibliography}{49}%
\makeatletter
\providecommand \@ifxundefined [1]{%
 \@ifx{#1\undefined}
}%
\providecommand \@ifnum [1]{%
 \ifnum #1\expandafter \@firstoftwo
 \else \expandafter \@secondoftwo
 \fi
}%
\providecommand \@ifx [1]{%
 \ifx #1\expandafter \@firstoftwo
 \else \expandafter \@secondoftwo
 \fi
}%
\providecommand \natexlab [1]{#1}%
\providecommand \enquote  [1]{``#1''}%
\providecommand \bibnamefont  [1]{#1}%
\providecommand \bibfnamefont [1]{#1}%
\providecommand \citenamefont [1]{#1}%
\providecommand \href@noop [0]{\@secondoftwo}%
\providecommand \href [0]{\begingroup \@sanitize@url \@href}%
\providecommand \@href[1]{\@@startlink{#1}\@@href}%
\providecommand \@@href[1]{\endgroup#1\@@endlink}%
\providecommand \@sanitize@url [0]{\catcode `\\12\catcode `\$12\catcode
  `\&12\catcode `\#12\catcode `\^12\catcode `\_12\catcode `\%12\relax}%
\providecommand \@@startlink[1]{}%
\providecommand \@@endlink[0]{}%
\providecommand \url  [0]{\begingroup\@sanitize@url \@url }%
\providecommand \@url [1]{\endgroup\@href {#1}{\urlprefix }}%
\providecommand \urlprefix  [0]{URL }%
\providecommand \Eprint [0]{\href }%
\providecommand \doibase [0]{http://dx.doi.org/}%
\providecommand \selectlanguage [0]{\@gobble}%
\providecommand \bibinfo  [0]{\@secondoftwo}%
\providecommand \bibfield  [0]{\@secondoftwo}%
\providecommand \translation [1]{[#1]}%
\providecommand \BibitemOpen [0]{}%
\providecommand \bibitemStop [0]{}%
\providecommand \bibitemNoStop [0]{.\EOS\space}%
\providecommand \EOS [0]{\spacefactor3000\relax}%
\providecommand \BibitemShut  [1]{\csname bibitem#1\endcsname}%
\let\auto@bib@innerbib\@empty
\bibitem [{\citenamefont {Abbott}\ \emph
  {et~al.}(2016{\natexlab{a}})\citenamefont {Abbott} \emph
  {et~al.}}]{Abbott:2016blz}%
  \BibitemOpen
  \bibfield  {author} {\bibinfo {author} {\bibfnamefont {B.~P.}\ \bibnamefont
  {Abbott}} \emph {et~al.} (\bibinfo {collaboration} {LIGO Scientific,
  Virgo}),\ }\bibfield  {title} {\enquote {\bibinfo {title} {{Observation of
  Gravitational Waves from a Binary Black Hole Merger}},}\ }\href {\doibase
  10.1103/PhysRevLett.116.061102} {\bibfield  {journal} {\bibinfo  {journal}
  {Phys. Rev. Lett.}\ }\textbf {\bibinfo {volume} {116}},\ \bibinfo {pages}
  {061102} (\bibinfo {year} {2016}{\natexlab{a}})},\ \Eprint
  {http://arxiv.org/abs/1602.03837} {arXiv:1602.03837 [gr-qc]} \BibitemShut
  {NoStop}%
\bibitem [{\citenamefont {Abbott}\ \emph
  {et~al.}(2016{\natexlab{b}})\citenamefont {Abbott} \emph
  {et~al.}}]{TheLIGOScientific:2016pea}%
  \BibitemOpen
  \bibfield  {author} {\bibinfo {author} {\bibfnamefont {B.~P.}\ \bibnamefont
  {Abbott}} \emph {et~al.} (\bibinfo {collaboration} {LIGO Scientific,
  Virgo}),\ }\bibfield  {title} {\enquote {\bibinfo {title} {{Binary Black Hole
  Mergers in the first Advanced LIGO Observing Run}},}\ }\href {\doibase
  10.1103/PhysRevX.6.041015} {\bibfield  {journal} {\bibinfo  {journal} {Phys.
  Rev. X}\ }\textbf {\bibinfo {volume} {6}},\ \bibinfo {pages} {041015}
  (\bibinfo {year} {2016}{\natexlab{b}})},\ \bibinfo {note} {[Erratum:
  Phys.Rev.X 8, 039903 (2018)]},\ \Eprint {http://arxiv.org/abs/1606.04856}
  {arXiv:1606.04856 [gr-qc]} \BibitemShut {NoStop}%
\bibitem [{\citenamefont {Abbott}\ \emph {et~al.}(2017)\citenamefont {Abbott}
  \emph {et~al.}}]{TheLIGOScientific:2017qsa}%
  \BibitemOpen
  \bibfield  {author} {\bibinfo {author} {\bibfnamefont {B.~P.}\ \bibnamefont
  {Abbott}} \emph {et~al.} (\bibinfo {collaboration} {LIGO Scientific,
  Virgo}),\ }\bibfield  {title} {\enquote {\bibinfo {title} {{GW170817:
  Observation of Gravitational Waves from a Binary Neutron Star Inspiral}},}\
  }\href {\doibase 10.1103/PhysRevLett.119.161101} {\bibfield  {journal}
  {\bibinfo  {journal} {Phys. Rev. Lett.}\ }\textbf {\bibinfo {volume} {119}},\
  \bibinfo {pages} {161101} (\bibinfo {year} {2017})},\ \Eprint
  {http://arxiv.org/abs/1710.05832} {arXiv:1710.05832 [gr-qc]} \BibitemShut
  {NoStop}%
\bibitem [{\citenamefont {Abbott}\ \emph {et~al.}(2019)\citenamefont {Abbott}
  \emph {et~al.}}]{LIGOScientific:2018mvr}%
  \BibitemOpen
  \bibfield  {author} {\bibinfo {author} {\bibfnamefont {B.~P.}\ \bibnamefont
  {Abbott}} \emph {et~al.} (\bibinfo {collaboration} {LIGO Scientific,
  Virgo}),\ }\bibfield  {title} {\enquote {\bibinfo {title} {{GWTC-1: A
  Gravitational-Wave Transient Catalog of Compact Binary Mergers Observed by
  LIGO and Virgo during the First and Second Observing Runs}},}\ }\href
  {\doibase 10.1103/PhysRevX.9.031040} {\bibfield  {journal} {\bibinfo
  {journal} {Phys. Rev. X}\ }\textbf {\bibinfo {volume} {9}},\ \bibinfo {pages}
  {031040} (\bibinfo {year} {2019})},\ \Eprint
  {http://arxiv.org/abs/1811.12907} {arXiv:1811.12907 [astro-ph.HE]}
  \BibitemShut {NoStop}%
\bibitem [{\citenamefont {Abbott}\ \emph {et~al.}(2020)\citenamefont {Abbott}
  \emph {et~al.}}]{Abbott:2020niy}%
  \BibitemOpen
  \bibfield  {author} {\bibinfo {author} {\bibfnamefont {R.}~\bibnamefont
  {Abbott}} \emph {et~al.} (\bibinfo {collaboration} {LIGO Scientific,
  Virgo}),\ }\bibfield  {title} {\enquote {\bibinfo {title} {{GWTC-2: Compact
  Binary Coalescences Observed by LIGO and Virgo During the First Half of the
  Third Observing Run}},}\ }\href@noop {} {\  (\bibinfo {year} {2020})},\
  \Eprint {http://arxiv.org/abs/2010.14527} {arXiv:2010.14527 [gr-qc]}
  \BibitemShut {NoStop}%
\bibitem [{\citenamefont {Punturo}\ \emph {et~al.}(2010)\citenamefont {Punturo}
  \emph {et~al.}}]{Punturo:2010zz}%
  \BibitemOpen
  \bibfield  {author} {\bibinfo {author} {\bibfnamefont {M.}~\bibnamefont
  {Punturo}} \emph {et~al.},\ }\bibfield  {title} {\enquote {\bibinfo {title}
  {{The Einstein Telescope: A third-generation gravitational wave
  observatory}},}\ }\href {\doibase 10.1088/0264-9381/27/19/194002} {\bibfield
  {journal} {\bibinfo  {journal} {Class. Quant. Grav.}\ }\textbf {\bibinfo
  {volume} {27}},\ \bibinfo {pages} {194002} (\bibinfo {year}
  {2010})}\BibitemShut {NoStop}%
\bibitem [{\citenamefont {Amaro-Seoane}\ \emph {et~al.}(2017)\citenamefont
  {Amaro-Seoane} \emph {et~al.}}]{Audley:2017drz}%
  \BibitemOpen
  \bibfield  {author} {\bibinfo {author} {\bibfnamefont {Pau}\ \bibnamefont
  {Amaro-Seoane}} \emph {et~al.} (\bibinfo {collaboration} {LISA}),\ }\bibfield
   {title} {\enquote {\bibinfo {title} {{Laser Interferometer Space
  Antenna}},}\ }\href@noop {} {\  (\bibinfo {year} {2017})},\ \Eprint
  {http://arxiv.org/abs/1702.00786} {arXiv:1702.00786 [astro-ph.IM]}
  \BibitemShut {NoStop}%
\bibitem [{\citenamefont {Reitze}\ \emph {et~al.}(2019)\citenamefont {Reitze}
  \emph {et~al.}}]{Reitze:2019iox}%
  \BibitemOpen
  \bibfield  {author} {\bibinfo {author} {\bibfnamefont {David}\ \bibnamefont
  {Reitze}} \emph {et~al.},\ }\bibfield  {title} {\enquote {\bibinfo {title}
  {{Cosmic Explorer: The U.S. Contribution to Gravitational-Wave Astronomy
  beyond LIGO}},}\ }\href@noop {} {\bibfield  {journal} {\bibinfo  {journal}
  {Bull. Am. Astron. Soc.}\ }\textbf {\bibinfo {volume} {51}},\ \bibinfo
  {pages} {035} (\bibinfo {year} {2019})},\ \Eprint
  {http://arxiv.org/abs/1907.04833} {arXiv:1907.04833 [astro-ph.IM]}
  \BibitemShut {NoStop}%
\bibitem [{\citenamefont {Kawai}\ \emph {et~al.}(1986)\citenamefont {Kawai},
  \citenamefont {Lewellen},\ and\ \citenamefont {Tye}}]{Kawai:1985xq}%
  \BibitemOpen
  \bibfield  {author} {\bibinfo {author} {\bibfnamefont {H.}~\bibnamefont
  {Kawai}}, \bibinfo {author} {\bibfnamefont {D.~C.}\ \bibnamefont {Lewellen}},
  \ and\ \bibinfo {author} {\bibfnamefont {S.~H.~H.}\ \bibnamefont {Tye}},\
  }\bibfield  {title} {\enquote {\bibinfo {title} {{A Relation Between Tree
  Amplitudes of Closed and Open Strings}},}\ }\href {\doibase
  10.1016/0550-3213(86)90362-7} {\bibfield  {journal} {\bibinfo  {journal}
  {Nucl. Phys.}\ }\textbf {\bibinfo {volume} {B269}},\ \bibinfo {pages} {1--23}
  (\bibinfo {year} {1986})}\BibitemShut {NoStop}%
\bibitem [{\citenamefont {Bern}\ \emph {et~al.}(2008)\citenamefont {Bern},
  \citenamefont {Carrasco},\ and\ \citenamefont {Johansson}}]{Bern:2008qj}%
  \BibitemOpen
  \bibfield  {author} {\bibinfo {author} {\bibfnamefont {Z.}~\bibnamefont
  {Bern}}, \bibinfo {author} {\bibfnamefont {J.~J.~M.}\ \bibnamefont
  {Carrasco}}, \ and\ \bibinfo {author} {\bibfnamefont {Henrik}\ \bibnamefont
  {Johansson}},\ }\bibfield  {title} {\enquote {\bibinfo {title} {{New
  Relations for Gauge-Theory Amplitudes}},}\ }\href {\doibase
  10.1103/PhysRevD.78.085011} {\bibfield  {journal} {\bibinfo  {journal} {Phys.
  Rev. D}\ }\textbf {\bibinfo {volume} {78}},\ \bibinfo {pages} {085011}
  (\bibinfo {year} {2008})},\ \Eprint {http://arxiv.org/abs/0805.3993}
  {arXiv:0805.3993 [hep-ph]} \BibitemShut {NoStop}%
\bibitem [{\citenamefont {Bern}\ \emph {et~al.}(2010)\citenamefont {Bern},
  \citenamefont {Carrasco},\ and\ \citenamefont {Johansson}}]{Bern:2010ue}%
  \BibitemOpen
  \bibfield  {author} {\bibinfo {author} {\bibfnamefont {Zvi}\ \bibnamefont
  {Bern}}, \bibinfo {author} {\bibfnamefont {John Joseph~M.}\ \bibnamefont
  {Carrasco}}, \ and\ \bibinfo {author} {\bibfnamefont {Henrik}\ \bibnamefont
  {Johansson}},\ }\bibfield  {title} {\enquote {\bibinfo {title} {{Perturbative
  Quantum Gravity as a Double Copy of Gauge Theory}},}\ }\href {\doibase
  10.1103/PhysRevLett.105.061602} {\bibfield  {journal} {\bibinfo  {journal}
  {Phys. Rev. Lett.}\ }\textbf {\bibinfo {volume} {105}},\ \bibinfo {pages}
  {061602} (\bibinfo {year} {2010})},\ \Eprint {http://arxiv.org/abs/1004.0476}
  {arXiv:1004.0476 [hep-th]} \BibitemShut {NoStop}%
\bibitem [{\citenamefont {Bern}\ \emph
  {et~al.}(2019{\natexlab{a}})\citenamefont {Bern}, \citenamefont {Carrasco},
  \citenamefont {Chiodaroli}, \citenamefont {Johansson},\ and\ \citenamefont
  {Roiban}}]{Bern:2019prr}%
  \BibitemOpen
  \bibfield  {author} {\bibinfo {author} {\bibfnamefont {Zvi}\ \bibnamefont
  {Bern}}, \bibinfo {author} {\bibfnamefont {John~Joseph}\ \bibnamefont
  {Carrasco}}, \bibinfo {author} {\bibfnamefont {Marco}\ \bibnamefont
  {Chiodaroli}}, \bibinfo {author} {\bibfnamefont {Henrik}\ \bibnamefont
  {Johansson}}, \ and\ \bibinfo {author} {\bibfnamefont {Radu}\ \bibnamefont
  {Roiban}},\ }\bibfield  {title} {\enquote {\bibinfo {title} {{The Duality
  Between Color and Kinematics and its Applications}},}\ }\href@noop {} {\
  (\bibinfo {year} {2019}{\natexlab{a}})},\ \Eprint
  {http://arxiv.org/abs/1909.01358} {arXiv:1909.01358 [hep-th]} \BibitemShut
  {NoStop}%
\bibitem [{\citenamefont {Damour}(2018)}]{Damour:2017zjx}%
  \BibitemOpen
  \bibfield  {author} {\bibinfo {author} {\bibfnamefont {Thibault}\
  \bibnamefont {Damour}},\ }\bibfield  {title} {\enquote {\bibinfo {title}
  {{High-energy gravitational scattering and the general relativistic two-body
  problem}},}\ }\href {\doibase 10.1103/PhysRevD.97.044038} {\bibfield
  {journal} {\bibinfo  {journal} {Phys. Rev. D}\ }\textbf {\bibinfo {volume}
  {97}},\ \bibinfo {pages} {044038} (\bibinfo {year} {2018})},\ \Eprint
  {http://arxiv.org/abs/1710.10599} {arXiv:1710.10599 [gr-qc]} \BibitemShut
  {NoStop}%
\bibitem [{\citenamefont {Bjerrum-Bohr}\ \emph {et~al.}(2018)\citenamefont
  {Bjerrum-Bohr}, \citenamefont {Damgaard}, \citenamefont {Festuccia},
  \citenamefont {Planté},\ and\ \citenamefont
  {Vanhove}}]{Bjerrum-Bohr:2018xdl}%
  \BibitemOpen
  \bibfield  {author} {\bibinfo {author} {\bibfnamefont {N.~E.~J.}\
  \bibnamefont {Bjerrum-Bohr}}, \bibinfo {author} {\bibfnamefont {Poul~H.}\
  \bibnamefont {Damgaard}}, \bibinfo {author} {\bibfnamefont {Guido}\
  \bibnamefont {Festuccia}}, \bibinfo {author} {\bibfnamefont {Ludovic}\
  \bibnamefont {Planté}}, \ and\ \bibinfo {author} {\bibfnamefont {Pierre}\
  \bibnamefont {Vanhove}},\ }\bibfield  {title} {\enquote {\bibinfo {title}
  {{General Relativity from Scattering Amplitudes}},}\ }\href {\doibase
  10.1103/PhysRevLett.121.171601} {\bibfield  {journal} {\bibinfo  {journal}
  {Phys. Rev. Lett.}\ }\textbf {\bibinfo {volume} {121}},\ \bibinfo {pages}
  {171601} (\bibinfo {year} {2018})},\ \Eprint
  {http://arxiv.org/abs/1806.04920} {arXiv:1806.04920 [hep-th]} \BibitemShut
  {NoStop}%
\bibitem [{\citenamefont {Cheung}\ \emph {et~al.}(2018)\citenamefont {Cheung},
  \citenamefont {Rothstein},\ and\ \citenamefont {Solon}}]{Cheung:2018wkq}%
  \BibitemOpen
  \bibfield  {author} {\bibinfo {author} {\bibfnamefont {Clifford}\
  \bibnamefont {Cheung}}, \bibinfo {author} {\bibfnamefont {Ira~Z.}\
  \bibnamefont {Rothstein}}, \ and\ \bibinfo {author} {\bibfnamefont
  {Mikhail~P.}\ \bibnamefont {Solon}},\ }\bibfield  {title} {\enquote {\bibinfo
  {title} {{From Scattering Amplitudes to Classical Potentials in the
  Post-Minkowskian Expansion}},}\ }\href {\doibase
  10.1103/PhysRevLett.121.251101} {\bibfield  {journal} {\bibinfo  {journal}
  {Phys. Rev. Lett.}\ }\textbf {\bibinfo {volume} {121}},\ \bibinfo {pages}
  {251101} (\bibinfo {year} {2018})},\ \Eprint
  {http://arxiv.org/abs/1808.02489} {arXiv:1808.02489 [hep-th]} \BibitemShut
  {NoStop}%
\bibitem [{\citenamefont {Bern}\ \emph
  {et~al.}(2019{\natexlab{b}})\citenamefont {Bern}, \citenamefont {Cheung},
  \citenamefont {Roiban}, \citenamefont {Shen}, \citenamefont {Solon},\ and\
  \citenamefont {Zeng}}]{Bern:2019nnu}%
  \BibitemOpen
  \bibfield  {author} {\bibinfo {author} {\bibfnamefont {Zvi}\ \bibnamefont
  {Bern}}, \bibinfo {author} {\bibfnamefont {Clifford}\ \bibnamefont {Cheung}},
  \bibinfo {author} {\bibfnamefont {Radu}\ \bibnamefont {Roiban}}, \bibinfo
  {author} {\bibfnamefont {Chia-Hsien}\ \bibnamefont {Shen}}, \bibinfo {author}
  {\bibfnamefont {Mikhail~P.}\ \bibnamefont {Solon}}, \ and\ \bibinfo {author}
  {\bibfnamefont {Mao}\ \bibnamefont {Zeng}},\ }\bibfield  {title} {\enquote
  {\bibinfo {title} {{Scattering Amplitudes and the Conservative Hamiltonian
  for Binary Systems at Third Post-Minkowskian Order}},}\ }\href {\doibase
  10.1103/PhysRevLett.122.201603} {\bibfield  {journal} {\bibinfo  {journal}
  {Phys. Rev. Lett.}\ }\textbf {\bibinfo {volume} {122}},\ \bibinfo {pages}
  {201603} (\bibinfo {year} {2019}{\natexlab{b}})},\ \Eprint
  {http://arxiv.org/abs/1901.04424} {arXiv:1901.04424 [hep-th]} \BibitemShut
  {NoStop}%
\bibitem [{\citenamefont {Bern}\ \emph
  {et~al.}(2019{\natexlab{c}})\citenamefont {Bern}, \citenamefont {Cheung},
  \citenamefont {Roiban}, \citenamefont {Shen}, \citenamefont {Solon},\ and\
  \citenamefont {Zeng}}]{Bern:2019crd}%
  \BibitemOpen
  \bibfield  {author} {\bibinfo {author} {\bibfnamefont {Zvi}\ \bibnamefont
  {Bern}}, \bibinfo {author} {\bibfnamefont {Clifford}\ \bibnamefont {Cheung}},
  \bibinfo {author} {\bibfnamefont {Radu}\ \bibnamefont {Roiban}}, \bibinfo
  {author} {\bibfnamefont {Chia-Hsien}\ \bibnamefont {Shen}}, \bibinfo {author}
  {\bibfnamefont {Mikhail~P.}\ \bibnamefont {Solon}}, \ and\ \bibinfo {author}
  {\bibfnamefont {Mao}\ \bibnamefont {Zeng}},\ }\bibfield  {title} {\enquote
  {\bibinfo {title} {{Black Hole Binary Dynamics from the Double Copy and
  Effective Theory}},}\ }\href {\doibase 10.1007/JHEP10(2019)206} {\bibfield
  {journal} {\bibinfo  {journal} {JHEP}\ }\textbf {\bibinfo {volume} {10}},\
  \bibinfo {pages} {206} (\bibinfo {year} {2019}{\natexlab{c}})},\ \Eprint
  {http://arxiv.org/abs/1908.01493} {arXiv:1908.01493 [hep-th]} \BibitemShut
  {NoStop}%
\bibitem [{\citenamefont {Bern}\ \emph {et~al.}(1994)\citenamefont {Bern},
  \citenamefont {Dixon}, \citenamefont {Dunbar},\ and\ \citenamefont
  {Kosower}}]{Bern:1994zx}%
  \BibitemOpen
  \bibfield  {author} {\bibinfo {author} {\bibfnamefont {Zvi}\ \bibnamefont
  {Bern}}, \bibinfo {author} {\bibfnamefont {Lance~J.}\ \bibnamefont {Dixon}},
  \bibinfo {author} {\bibfnamefont {David~C.}\ \bibnamefont {Dunbar}}, \ and\
  \bibinfo {author} {\bibfnamefont {David~A.}\ \bibnamefont {Kosower}},\
  }\bibfield  {title} {\enquote {\bibinfo {title} {{One loop n point gauge
  theory amplitudes, unitarity and collinear limits}},}\ }\href {\doibase
  10.1016/0550-3213(94)90179-1} {\bibfield  {journal} {\bibinfo  {journal}
  {Nucl. Phys. B}\ }\textbf {\bibinfo {volume} {425}},\ \bibinfo {pages}
  {217--260} (\bibinfo {year} {1994})},\ \Eprint
  {http://arxiv.org/abs/hep-ph/9403226} {arXiv:hep-ph/9403226} \BibitemShut
  {NoStop}%
\bibitem [{\citenamefont {Bern}\ \emph {et~al.}(1995)\citenamefont {Bern},
  \citenamefont {Dixon}, \citenamefont {Dunbar},\ and\ \citenamefont
  {Kosower}}]{Bern:1994cg}%
  \BibitemOpen
  \bibfield  {author} {\bibinfo {author} {\bibfnamefont {Zvi}\ \bibnamefont
  {Bern}}, \bibinfo {author} {\bibfnamefont {Lance~J.}\ \bibnamefont {Dixon}},
  \bibinfo {author} {\bibfnamefont {David~C.}\ \bibnamefont {Dunbar}}, \ and\
  \bibinfo {author} {\bibfnamefont {David~A.}\ \bibnamefont {Kosower}},\
  }\bibfield  {title} {\enquote {\bibinfo {title} {{Fusing gauge theory tree
  amplitudes into loop amplitudes}},}\ }\href {\doibase
  10.1016/0550-3213(94)00488-Z} {\bibfield  {journal} {\bibinfo  {journal}
  {Nucl. Phys. B}\ }\textbf {\bibinfo {volume} {435}},\ \bibinfo {pages}
  {59--101} (\bibinfo {year} {1995})},\ \Eprint
  {http://arxiv.org/abs/hep-ph/9409265} {arXiv:hep-ph/9409265} \BibitemShut
  {NoStop}%
\bibitem [{\citenamefont {Britto}\ \emph {et~al.}(2005)\citenamefont {Britto},
  \citenamefont {Cachazo},\ and\ \citenamefont {Feng}}]{Britto:2004nc}%
  \BibitemOpen
  \bibfield  {author} {\bibinfo {author} {\bibfnamefont {Ruth}\ \bibnamefont
  {Britto}}, \bibinfo {author} {\bibfnamefont {Freddy}\ \bibnamefont
  {Cachazo}}, \ and\ \bibinfo {author} {\bibfnamefont {Bo}~\bibnamefont
  {Feng}},\ }\bibfield  {title} {\enquote {\bibinfo {title} {{Generalized
  unitarity and one-loop amplitudes in N=4 super-Yang-Mills}},}\ }\href
  {\doibase 10.1016/j.nuclphysb.2005.07.014} {\bibfield  {journal} {\bibinfo
  {journal} {Nucl. Phys. B}\ }\textbf {\bibinfo {volume} {725}},\ \bibinfo
  {pages} {275--305} (\bibinfo {year} {2005})},\ \Eprint
  {http://arxiv.org/abs/hep-th/0412103} {arXiv:hep-th/0412103} \BibitemShut
  {NoStop}%
\bibitem [{\citenamefont {Cheung}\ and\ \citenamefont
  {Solon}(2020)}]{Cheung:2020gyp}%
  \BibitemOpen
  \bibfield  {author} {\bibinfo {author} {\bibfnamefont {Clifford}\
  \bibnamefont {Cheung}}\ and\ \bibinfo {author} {\bibfnamefont {Mikhail~P.}\
  \bibnamefont {Solon}},\ }\bibfield  {title} {\enquote {\bibinfo {title}
  {{Classical gravitational scattering at $ \mathcal{O} $(G$^{3}$) from Feynman
  diagrams}},}\ }\href {\doibase 10.1007/JHEP06(2020)144} {\bibfield  {journal}
  {\bibinfo  {journal} {JHEP}\ }\textbf {\bibinfo {volume} {06}},\ \bibinfo
  {pages} {144} (\bibinfo {year} {2020})},\ \Eprint
  {http://arxiv.org/abs/2003.08351} {arXiv:2003.08351 [hep-th]} \BibitemShut
  {NoStop}%
\bibitem [{\citenamefont {K\"alin}\ \emph {et~al.}(2020)\citenamefont
  {K\"alin}, \citenamefont {Liu},\ and\ \citenamefont {Porto}}]{Kalin:2020fhe}%
  \BibitemOpen
  \bibfield  {author} {\bibinfo {author} {\bibfnamefont {Gregor}\ \bibnamefont
  {K\"alin}}, \bibinfo {author} {\bibfnamefont {Zhengwen}\ \bibnamefont {Liu}},
  \ and\ \bibinfo {author} {\bibfnamefont {Rafael~A.}\ \bibnamefont {Porto}},\
  }\bibfield  {title} {\enquote {\bibinfo {title} {{Conservative Dynamics of
  Binary Systems to Third Post-Minkowskian Order from the Effective Field
  Theory Approach}},}\ }\href {\doibase 10.1103/PhysRevLett.125.261103}
  {\bibfield  {journal} {\bibinfo  {journal} {Phys. Rev. Lett.}\ }\textbf
  {\bibinfo {volume} {125}},\ \bibinfo {pages} {261103} (\bibinfo {year}
  {2020})},\ \Eprint {http://arxiv.org/abs/2007.04977} {arXiv:2007.04977
  [hep-th]} \BibitemShut {NoStop}%
\bibitem [{\citenamefont {Di~Vecchia}\ \emph {et~al.}(2021)\citenamefont
  {Di~Vecchia}, \citenamefont {Heissenberg}, \citenamefont {Russo},\ and\
  \citenamefont {Veneziano}}]{DiVecchia:2021bdo}%
  \BibitemOpen
  \bibfield  {author} {\bibinfo {author} {\bibfnamefont {Paolo}\ \bibnamefont
  {Di~Vecchia}}, \bibinfo {author} {\bibfnamefont {Carlo}\ \bibnamefont
  {Heissenberg}}, \bibinfo {author} {\bibfnamefont {Rodolfo}\ \bibnamefont
  {Russo}}, \ and\ \bibinfo {author} {\bibfnamefont {Gabriele}\ \bibnamefont
  {Veneziano}},\ }\bibfield  {title} {\enquote {\bibinfo {title} {{The Eikonal
  Approach to Gravitational Scattering and Radiation at $\mathcal O(G^3)$}},}\
  }\href@noop {} {\  (\bibinfo {year} {2021})},\ \Eprint
  {http://arxiv.org/abs/2104.03256} {arXiv:2104.03256 [hep-th]} \BibitemShut
  {NoStop}%
\bibitem [{\citenamefont {Bjerrum-Bohr}\ \emph {et~al.}(2021)\citenamefont
  {Bjerrum-Bohr}, \citenamefont {Damgaard}, \citenamefont {Plant\'e},\ and\
  \citenamefont {Vanhove}}]{Bjerrum-Bohr:2021din}%
  \BibitemOpen
  \bibfield  {author} {\bibinfo {author} {\bibfnamefont {N.~E.~J.}\
  \bibnamefont {Bjerrum-Bohr}}, \bibinfo {author} {\bibfnamefont {P.~H.}\
  \bibnamefont {Damgaard}}, \bibinfo {author} {\bibfnamefont {L.}~\bibnamefont
  {Plant\'e}}, \ and\ \bibinfo {author} {\bibfnamefont {P.}~\bibnamefont
  {Vanhove}},\ }\bibfield  {title} {\enquote {\bibinfo {title} {{The Amplitude
  for Classical Gravitational Scattering at Third Post-Minkowskian Order}},}\
  }\href@noop {} {\  (\bibinfo {year} {2021})},\ \Eprint
  {http://arxiv.org/abs/2105.05218} {arXiv:2105.05218 [hep-th]} \BibitemShut
  {NoStop}%
\bibitem [{\citenamefont {Damour}(2020{\natexlab{a}})}]{Damour:2019lcq}%
  \BibitemOpen
  \bibfield  {author} {\bibinfo {author} {\bibfnamefont {Thibault}\
  \bibnamefont {Damour}},\ }\bibfield  {title} {\enquote {\bibinfo {title}
  {{Classical and quantum scattering in post-Minkowskian gravity}},}\ }\href
  {\doibase 10.1103/PhysRevD.102.024060} {\bibfield  {journal} {\bibinfo
  {journal} {Phys. Rev. D}\ }\textbf {\bibinfo {volume} {102}},\ \bibinfo
  {pages} {024060} (\bibinfo {year} {2020}{\natexlab{a}})},\ \Eprint
  {http://arxiv.org/abs/1912.02139} {arXiv:1912.02139 [gr-qc]} \BibitemShut
  {NoStop}%
\bibitem [{\citenamefont {Amati}\ \emph {et~al.}(1990)\citenamefont {Amati},
  \citenamefont {Ciafaloni},\ and\ \citenamefont {Veneziano}}]{Amati:1990xe}%
  \BibitemOpen
  \bibfield  {author} {\bibinfo {author} {\bibfnamefont {D.}~\bibnamefont
  {Amati}}, \bibinfo {author} {\bibfnamefont {M.}~\bibnamefont {Ciafaloni}}, \
  and\ \bibinfo {author} {\bibfnamefont {G.}~\bibnamefont {Veneziano}},\
  }\bibfield  {title} {\enquote {\bibinfo {title} {{Higher Order Gravitational
  Deflection and Soft Bremsstrahlung in Planckian Energy Superstring
  Collisions}},}\ }\href {\doibase 10.1016/0550-3213(90)90375-N} {\bibfield
  {journal} {\bibinfo  {journal} {Nucl. Phys. B}\ }\textbf {\bibinfo {volume}
  {347}},\ \bibinfo {pages} {550--580} (\bibinfo {year} {1990})}\BibitemShut
  {NoStop}%
\bibitem [{\citenamefont {Di~Vecchia}\ \emph
  {et~al.}(2020{\natexlab{a}})\citenamefont {Di~Vecchia}, \citenamefont
  {Naculich}, \citenamefont {Russo}, \citenamefont {Veneziano},\ and\
  \citenamefont {White}}]{DiVecchia:2019kta}%
  \BibitemOpen
  \bibfield  {author} {\bibinfo {author} {\bibfnamefont {Paolo}\ \bibnamefont
  {Di~Vecchia}}, \bibinfo {author} {\bibfnamefont {Stephen~G.}\ \bibnamefont
  {Naculich}}, \bibinfo {author} {\bibfnamefont {Rodolfo}\ \bibnamefont
  {Russo}}, \bibinfo {author} {\bibfnamefont {Gabriele}\ \bibnamefont
  {Veneziano}}, \ and\ \bibinfo {author} {\bibfnamefont {Chris~D.}\
  \bibnamefont {White}},\ }\bibfield  {title} {\enquote {\bibinfo {title} {{A
  tale of two exponentiations in $ \mathcal{N} $ = 8 supergravity at subleading
  level}},}\ }\href {\doibase 10.1007/JHEP03(2020)173} {\bibfield  {journal}
  {\bibinfo  {journal} {JHEP}\ }\textbf {\bibinfo {volume} {03}},\ \bibinfo
  {pages} {173} (\bibinfo {year} {2020}{\natexlab{a}})},\ \Eprint
  {http://arxiv.org/abs/1911.11716} {arXiv:1911.11716 [hep-th]} \BibitemShut
  {NoStop}%
\bibitem [{\citenamefont {Bern}\ \emph
  {et~al.}(2020{\natexlab{a}})\citenamefont {Bern}, \citenamefont {Ita},
  \citenamefont {Parra-Martinez},\ and\ \citenamefont {Ruf}}]{Bern:2020gjj}%
  \BibitemOpen
  \bibfield  {author} {\bibinfo {author} {\bibfnamefont {Zvi}\ \bibnamefont
  {Bern}}, \bibinfo {author} {\bibfnamefont {Harald}\ \bibnamefont {Ita}},
  \bibinfo {author} {\bibfnamefont {Julio}\ \bibnamefont {Parra-Martinez}}, \
  and\ \bibinfo {author} {\bibfnamefont {Michael~S.}\ \bibnamefont {Ruf}},\
  }\bibfield  {title} {\enquote {\bibinfo {title} {{Universality in the
  classical limit of massless gravitational scattering}},}\ }\href {\doibase
  10.1103/PhysRevLett.125.031601} {\bibfield  {journal} {\bibinfo  {journal}
  {Phys. Rev. Lett.}\ }\textbf {\bibinfo {volume} {125}},\ \bibinfo {pages}
  {031601} (\bibinfo {year} {2020}{\natexlab{a}})},\ \Eprint
  {http://arxiv.org/abs/2002.02459} {arXiv:2002.02459 [hep-th]} \BibitemShut
  {NoStop}%
\bibitem [{\citenamefont {Di~Vecchia}\ \emph
  {et~al.}(2020{\natexlab{b}})\citenamefont {Di~Vecchia}, \citenamefont
  {Heissenberg}, \citenamefont {Russo},\ and\ \citenamefont
  {Veneziano}}]{DiVecchia:2020ymx}%
  \BibitemOpen
  \bibfield  {author} {\bibinfo {author} {\bibfnamefont {Paolo}\ \bibnamefont
  {Di~Vecchia}}, \bibinfo {author} {\bibfnamefont {Carlo}\ \bibnamefont
  {Heissenberg}}, \bibinfo {author} {\bibfnamefont {Rodolfo}\ \bibnamefont
  {Russo}}, \ and\ \bibinfo {author} {\bibfnamefont {Gabriele}\ \bibnamefont
  {Veneziano}},\ }\bibfield  {title} {\enquote {\bibinfo {title} {{Universality
  of ultra-relativistic gravitational scattering}},}\ }\href {\doibase
  10.1016/j.physletb.2020.135924} {\bibfield  {journal} {\bibinfo  {journal}
  {Phys. Lett. B}\ }\textbf {\bibinfo {volume} {811}},\ \bibinfo {pages}
  {135924} (\bibinfo {year} {2020}{\natexlab{b}})},\ \Eprint
  {http://arxiv.org/abs/2008.12743} {arXiv:2008.12743 [hep-th]} \BibitemShut
  {NoStop}%
\bibitem [{\citenamefont {Damour}(2020{\natexlab{b}})}]{Damour:2020tta}%
  \BibitemOpen
  \bibfield  {author} {\bibinfo {author} {\bibfnamefont {Thibault}\
  \bibnamefont {Damour}},\ }\bibfield  {title} {\enquote {\bibinfo {title}
  {{Radiative contribution to classical gravitational scattering at the third
  order in $G$}},}\ }\href {\doibase 10.1103/PhysRevD.102.124008} {\bibfield
  {journal} {\bibinfo  {journal} {Phys. Rev. D}\ }\textbf {\bibinfo {volume}
  {102}},\ \bibinfo {pages} {124008} (\bibinfo {year} {2020}{\natexlab{b}})},\
  \Eprint {http://arxiv.org/abs/2010.01641} {arXiv:2010.01641 [gr-qc]}
  \BibitemShut {NoStop}%
\bibitem [{\citenamefont {Bini}\ and\ \citenamefont
  {Damour}(2012)}]{Bini:2012ji}%
  \BibitemOpen
  \bibfield  {author} {\bibinfo {author} {\bibfnamefont {Donato}\ \bibnamefont
  {Bini}}\ and\ \bibinfo {author} {\bibfnamefont {Thibault}\ \bibnamefont
  {Damour}},\ }\bibfield  {title} {\enquote {\bibinfo {title} {{Gravitational
  radiation reaction along general orbits in the effective one-body
  formalism}},}\ }\href {\doibase 10.1103/PhysRevD.86.124012} {\bibfield
  {journal} {\bibinfo  {journal} {Phys. Rev. D}\ }\textbf {\bibinfo {volume}
  {86}},\ \bibinfo {pages} {124012} (\bibinfo {year} {2012})},\ \Eprint
  {http://arxiv.org/abs/1210.2834} {arXiv:1210.2834 [gr-qc]} \BibitemShut
  {NoStop}%
\bibitem [{\citenamefont {Herrmann}\ \emph
  {et~al.}(2021{\natexlab{a}})\citenamefont {Herrmann}, \citenamefont
  {Parra-Martinez}, \citenamefont {Ruf},\ and\ \citenamefont
  {Zeng}}]{Herrmann:2021lqe}%
  \BibitemOpen
  \bibfield  {author} {\bibinfo {author} {\bibfnamefont {Enrico}\ \bibnamefont
  {Herrmann}}, \bibinfo {author} {\bibfnamefont {Julio}\ \bibnamefont
  {Parra-Martinez}}, \bibinfo {author} {\bibfnamefont {Michael~S.}\
  \bibnamefont {Ruf}}, \ and\ \bibinfo {author} {\bibfnamefont {Mao}\
  \bibnamefont {Zeng}},\ }\bibfield  {title} {\enquote {\bibinfo {title}
  {{Gravitational Bremsstrahlung from Reverse Unitarity}},}\ }\href@noop {} {\
  (\bibinfo {year} {2021}{\natexlab{a}})},\ \Eprint
  {http://arxiv.org/abs/2101.07255} {arXiv:2101.07255 [hep-th]} \BibitemShut
  {NoStop}%
\bibitem [{\citenamefont {Kosower}\ \emph {et~al.}(2019)\citenamefont
  {Kosower}, \citenamefont {Maybee},\ and\ \citenamefont
  {O'Connell}}]{Kosower:2018adc}%
  \BibitemOpen
  \bibfield  {author} {\bibinfo {author} {\bibfnamefont {David~A.}\
  \bibnamefont {Kosower}}, \bibinfo {author} {\bibfnamefont {Ben}\ \bibnamefont
  {Maybee}}, \ and\ \bibinfo {author} {\bibfnamefont {Donal}\ \bibnamefont
  {O'Connell}},\ }\bibfield  {title} {\enquote {\bibinfo {title} {{Amplitudes,
  Observables, and Classical Scattering}},}\ }\href {\doibase
  10.1007/JHEP02(2019)137} {\bibfield  {journal} {\bibinfo  {journal} {JHEP}\
  }\textbf {\bibinfo {volume} {02}},\ \bibinfo {pages} {137} (\bibinfo {year}
  {2019})},\ \Eprint {http://arxiv.org/abs/1811.10950} {arXiv:1811.10950
  [hep-th]} \BibitemShut {NoStop}%
\bibitem [{\citenamefont {Herrmann}\ \emph
  {et~al.}(2021{\natexlab{b}})\citenamefont {Herrmann}, \citenamefont
  {Parra-Martinez}, \citenamefont {Ruf},\ and\ \citenamefont
  {Zeng}}]{Herrmann:2021tct}%
  \BibitemOpen
  \bibfield  {author} {\bibinfo {author} {\bibfnamefont {Enrico}\ \bibnamefont
  {Herrmann}}, \bibinfo {author} {\bibfnamefont {Julio}\ \bibnamefont
  {Parra-Martinez}}, \bibinfo {author} {\bibfnamefont {Michael~S.}\
  \bibnamefont {Ruf}}, \ and\ \bibinfo {author} {\bibfnamefont {Mao}\
  \bibnamefont {Zeng}},\ }\bibfield  {title} {\enquote {\bibinfo {title}
  {{Radiative Classical Gravitational Observables at $\mathcal O(G^3)$ from
  Scattering Amplitudes}},}\ }\href@noop {} {\  (\bibinfo {year}
  {2021}{\natexlab{b}})},\ \Eprint {http://arxiv.org/abs/2104.03957}
  {arXiv:2104.03957 [hep-th]} \BibitemShut {NoStop}%
\bibitem [{\citenamefont {Bini}\ \emph {et~al.}(2021)\citenamefont {Bini},
  \citenamefont {Damour},\ and\ \citenamefont {Geralico}}]{Bini:2021gat}%
  \BibitemOpen
  \bibfield  {author} {\bibinfo {author} {\bibfnamefont {Donato}\ \bibnamefont
  {Bini}}, \bibinfo {author} {\bibfnamefont {Thibault}\ \bibnamefont {Damour}},
  \ and\ \bibinfo {author} {\bibfnamefont {Andrea}\ \bibnamefont {Geralico}},\
  }\bibfield  {title} {\enquote {\bibinfo {title} {{Radiative contributions to
  gravitational scattering}},}\ }\href@noop {} {\  (\bibinfo {year} {2021})},\
  \Eprint {http://arxiv.org/abs/2107.08896} {arXiv:2107.08896 [gr-qc]}
  \BibitemShut {NoStop}%
\bibitem [{\citenamefont {Brandhuber}\ \emph {et~al.}(2021)\citenamefont
  {Brandhuber}, \citenamefont {Chen}, \citenamefont {Travaglini},\ and\
  \citenamefont {Wen}}]{Brandhuber:2021eyq}%
  \BibitemOpen
  \bibfield  {author} {\bibinfo {author} {\bibfnamefont {Andreas}\ \bibnamefont
  {Brandhuber}}, \bibinfo {author} {\bibfnamefont {Gang}\ \bibnamefont {Chen}},
  \bibinfo {author} {\bibfnamefont {Gabriele}\ \bibnamefont {Travaglini}}, \
  and\ \bibinfo {author} {\bibfnamefont {Congkao}\ \bibnamefont {Wen}},\
  }\bibfield  {title} {\enquote {\bibinfo {title} {{Classical gravitational
  scattering from a gauge-invariant double copy}},}\ }\href@noop {} {\
  (\bibinfo {year} {2021})},\ \Eprint {http://arxiv.org/abs/2108.04216}
  {arXiv:2108.04216 [hep-th]} \BibitemShut {NoStop}%
\bibitem [{\citenamefont {Bern}\ \emph {et~al.}(2021)\citenamefont {Bern},
  \citenamefont {Parra-Martinez}, \citenamefont {Roiban}, \citenamefont {Ruf},
  \citenamefont {Shen}, \citenamefont {Solon},\ and\ \citenamefont
  {Zeng}}]{Bern:2021dqo}%
  \BibitemOpen
  \bibfield  {author} {\bibinfo {author} {\bibfnamefont {Zvi}\ \bibnamefont
  {Bern}}, \bibinfo {author} {\bibfnamefont {Julio}\ \bibnamefont
  {Parra-Martinez}}, \bibinfo {author} {\bibfnamefont {Radu}\ \bibnamefont
  {Roiban}}, \bibinfo {author} {\bibfnamefont {Michael~S.}\ \bibnamefont
  {Ruf}}, \bibinfo {author} {\bibfnamefont {Chia-Hsien}\ \bibnamefont {Shen}},
  \bibinfo {author} {\bibfnamefont {Mikhail~P.}\ \bibnamefont {Solon}}, \ and\
  \bibinfo {author} {\bibfnamefont {Mao}\ \bibnamefont {Zeng}},\ }\bibfield
  {title} {\enquote {\bibinfo {title} {{Scattering Amplitudes and Conservative
  Binary Dynamics at ${\cal O}(G^4)$}},}\ }\href {\doibase
  10.1103/PhysRevLett.126.171601} {\bibfield  {journal} {\bibinfo  {journal}
  {Phys. Rev. Lett.}\ }\textbf {\bibinfo {volume} {126}},\ \bibinfo {pages}
  {171601} (\bibinfo {year} {2021})},\ \Eprint
  {http://arxiv.org/abs/2101.07254} {arXiv:2101.07254 [hep-th]} \BibitemShut
  {NoStop}%
\bibitem [{\citenamefont {Dlapa}\ \emph {et~al.}(2021)\citenamefont {Dlapa},
  \citenamefont {K\"alin}, \citenamefont {Liu},\ and\ \citenamefont
  {Porto}}]{Dlapa:2021npj}%
  \BibitemOpen
  \bibfield  {author} {\bibinfo {author} {\bibfnamefont {Christoph}\
  \bibnamefont {Dlapa}}, \bibinfo {author} {\bibfnamefont {Gregor}\
  \bibnamefont {K\"alin}}, \bibinfo {author} {\bibfnamefont {Zhengwen}\
  \bibnamefont {Liu}}, \ and\ \bibinfo {author} {\bibfnamefont {Rafael~A.}\
  \bibnamefont {Porto}},\ }\bibfield  {title} {\enquote {\bibinfo {title}
  {{Dynamics of Binary Systems to Fourth Post-Minkowskian Order from the
  Effective Field Theory Approach}},}\ }\href@noop {} {\  (\bibinfo {year}
  {2021})},\ \Eprint {http://arxiv.org/abs/2106.08276} {arXiv:2106.08276
  [hep-th]} \BibitemShut {NoStop}%
\bibitem [{\citenamefont {Westpfahl}(1985)}]{Westpfahl:1985tsl}%
  \BibitemOpen
  \bibfield  {author} {\bibinfo {author} {\bibfnamefont {Konradin}\
  \bibnamefont {Westpfahl}},\ }\bibfield  {title} {\enquote {\bibinfo {title}
  {{High-Speed Scattering of Charged and Uncharged Particles in General
  Relativity}},}\ }\href {\doibase 10.1002/prop.2190330802} {\bibfield
  {journal} {\bibinfo  {journal} {Fortsch. Phys.}\ }\textbf {\bibinfo {volume}
  {33}},\ \bibinfo {pages} {417--493} (\bibinfo {year} {1985})}\BibitemShut
  {NoStop}%
\bibitem [{\citenamefont {Mogull}\ \emph {et~al.}(2021)\citenamefont {Mogull},
  \citenamefont {Plefka},\ and\ \citenamefont {Steinhoff}}]{Mogull:2020sak}%
  \BibitemOpen
  \bibfield  {author} {\bibinfo {author} {\bibfnamefont {Gustav}\ \bibnamefont
  {Mogull}}, \bibinfo {author} {\bibfnamefont {Jan}\ \bibnamefont {Plefka}}, \
  and\ \bibinfo {author} {\bibfnamefont {Jan}\ \bibnamefont {Steinhoff}},\
  }\bibfield  {title} {\enquote {\bibinfo {title} {{Classical black hole
  scattering from a worldline quantum field theory}},}\ }\href {\doibase
  10.1007/JHEP02(2021)048} {\bibfield  {journal} {\bibinfo  {journal} {JHEP}\
  }\textbf {\bibinfo {volume} {02}},\ \bibinfo {pages} {048} (\bibinfo {year}
  {2021})},\ \Eprint {http://arxiv.org/abs/2010.02865} {arXiv:2010.02865
  [hep-th]} \BibitemShut {NoStop}%
\bibitem [{\citenamefont {Jakobsen}\ \emph {et~al.}(2021)\citenamefont
  {Jakobsen}, \citenamefont {Mogull}, \citenamefont {Plefka},\ and\
  \citenamefont {Steinhoff}}]{Jakobsen:2021smu}%
  \BibitemOpen
  \bibfield  {author} {\bibinfo {author} {\bibfnamefont {Gustav~Uhre}\
  \bibnamefont {Jakobsen}}, \bibinfo {author} {\bibfnamefont {Gustav}\
  \bibnamefont {Mogull}}, \bibinfo {author} {\bibfnamefont {Jan}\ \bibnamefont
  {Plefka}}, \ and\ \bibinfo {author} {\bibfnamefont {Jan}\ \bibnamefont
  {Steinhoff}},\ }\bibfield  {title} {\enquote {\bibinfo {title} {{Classical
  Gravitational Bremsstrahlung from a Worldline Quantum Field Theory}},}\
  }\href {\doibase 10.1103/PhysRevLett.126.201103} {\bibfield  {journal}
  {\bibinfo  {journal} {Phys. Rev. Lett.}\ }\textbf {\bibinfo {volume} {126}},\
  \bibinfo {pages} {201103} (\bibinfo {year} {2021})},\ \Eprint
  {http://arxiv.org/abs/2101.12688} {arXiv:2101.12688 [gr-qc]} \BibitemShut
  {NoStop}%
\bibitem [{\citenamefont {K\"alin}\ and\ \citenamefont
  {Porto}(2020{\natexlab{a}})}]{Kalin:2019rwq}%
  \BibitemOpen
  \bibfield  {author} {\bibinfo {author} {\bibfnamefont {Gregor}\ \bibnamefont
  {K\"alin}}\ and\ \bibinfo {author} {\bibfnamefont {Rafael~A.}\ \bibnamefont
  {Porto}},\ }\bibfield  {title} {\enquote {\bibinfo {title} {{From Boundary
  Data to Bound States}},}\ }\href {\doibase 10.1007/JHEP01(2020)072}
  {\bibfield  {journal} {\bibinfo  {journal} {JHEP}\ }\textbf {\bibinfo
  {volume} {01}},\ \bibinfo {pages} {072} (\bibinfo {year}
  {2020}{\natexlab{a}})},\ \Eprint {http://arxiv.org/abs/1910.03008}
  {arXiv:1910.03008 [hep-th]} \BibitemShut {NoStop}%
\bibitem [{\citenamefont {K\"alin}\ and\ \citenamefont
  {Porto}(2020{\natexlab{b}})}]{Kalin:2019inp}%
  \BibitemOpen
  \bibfield  {author} {\bibinfo {author} {\bibfnamefont {Gregor}\ \bibnamefont
  {K\"alin}}\ and\ \bibinfo {author} {\bibfnamefont {Rafael~A.}\ \bibnamefont
  {Porto}},\ }\bibfield  {title} {\enquote {\bibinfo {title} {{From boundary
  data to bound states. Part II. Scattering angle to dynamical invariants (with
  twist)}},}\ }\href {\doibase 10.1007/JHEP02(2020)120} {\bibfield  {journal}
  {\bibinfo  {journal} {JHEP}\ }\textbf {\bibinfo {volume} {02}},\ \bibinfo
  {pages} {120} (\bibinfo {year} {2020}{\natexlab{b}})},\ \Eprint
  {http://arxiv.org/abs/1911.09130} {arXiv:1911.09130 [hep-th]} \BibitemShut
  {NoStop}%
\bibitem [{\citenamefont {Bini}\ \emph {et~al.}(2020)\citenamefont {Bini},
  \citenamefont {Damour},\ and\ \citenamefont {Geralico}}]{Bini:2020hmy}%
  \BibitemOpen
  \bibfield  {author} {\bibinfo {author} {\bibfnamefont {Donato}\ \bibnamefont
  {Bini}}, \bibinfo {author} {\bibfnamefont {Thibault}\ \bibnamefont {Damour}},
  \ and\ \bibinfo {author} {\bibfnamefont {Andrea}\ \bibnamefont {Geralico}},\
  }\bibfield  {title} {\enquote {\bibinfo {title} {{Sixth post-Newtonian
  nonlocal-in-time dynamics of binary systems}},}\ }\href {\doibase
  10.1103/PhysRevD.102.084047} {\bibfield  {journal} {\bibinfo  {journal}
  {Phys. Rev. D}\ }\textbf {\bibinfo {volume} {102}},\ \bibinfo {pages}
  {084047} (\bibinfo {year} {2020})},\ \Eprint
  {http://arxiv.org/abs/2007.11239} {arXiv:2007.11239 [gr-qc]} \BibitemShut
  {NoStop}%
\bibitem [{\citenamefont {Junker}\ and\ \citenamefont
  {Sch\"afer}(1992)}]{Junker:1992kle}%
  \BibitemOpen
  \bibfield  {author} {\bibinfo {author} {\bibfnamefont {Wolfgang}\
  \bibnamefont {Junker}}\ and\ \bibinfo {author} {\bibfnamefont {Gerhard}\
  \bibnamefont {Sch\"afer}},\ }\bibfield  {title} {\enquote {\bibinfo {title}
  {{Binary systems: higher order gravitational radiation damping and wave
  emission}},}\ }\href {\doibase 10.1093/mnras/254.1.146} {\bibfield  {journal}
  {\bibinfo  {journal} {Mon. Not. Roy. Astron. Soc.}\ }\textbf {\bibinfo
  {volume} {254}},\ \bibinfo {pages} {146--164} (\bibinfo {year}
  {1992})}\BibitemShut {NoStop}%
\bibitem [{\citenamefont {Bern}\ \emph
  {et~al.}(2020{\natexlab{b}})\citenamefont {Bern}, \citenamefont {Luna},\ and\
  \citenamefont {Gatica}}]{BernPrivate}%
  \BibitemOpen
  \bibfield  {author} {\bibinfo {author} {\bibfnamefont {Zvi}\ \bibnamefont
  {Bern}}, \bibinfo {author} {\bibfnamefont {Andr\'es}\ \bibnamefont {Luna}}, \
  and\ \bibinfo {author} {\bibfnamefont {Juan~Pablo}\ \bibnamefont {Gatica}},\
  }\bibfield  {title} {\enquote {\bibinfo {title} {{Private communication}},}\
  }\href@noop {} {\  (\bibinfo {year} {2020}{\natexlab{b}})}\BibitemShut
  {NoStop}%
\bibitem [{\citenamefont {Poisson}(1999)}]{Poisson:1999tv}%
  \BibitemOpen
  \bibfield  {author} {\bibinfo {author} {\bibfnamefont {Eric}\ \bibnamefont
  {Poisson}},\ }\bibfield  {title} {\enquote {\bibinfo {title} {{An
  Introduction to the Lorentz-Dirac equation}},}\ }\href@noop {} {\  (\bibinfo
  {year} {1999})},\ \Eprint {http://arxiv.org/abs/gr-qc/9912045}
  {arXiv:gr-qc/9912045} \BibitemShut {NoStop}%
\bibitem [{\citenamefont {Vines}\ \emph {et~al.}(2019)\citenamefont {Vines},
  \citenamefont {Steinhoff},\ and\ \citenamefont {Buonanno}}]{Vines:2018gqi}%
  \BibitemOpen
  \bibfield  {author} {\bibinfo {author} {\bibfnamefont {Justin}\ \bibnamefont
  {Vines}}, \bibinfo {author} {\bibfnamefont {Jan}\ \bibnamefont {Steinhoff}},
  \ and\ \bibinfo {author} {\bibfnamefont {Alessandra}\ \bibnamefont
  {Buonanno}},\ }\bibfield  {title} {\enquote {\bibinfo {title}
  {{Spinning-black-hole scattering and the test-black-hole limit at second
  post-Minkowskian order}},}\ }\href {\doibase 10.1103/PhysRevD.99.064054}
  {\bibfield  {journal} {\bibinfo  {journal} {Phys. Rev. D}\ }\textbf {\bibinfo
  {volume} {99}},\ \bibinfo {pages} {064054} (\bibinfo {year} {2019})},\
  \Eprint {http://arxiv.org/abs/1812.00956} {arXiv:1812.00956 [gr-qc]}
  \BibitemShut {NoStop}%
\bibitem [{\citenamefont {Blanchet}\ and\ \citenamefont
  {Schaefer}(1989)}]{1989MNRAS239845B}%
  \BibitemOpen
  \bibfield  {author} {\bibinfo {author} {\bibfnamefont {Luc}\ \bibnamefont
  {Blanchet}}\ and\ \bibinfo {author} {\bibfnamefont {Gerhard}\ \bibnamefont
  {Schaefer}},\ }\bibfield  {title} {\enquote {\bibinfo {title} {{Higher order
  gravitational radiation losses in binary systems}},}\ }\href {\doibase
  10.1093/mnras/239.3.845} {\bibfield  {journal} {\bibinfo  {journal} {Monthly
  notices of the royal astronomical society}\ }\textbf {\bibinfo {volume}
  {239}},\ \bibinfo {pages} {845--867} (\bibinfo {year} {1989})}\BibitemShut
  {NoStop}%
\end{thebibliography}

%

\end{document}